\documentstyle[12pt,psfig]{article}

\newtheorem{lemma}{Lemma}

\newtheorem{theor}{\large\bf Theorem}

\def\FF{\hbox to 8.33887pt{\rm I\hskip-1.8pt F}}
\def\NN{\hbox to 9.3111pt{\rm I\hskip-1.8pt N}}
\def\PP{\hbox to 8.61664pt{\rm I\hskip-1.8pt P}}
\def\QQ{\rlap {\raise 0.4ex \hbox{$\scriptscriptstyle |$}}
{\hskip -4.5pt Q}}
\def\RR{\hbox to 9.1722pt{\rm I\hskip-1.8pt R}}
\def\ZZ{\hbox to 8.2222pt{\rm Z\hskip-4pt \rm Z}}

\renewcommand{\thesection}{\Roman{section}}

\newcommand{\resetsect}{\setcounter{section}{1}}

\newcommand{\resetequ}{\setcounter{equation}{0}}
\newcommand{\tree}{{\cal T}}           
\newcommand{\C}{{\bf C}}           

\newcommand{\be}{\begin{equation}}
\newcommand{\ee}{\end{equation}}
\newcommand{\bqa}{\begin{eqnarray}}
\newcommand{\eqa}{\end{eqnarray}}
\newcommand{\ba}{\begin{array}}
\newcommand{\ea}{\end{array}}
\newcommand{\p}[1]{{\partial\over \partial{#1}}}
\newcommand{\no}{\nonumber}

\newcommand{\lp}{\left (}
\newcommand{\rp}{\right )}
\newcommand{\qed}{\hfill \rule {1ex}{1ex}}
\newcommand{\al}{\alpha}
\newcommand{\bt}{\beta}

\newcommand{\de}{\delta}
\newcommand{\vep}{\varepsilon}

\newcommand{\th}{\theta}

\newcommand{\la}{\lambda}

\newcommand{\Om}{\Omega}
\newcommand{\Si}{\Sigma}
\newcommand{\La}{\Lambda}
\newcommand{\Lazero}{{\Lambda_{0}}}

\newcommand{\Ga}{\Gamma}
\newcommand{\De}{\Delta}

\newcommand{\bpsi}{\bar{\psi}}

\begin{document}

\centerline{\large \bf Interacting Fermi liquid}
\centerline{\large \bf in two dimensions
at finite temperature}\centerline{\large \bf Part II: Renormalization}
\vskip 2cm

\centerline{M. Disertori and V. Rivasseau}

\centerline{Centre de Physique Th{\'e}orique, CNRS UPR 14}
\centerline{Ecole Polytechnique}
\centerline{91128 Palaiseau Cedex, FRANCE}

\vskip 1cm
\medskip
\noindent{\bf Abstract}

This is a companion paper to [DR1].
Using the method of continuous renormalization group around the Fermi
surface and the results of [DR1], 
we achieve the proof that a two-dimensional 
jellium system of interacting Fermions at low temperature $T$
is a Fermi liquid above the BCS temperature
Following [S1], this means proving 
analyticity in the coupling constant  $\lambda$ 
for $ |\lambda| |\log T| \le K  $ where $K$ is some numerical 
constant, and some uniform bounds on the derivatives of the self-energy.

\section{Introduction}
\resetequ

For general introduction we refer to the [DR1] paper. 
We assume all its results and notations. In  [DR1] 
the ``convergent contributions'' to the vertex functions of a two-dimensional
weakly interacting Fermi liquid were controlled hence
the results of [FMRT] were essentially reproduced but with a
continuous renormalization group analysis, as advocated in [S1]. 
In this paper we consider the complete sum 
of all graphs, perform renormalization of the two point subgraphs and obtain
our main theorem. This is not a trivial extension of the methods
of [FMRT] and [FT1-2], since renormalization has to be performed
in phase space, not momentum space. This raises a delicate point:
since angular sector decomposition has to be anisotropic [FMRT], it is not
obvious that one gains anything by renormalizing in phase space, if the
sector directions of the spanning tree used for spatial integration do
not match the sector directions of the external legs. 
This non-trivial problem is solved here by a somewhat delicate one-particle
irreducibility analysis for two point subgraphs that must respect the
determinant structure of the Fermionic loop variables and Gram's inequality.

Here we go.

\section{Renormalization}
\resetequ

We consider now the sum over all (not necessarily convergent)
attributions. By [DR1], eq.(IV.51-53)
the four point and two point subgraphs are convergent 
at finite temperature,
but  diverge logarithmically and linearly respectively when  
$T\rightarrow 0$. Remark that, as we keep $T\geq T_c >0$, we could avoid 
performing renormalization at all,  but in this case the estimation of
the convergence radius would be bad. Actually, we would have to bound 
a sum such as
\be
\sum_{n=1}^\infty \sum_{n_4+n_2\leq n}
|\la|^n K_2^{n} |\log w_T|^{n_4} \ w_T^{-{n_2\over 2}}
\ee
where $n_4$ and $n_2$ are the number of four point and two point 
subgraphs respectively. Since $n_{2}+n_{4}\le n$
it is easy to check that the convergence radius of 
this sum is defined by the upper bound on the critical temperature
\be
T^{upper}_c= \max \left [T_c^{(4)} ,T_c^{(2)} \right ] =  
\frac{1}{\pi\sqrt{2}} \max \left [
e^{-{1\over |\la|2K_2  }} , \;  (|\la|2 K_2)\right ] =  
\frac{|\la|2 K_2}{\pi\sqrt{2}}\ .
\ee
Actually one can do slightly better and find a bound in $|\la|^{2}$,
because tadpoles vanish, so that one has effectively  $n_{2}\le n/2$.
But we see that without renormalization of the two point subgraphs,
we cannot get an upper bound on the critical temperature
of the non-perturbative form predicted by the theory of 
superconductivity\footnote{We recall that in dimension $d=2$ by 
the Mermin-Wagner theorem
there is no continuous symmetry breaking at finite temperature, but there 
ought to be a critical temperature associated to a Kosterlitz Thouless
phase. At zero temperature, there are three non compact dimensions
(space plus imaginary time) and there should be a continuous symmetry 
breaking with an associated Goldstone boson.}, namely:
\be
T^{true}_c \simeq C_{1} e^{-{1\over C_{2} |\la |}}\ .
\label{Tc}\ee 
where $C_{1}$ and $C_{2}$ are two constants
related to the physical parameters of the model 
such as the Debye frequency, 
the electron mass, the interatomic distance, and the particular 
crystalline lattice structure.

Our goal in this paper is to prove an upper bound
on $T_{c}$ i.e. give a value of $T^{upper}_c$  
which is non-perturbative like (\ref{Tc})
but with different constants $K_{1}$ and $K_{2}$.
To obtain this behavior 
we need to perform renormalization, but only for two-point subgraphs,
which amounts to a computation of the flow of the chemical potential only 
\footnote{To find the exact constant 
$K_{1}= C_1$ in our Theorem 3 is trivial, 
but to find a bound with the exact constant $K_{2}= C_2$
requires to compute the flows of the coupling constant also. This is 
almost certainly also doable within the methods of this paper,
but introduces some painful complications, since there are really
infinitely many running coupling constants [FT2]}. 

Hence in this paper we will use the interacting action 
\be
S_V = \frac{\la}{2} \int_V d^3x\; \lp\sum_a \bpsi\psi\rp^2
\ + \ \de \mu^1_\La  \int_V d^3x\; \lp\sum_a \bpsi\psi\rp
\label{actionr}\ee
where $\la$ is the bare coupling constant and $\de \mu^1_\La$ is the bare 
chemical potential  counterterm, which is function of the 
ultraviolet cut-off $\La_0=1$ and the infrared cut-off 
$\La$. The free covariance is as usual
\be
\hat{C}_{ab}(k) =\de_{ab} \ \frac{1}{ik_0-\lp\vec{k}^2-\mu\rp}\ ,
\label{tfpropr}\ee
where $\mu=1$ is the renormalized chemical potential and we have taken 
$2m=1$.  
The BPHZ condition states

\be
\de\mu_{ren}(\La)\; = \;\de \mu^\La_\La \;=\;
\hat{\Si}^{\La}(k_F)\;
=\; \int d^3x \; e^{-ik_F x}\;\Si^{\La}(0,x)\; =\; 0
\ee
where $\Si^{\La}$ is the two point vertex function 
 $\Ga_2^{\La_0,\La}(x_1,x_2)$ ($\La_0=1$),
 $k_f$ is some vector as near as possible  to the Fermi surface, 
(the Fermi surface cannot be reached at finite temperature 
because of the antiperiodicity of  Fermions) hence with
$|k_{F0}|=\pi T$ and $|\vec{k}_F|=1$. 
This function actually coincides with the 1PI one, as the
Gevrey cut-off on internal lines fixes the 1PR contributions to zero.
 By rotation invariance, this condition
does not depend on the angular part of $\vec k_{F}$. On the other 
hand, to conserve the parity in the imaginary time direction we should take 
the mean value
$1/2[\hat{\Si}^\La(\pi T,\vec{k}_F)+\hat{\Si}^\La(-\pi T,\vec{k}_F)]$, but in 
our computations this is not necessary.
The main result of our paper is

\begin{theor}
The limit $\La\rightarrow 0$ of $\Ga_{2p}^{\La\La_0}(\phi_1,...\phi_{2p})$
is analytic in the bare coupling constant $\la$, for all values of
$\la\in \C$ such that $|\la| \le c$, with 
$c$ given by the equivalent relations
\[
T = K_1 e^{-{1\over c K_2 }}\quad ; \quad   
c= {1\over K_2 |\log T/K_{1}| }
\]
for some constants $K_1$ and $K_2$ (this relation being limited to the
interesting low temperature regime $ T/K_{1} < 1$).
\end{theor}
This theorem is in a sense a generalization
of [DR1], Theorem 1, and the remaining part of this section is devoted 
to its proof. 

With the new action (\ref{actionr}) the expression to bound 
becomes:
\bqa
&&\hspace{-0.5cm}\Ga_{2p}^{\La\Lazero}(\phi_1,...\phi_{2p})=
\sum_{{\bar n}\geq 1}
\frac{\la^n}{n!} \frac{\lp \de\mu^1_\La \rp^{n'}}{n'!}
\sum_{o-\tree}\sum_{E}\sum_{\Om}  \vep(\tree, \Om)\int
d^3x_1...d^3x_{\bar n}\label{sviluppo2r}\\
&&\hspace{-0.5cm}\phi^{\La_T}_1(x_{i_{1}})...\phi^{\La_T}_{2p}(x_{j_{p}})\; 
\int_{w_T\le w_{1} \le ...\le w_{{\bar n}-1}\le 1}
\left [\prod_{q=1}^{{\bar n}-1} \p{w_{q}}  C_{\La}^{\La(w_q)}
(x_{l_{q}}, {\bar x}_{l_{q}})dw_q \right]\det{\cal M}(E)\no
\eqa
where $n$ is the number of four point vertices (with coupling constant 
$\la$), $n'$ is the number of two point vertices (with coupling constant 
$\delta \mu^1_\La$) and we defined ${\bar n}= n+n'$.
Now, we can insert band attributions and classes exactly as in 
[DR1]. 

\subsection{Extracting loop lines.}
Before introducing sectors, we must perform an additional expansion  
of the loop determinant. This is necessary for two reasons:
\begin{itemize}
\item{}
to select the two-point subgraphs that really need renormalization;
\item{} to optimize sector counting by reducing 
the number of possible sector choices, in order to perform renormalization.
\end{itemize}

\begin{figure}
\centerline{\psfig{figure=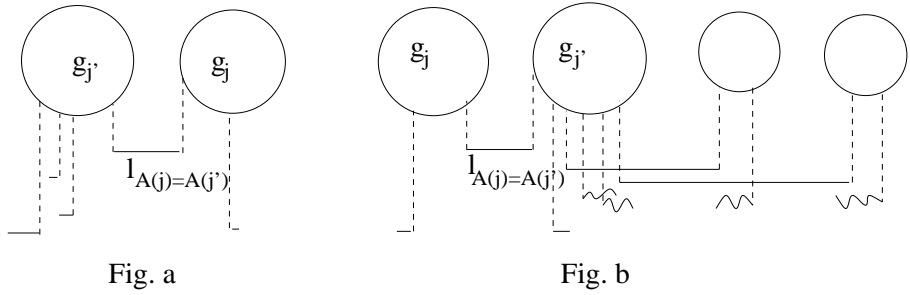,width=12cm}}
\caption{examples of subgraphs not belonging to $D_{\cal C}$; tree lines 
are solid and 
loop fields are wavy.}\label{rinorm2}
\end{figure}

We introduce some notations. 
For any class ${\cal C}$  we define $D_{\cal C}$ as the set of 
``potentially dangerous'' two-point 
subgraphs $g_i$. They are defined by the following property: 
by cutting a single tree line on the path 
joining the two external vertices of $g_i$ we cannot separate  $g_i$ 
into two disconnected subgraphs 
$g_j({\cal C})$ and $g_{j'}({\cal C})$, one of them, say $g_j({\cal C})$, 
being a two point subgraph.   This property is similar but not equal to 1PI 
(one particle irreducibility).  In Fig.\ref{rinorm2} there are some examples 
of subgraphs not belonging to $D_{\cal C}$.

\begin{figure}
\centerline{\psfig{figure=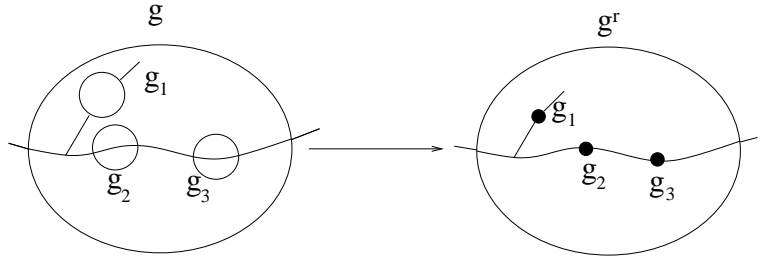,width=10cm}}
\caption{a subgraph $g$ and the  reduced correspondent subgraph $g^r$;
$g_1$, $g_2$ and $g_3$ belong to ${\cal A}(g)$\label{fermi19}}
\end{figure}

By the relation of partial order in the $CTS$,  $D_{\cal C}$ has a forest 
structure (see [R]). This means that for any pair $g$ and $g'\in D_{\cal C}$ 
we have $g\cap g'=\emptyset$ or $g\subseteq g'$ or $g'\subseteq g$. 
Now, for any $g\in D_{\cal C}$, we define the set ${\cal A}(g)$ of maximal 
subgraphs $g'\in D_{\cal C}$, $g'\subset g$. The 
loop determinant is then factorized on the product of several terms: one
for each set  $il_j({\cal C})$, $g_j\in {\cal A}(g)$, one containing  
the remaining internal loop fields in $g_i$, and a last term containing all 
the other loop fields. Then, the good object to study is not $g$, but 
the reduced graph $g^r:= g/D_{\cal C}$, where each $g_j\in {\cal A}(g)$ has
been reduced to a single point (see Fig.\ref{fermi19}).
For each $g_i^r$ we denote the set of internal loop half-lines by $il_i^r$ 
and the set of vertices by $V_i^r$. 

\begin{figure}
\centerline{\psfig{figure=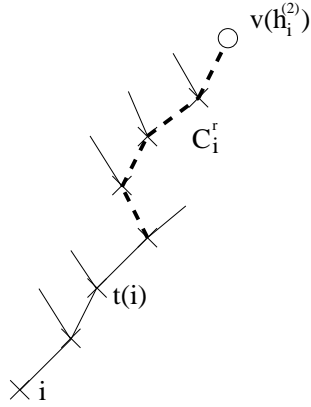,width=4cm}}
\caption{example of $C_i^r$: the dashed lines belong to the chain}
\label{fermi9}
\end{figure}

For each $g^r_i$, $g_i\in D_{\cal C}$, we
call $h^{(1)}_i$ the external half-line $h^{root}_i$ and 
$h^{(2)}_i$ the other external half-line. In the same way we define 
$v^{(1)}_i$ and  $v^{(2)}_i$.  With these definitions, 
we introduce the chain $C^r_{i}$ which
joins the dot vertex $v_{h^{(2)}_{i}}$ to the cross vertex just 
above the cross $t(i)$ (see Fig.\ref{fermi9}). 

\begin{figure}
\centerline{\psfig{figure=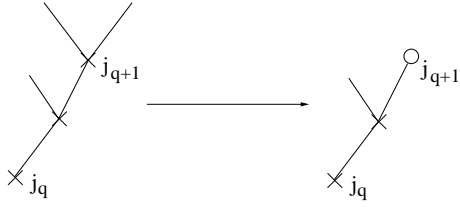,width=6cm}}
\caption{reduction of $g_{j_{q+1}}$ in  $g_{j_{q}}$}
\label{fermi20}
\end{figure}

On this chain we define the set $J_i$ of crosses 
(and eventually one dot) indices $j$ corresponding to
four-point subgraphs $|eg_j({\cal C})|=4$.  We order them starting from the
lowest index $j_1$ and going up to the highest $j_{|J_i|}$.  
Remark that, by definition of $D_{\cal C}$, there is no
index $j$ on the chain with   $|eg_j({\cal C})|=2$. 
Again we introduce  the reduced 
subgraphs $g_{j_q}^r({\cal C}):= g_{j_q}({\cal C})/g_{j_{q+1}}({\cal C})$
(see Fig.\ref{fermi20}),
the set of internal loop half-lines of $g_{j_q}^r$, $il_{j_q}^r$, and
that of internal vertices, $V_{j_q}^r$. Then 
the corresponding  loop determinant is factorized
\be
\det( il_{j_q}) = \det(il^r_{j_q}) \det (il_{j_{q+1}}^r)\ . \label{factordet}
\ee
For the first step of the induction 
we define $j_0=i$, $g_{j_0}:=g^r_i$  (it is a two point subgraph!) and
$g^{r}_{j_0}:= g^{r}_{j_0}/g^{r}_{j_1}$. 

For each $g^r_{j_q}$, $q=1,..,|J_i|$ we call $h^{(1)}_{j_q}$ the 
external tree half-line $h^{root}_{j_q}$, $l_{j_q}^{root}$ 
the corresponding tree line, $h^{(2)}_{j_q}=h^{(2)}_i$ 
(remark that $h^{root}_{j_q}$ can never 
coincide with $h^{(2)}_i$ by construction), and $h^{(3)}_{j_q}$, 
$h^{(4)}_{j_q}$ the remaining two external  half-lines.  
The line $l_{j_q}^{root}$ cuts the  tree $t_i$ into
two connected components. We call $\tree_{j_q}^L(i)$ the
component that  contains the  vertex $x^{(1)}_{i}$, and   
$\tree^R_{j_q}(i)$ the other component, 
that  contains the  vertex $x^{(2)}_{i}$. 
Remark that all vertices in $g^r_{j_q}$,  belong to
$\tree^R_{j_q}(i)$.

For each $q=1,..|J_i|$ (starting from the lowest and going up) 
we test if there is some loop line  $l_{fg}$ with 
$f,g\in il^r_{i}$ connecting $\tree^L_{j_q}(i)$ with  $\tree^R_{j_q}(i)$. 
If for some $j_q\in J_i$ there is no loop line $g_i$ is actually 
1PR (one particle reducible) and, by momentum conservation, it does not need 
to be renormalized (as it is shown below). On the other hand, if  
$\forall j\in J_i$ we can find 
a loop line, then $g_i$ is 1PI and it must be renormalized.

We perform this test inductively. At each subgraph $g_{j_q}$ we define 
\bqa
L^R_{j_q}(i) &:=& \{ a\in il_{j_{q-1}}^r| a\in \tree^R_{j_q}(i) ), 
m(a,{\cal C})\leq i(l_{j_q}^{root}) \leq {\cal A}(j_{q})\} \no\\
L^L_{j_q}(i) &:=& \{ a\in il_{j_{q-1}}^r| a\in \tree^L_{j_q}(i) ), 
m(a,{\cal C})\leq i(l_{j_q}^{root}) \leq {\cal A}(j_{q})\} \ , \no\\
\eqa
(where we recall that ${\cal A}(j_{q})$) (defined in [DR1]) is the index
of the highest external tree line of $g_{j_{q}}^{r}$).
Actually, $L^R_{j_q}(i)$ is the set of internal loop
half-lines of $g^r_{j_{q-1}}$ which are hooked to  $\tree^R_{j_q}(i)$ 
and may connect somewhere in $\tree^L_{j_q}(i)$.  
By construction, no internal loop  half-line
of $g^r_{j_q}$ and no external loop half-line of $g^r_{j_{q-1}}$ 
belongs to  $L^L_{j_q}(i)\cup L^R_{j_q}(i)$. 
This is the main reason for which this expansion
does not develop any new factorial. 

We distinguish three  situations:

\paragraph{1.} 
 $h^{(3)}_{j_q}$ and $h^{(4)}_{j_q}\in L$
(see Fig.\ref{fermi13}). Then $l^{root}_{j_q}=l_{{\cal A}(j_q)}$,  
$L^R_{j_q}(i)$ is reduced to two elements and 
we develop the determinant to chose where they contract, applying 
two times the following formula:
\be
\det {\cal M}' = \sum_{a} {\cal M}'_{h^{(3)}_{j_q},{a}}  
\vep(h^{(3)}_{j_q},a) \det {\cal M}'_{red}. \label{loop1}
\ee
where $\vep(h^{(3)}_{j_q},a)$ is a sign and  
$\det {\cal M}'_{red}$ is the determinant of the reduced matrix obtained 
taking away a row and a column.  
If they contract together $g_i$ is 1PR. If not,  we have $|L^L_{j_q}(i)|^2$ 
choices to contract them. 
Remark that  if $h^{(3)}_{j_q}$ or $h^{(4)}_{j_q}$, or both are external
lines at some $g_{j_{q'}}$, with $q'<q$, then they have already 
been extracted from the determinant and we do not touch them. 

\paragraph{2.} 
 $h^{(3)}_{j_q}\in L$ and   $h^{(4)}_{j_q}\in t_i$
(see Fig.\ref{fermi14}). 

If $h^{(3)}_{j_q}$ has not been already contracted at some 
lower scale, we develop the determinant as before
to chose where  $h^{(3)}_{j_q}$ contracts. 
If $h^{(3)}_{j_q}$ has already been  contracted at some 
lower scale we do not touch it. 

In any case, if $h^{(3)}_{j_q}$  
contracts with some element of  $L^L_{j_q}(i)$, 
then 1PI is assured and we go to the step $q+1$. 
If not (Fig.\ref{fermi15}a), we test the loop determinant in the 
following way:
\be
\det{\cal M}'({\cal C}) = \det{\cal M}'({\cal C})(0) + 
\int_0^1 ds_{j_q} \;\frac{d}{ds_{j_q}}  
\det {\cal M}'({\cal C})(s_{j_q}).
\label{test1}\ee
where we defined 
\be
{\cal M}'_{x_f,x_g}({\cal C})(s_{j_q}) = 
s_{j_q}\; {\cal M}'_{x_f,x_g}({\cal C})\qquad 
s_{j_q}\in [0,1]
\label{test2}\ee
if $(f,g)$ or $(g,f)$ belong to $ L^R_{j_q}(i)\times L^L_{j_q}(i)$ and 
\be
{\cal M}'_{x_f,x_g}({\cal C})(s_{j_q}) = {\cal M}'_{x_f,x_g}({\cal C})
\ee
otherwise. The term  $s\neq 0$ extracts from the determinant  
the loop line we wanted (see Fig.\ref{fermi15}a).
The term $s=0$ means that  $g_i$ is 1PR. 
The number of choices is bounded by $|L^R_{j_q}(i)|^2|L^L_{j_q}(i)| $. 

\paragraph{3.} 
$h^{(3)}_{j_q}$ and   $h^{(4)}_{j_q}\in t_i$. 
Then we apply directly the interpolation formulas (\ref{test1})-(\ref{test2}). 
Again we distinguish the case $s=0$, that corresponds to $g_i$ 1PR, 
and the case 
$s\neq 0$ that corresponds to $g_i$ 1PI and  has at most 
$|L^L_{j_q}(i)||L^R_{j_q}(i)| $ terms (see Fig.\ref{fermi16}a). 

Repeating the same procedure for all $j\in J_i$ we extract from the
loop determinant  at most $2|J_i|$ internal loop line propagators. 
For each class ${\cal C}$, the  process $J^0_i$ specifies
the set of  $j_q\in J_i$ for which one or two loop lines 
have been extracted simply developing the determinant, $J^1_i$ specifies
the set of  $j_q\in J_i$ for which one loop line 
has been extracted  applying (\ref{test2}). In the same way 
the process $P_0$ and $P_1$ specifies which  
loop fields are contracted in  $J^0_i$ and  $J^1_i$ for all $i$. Then
\bqa
&&\det {\cal M}'({\cal C}) = \sum_{J}
\sum_{P} \prod_{g_i\in D_{\cal C}}\no\\
&&\quad \left [
\lp\prod_{fg\in il^r_i\atop l_{fg}\in P} 
 {\cal M}_{f,g}'({\cal C}) \rp
 \prod_{ j_q\in J^1_i} 
\int_0^1 ds_{j_q} \;\;\det {\cal M}'({\cal C})(\{s_{j_q}\})\right ].
\eqa
where $J$ defines the sets $J^0_i$ and  $J^1_i$ for all $i$.
For each loop line $l_{fg}$ extracted, the set of band indices 
accessible for both $f$ and $g$  is reduced to
\bqa
M^r(f,{\cal C})&=& M^r(g,{\cal C})= \min [M(f,{\cal C}),M(g,{\cal C}) ] \no\\
m^r(f,{\cal C})&=& m^r(g,{\cal C})= \max [m(f,{\cal C}),m(g,{\cal C}) ] 
\eqa
We have to verify that the 
new matrix ${\cal M}'({\cal C})(\{s_{j_q}\})$
 still satisfies  a Gram's inequality and 
 that the sum over processes does not develop a factorial. This is done in the
following two lemmas. 
Remark that the sum over $J$  is not dangerous. Actually
at each $j_q$ we have two choices, hence $|J|\leq 2^{\bar n}$.  
\begin{lemma}
${\cal M}'({\cal C})(\{s_{j_q}\})$ satisfies the same Gram inequality 
as ${\cal M}'({\cal C})$ in [DR1], (IV.4), which does not depend on the 
parameters $s_{fg}$.
\end{lemma}
\paragraph{Proof} The proof is identical to that of [DR1], Lemma 4. The only
difference is that now ${\cal W}^k_{v,a;v',a'}$ 
contains an additional $s$ dependent factor ${\cal S}^k_{v,a;v',a'}$. 
By (\ref{factordet}) or 
(\ref{factordet1}) below, we recall that the determinant for the set 
$L^R_{j_q}(i)\cup L^L_{j_q}(i)$ of fields and antifields which may be
concerned by the $s_{j_{q}}$ interpolation step factorize in the big
loop determinant, so we need only to consider a single such factor
${\cal S}^{k,j_{q}}_{v,a;v',a'}$, and prove that it is still a positive
matrix. This is obvious if we reason on the index space for the
vertices $v$ and $v'$ to which the fields and antifields hook (and not
on the fields or antifields indices $a$ and $a'$ themselves). Indeed 
${\cal S}^{k,j_{q}}_{v,a;v',a'}$ is 
$\chi_a^v. \chi_{a'}^{v'}$ (the positive matrix which is 1 if $a$ hooks to $v$
and $a'$ hooks to $v'$, and 0 otherwise) times the combination with positive
coefficients $s_{j_{q}}M_{v,v'} + (1-s_{j_{q}})N_{v,v'}$ 
of the positive matrix $M$ which has each coefficient equal to 1
and the positive block matrix $N$ which has $N_{v,v'}=1$ if $v$
and $v'$ belong both to $\tree^{R} _{j_{q}}$ or both to $\tree^{L} _{j_{q}}$
and $N_{v,v'}=0$ otherwise. 
Therefore the matrix ${\cal S}^{k,j_{q}}_{v,a;v',a'}$
is positive in the big tensor space spanned by pairs of indices $v,a$, it 
has a diagonal bounded by 1,
and we can complete the proof as in [DR1], Lemma 4. The conclusion is that
the additional interpolation parameters  $s_{j_{q}}$ do not change the 
Gram estimate and the norms of $F_f$ and $G_g$ given in [DR1]. 
   
\begin{lemma}
The cardinal of  $P$ is bounded 
by $K^{\bar n}$ for some constant $K$.
\end{lemma}
\paragraph{Proof} The loop determinant is factorized
on determinants restricted to
 each reduced two-point subgraph in $D_{\cal C}$:
\be
\det {M}' = \prod_{g^r_i\in D_{\cal C}/{\cal A}(g_i)} \det {M}'(il^r_i). 
\ee
Each determinant $\det {M}'(il(g^r_i))$ is in turn factorized
on  determinants restricted to
internal loop fields for each reduced  subgraph $g^r_{j_q}$, $q=0,..,|J_i|$:
\be
\det {M}'(il(g^r_i))= \prod_{q=0}^{|J_i|}
\det {M}'(il^r_{j_q}).\label{factordet1} 
\ee
We have seen that for each $g_{j_q}$ the number of terms in $P$ 
is bounded
by $|L^R_{j_q}(i)|^2|L^L_{j_q}(i)| $.  Then 
\bqa
|P| &\leq & \prod_{g^r_i\in D_{\cal C}/{\cal A}(g_i)} 
\prod_{q=0}^{|J_i|} (|L^L_{j_q}(i)|\;|L^R_{j_q}(i)|^2)\no\\
 &\leq &  
2^{\bar n}\; e^{ \sum_{g^r_i\in D_{\cal C}/{\cal A}(g_i)}
\sum_{q=0}^{|J_i|} (|L^R_{j_q}(i)|+|L^L_{j_q}(i)|)}\no\\
&\leq &  
2^{\bar n}\; 
e^{4 \sum_{g^r_i\in D_{\cal C}/{\cal A}(g_i)} |V_i^r|} \leq K^{\bar n}\ ,
\eqa
where we applied 
\be
 \sum_{ q=0}^{|J_i|} 
(|L^R_{j_q}(i)|+|L^L_{j_q}(i)|) \leq 4 |V_i^r| \quad ; \quad   
\sum_{g^r_i\in D_{\cal C}/{\cal A}(g_i)} |V_i^r|  \leq {\bar n}\ .
\ee
This completes the  proof.
\qed

Now we can insert sector decouplings
exactly as we did in [DR1], but with a few additional operations.

\subsection{Sector refinement.} 
For each $g_i\in D_{\cal C}$ and  1PI we introduce one more sector  
decomposition on $h^{(2)}_{i}$, in order to optimize the bounds
from renormalization (Sec.II.7).   
Actually, the finest sector of size 
 $\La^{1\over 2}(w_{i(h^{(2)}_{i})})$  is further decomposed 
in a smaller sector of size
\be 
\La^{1\over 2}(w_{j_{h^{(2)}_{i},1}}):=
\La^{1\over 2}(w_{i(h^{(2)}_{i})})\; z_i 
\ee
where   $i(h^{(2)}_{i})\leq {\cal A}(i)$
is the band index of $h^{(2)}_{i}$ and $0<z_i\leq1$ is a factor
to be chosen.  
This sector is introduced applying the
identity [DR1](III.11) with $\al_s=\al_{j_{h,1}}$ defined by
$\al^{-{1\over 4}}_{j_{h,1}}:=\La^{1\over 2}(w_{j_{h,1}})$. All the other
larger sectors are introduced through the identity [DR1](III.13).  
The effect of last refinement is an additional factor
$1/z_i$ from sector counting and spatial integration 
(this only if the refined line is a tree one), and a
 factor $z_i$ from the volume in impulsion space. Then, 
we have are left with a global factor $1/z_i$. The optimal value
is $z_i = \La^{1\over 2}(w_{t(i)})$, as will be explained at the
end of Sec. II.7.

The expression to bound is then similar to [DR1] (III.16):

\bqa
\lefteqn{\Ga^{\La\La_0}_{2p}(\phi_1^{\La_T},...,\phi_{2p}^{\La_T})=
\sum_{{\bar n}\geq 1} \frac{\la^n}{n!} 
\frac{\lp\de\mu^1_\La\rp^{n'}}{n'!}}\label{conv1r}\\
&& \sum_{CTS}
\sum_{u-\tree}\sum_{\cal L}\sum_{E\Om}\sum_{{\cal C}}\sum_{J,P}
\vep(\tree, \Om)\int_{w_T\le w_{{\cal A}(i)} \le  w_{i}\le 1}
\prod_{q=1}^{{\bar n}-1} dw_q\no\\
&&\prod_{h\in L\cup\tree_L\cup E}
\left\{ [{\scriptstyle {4 \over 3} \La^{-{1\over 2}}(w_{j_{h,n_h}})}]
\int_0^{2\pi}  d\th_{h,n_h}
[ {\scriptstyle{4 \over 3}\La^{-{1\over 2}}(w_{j_{h,n_h-1}})}]
\int_{\Si_{j_{h,n_h}}}
d\th_{h,n_h-1} \right .\no\\
&&...\;\;[{\scriptstyle {4 \over 3}\La^{-{1\over 2}}(w_{j_{h,1}})}]
\left .\int_{\Si_{j_{h,2}}}
d\th_{h,1}
\left [\prod_{r=2}^{n_{h}}
\chi^{\th_{h,r}}_{\al_{j_{h,r}}}(\th_{h,1})\right ]
 \right \}\no\\
&& \prod_{g_i|\ i=r\ {\rm or}\atop |eg_i({\cal C}|)\leq 8}
\Upsilon\lp\th_i^{root},\{\th_{h,r(i)}\}_{h\in eg^\ast_i}
 \rp \;
\prod_{v\in V\cup V'}  \Upsilon\lp\th_{h_v^{root}},
\{\th_{h,n_h}\}_{ h\in H^\ast(v)} \rp\no\\
&&\int
d^3x_1...d^3x_{\bar n}\;\;
\phi^{\La_T}_1(x_{i_{1}},\th_{e_1,1})\;...\;
\phi^{\La_T}_{2p}(x_{j_{p}},\th_{e_{2p},1})
\left[\prod_{q=1}^{{\bar n}-1}  C^{w_q}
( x_{q}, {\bar x}_{q}, \th_{h,1}) \right]\no\\
&& \left [ \prod_{l_{fg}\in P}
{\cal M}_{f,g}'({\cal C},E ,\{\th_{a,1}\})\right ] 
\;\left [  \prod_{ j_q\in J^1} 
\int_0^1 ds_{j_q} \right ] 
\det {\cal M}'({\cal C},E,\{\th_{a,1}\},\{s_{j_q}\}) \ ,
\no\eqa
where we defined $V$ and $V'$ as the set of four point and 
two point vertex respectively.
To perform renormalization we   apply to the amplitude of 
each two point  subgraph $g$ the operator $(1-\tau_g)+\tau_g$, 
where $\tau_g$ selects the linearly divergent term in $g$ 
giving a local counterterm for  $\de\mu$
that depends on the scale of the external lines of $g$. We start the 
renormalization from the leaves of the $CTS$ (hence
from the smallest subgraphs at highest scale) and go down.
 
\subsection{Momentum space} 
The Taylor expansion of $\hat{g}(k)$ 
around a vector $k_F$ near the Fermi surface  gives 
two possible sources of counterterms. The term of order 0 in the 
Taylor expansion is linearly divergent and gives rise to a chemical potential 
counterterm;
the term of  order 1 is logarithmic and would give rise to
wave function counterterms
(in fact proportional to $k_{0}$ and $\vec k ^{2}$), that we do not need
to consider for our upper bound, Theorem 1. As we said, for this kind of bound
we need only to perform  the linearly divergent renormalization.
Therefore we define the localization operator
acting on a two-point function  as:
\be
\delta(k_1+k_2)\; \;\tau_g \hat{g}(k_2) =\delta(k_1+k_2) \;\;
\hat{g}(k_F).
\ee

Remark that by rotational invariance there is no ambiguity in the choice 
of the spatial component of $k_F$. For the temporal component 
we choice $k_{F0}=\pi T$, to simplify  computations. 
This choice breaks  parity in the imaginary time direction, but  in 
our context this is not essential.

\subsubsection{Not Dangerous subgraphs.} 
We do not need to renormalize all two point subgraphs 
but only the subset 
\be
D({\cal C},P) := \{ g_i | \;|eg_i({\cal C})|=2, \; g_i \;{\rm 1PI}\}
\ee
in the sense explained in the section II.1.
By momentum conservation it is easy to see that, if $g_i(\cal C)$ 
is 1PR  and $g_j({\cal C})$ is the two-point subgraph we obtain 
cutting one tree line of $g_i$ 
\be
\tau_{g_i({\cal C})}\lp 1-\tau_{g_j({\cal C})}\rp =
\lp 1-\tau_{g_i({\cal C})}\rp \tau_{g_j({\cal C})} = 0
\ee 
hence the renormalization of $g_i({\cal C})$ is ensured by that of  
$g_j({\cal C})$.  
Remark that, by the relation of partial order in the $CTS$,   
$D({\cal C},P)$ has a forest structure (see [R] and [DR2]). 

We denote by $ND({\cal C},P)$ (not-dangerous\dots ) 
the set of two point subgraphs which are 
1PR, hence are not renormalized. It is the union of the
set of two point subgraphs not in $D_{\cal C}$, for which we knew one 
particle reducibility from start, and the set  
$D_{\cal C}\backslash D({\cal C},P)$, for which we learnt it after
the loop extraction process.
For any $g_i\in ND({\cal C},P)$  
one internal line $l_j$ must have the same momentum as the external
line $l_{{\cal A}(i)}$. Then the internal and external 
scales of $g_i$ cannot  be far; this imposes a constraint on the 
integral over the parameter $w_i$ that allows to avoid renormalizing
these subgraphs.

\subsection{Real space}

The formulation of renormalization in momentum space is the one of [FT2]
and is sufficient for perturbative results. 
In this formulation the localization operator is rotation invariant.
However for constructive bounds 
we need a phase space analysis, hence a direct space ``dual version'' of this
operator [R]. 

In the space of positions, the dual localization operator, which
acts on the external lines of the subgraph, 
is never unique. In relativistic euclidean field theory it depends on the
choice of an arbitrary localization point (see [R]), a 
convenient choice being the position of one of the external
vertices. Here, in condensed matter, this dual operator depends on an
additional choice, namely a direction on the Fermi surface. 
A convenient choice is found  thanks to the sector decomposition. Actually, 
before performing the sum over sector attributions, 
the two external propagators 
of a graph $g_i$ belong to 
well defined sectors $\Si(\al_{j(1)},\th_1)$ and $\Si(\al_{j(2)},\th_2)$ 
with sector center on the vectors $(0,\vec{r}_k)$, $k=1,2$, where
$j(1),j(2)\leq {\cal A}(i)$.
Therefore we define the  operator $\tau_g$ 
as a first order Taylor expansion around the momentum
$k_2=-r_2=(-\pi T,-\vec{r}_2)$ (the minus sign
corresponding to integration by parts). The dual $x$-space operator 
 $\tau_g^*$ acts on the product of external 
propagators $C_{\th_1}(x_1,y_1)C_{\th_2}(x_2,y_2)$ by
\be
\tau_g^* \;C_{\th_1}(x_1,y_1)\; C_{\th_2}(x_2,y_2)\;= \; e^{i r_2(x_2-x_1)}\;
 C_{\th_1}(x_1,y_1) \; C_{\th_2}(x_1,y_2)\ .
\label{local}\ee
This formula does not coincide with the usual one (see [R]) and can be 
justified observing that $C_{\th_s}(x,y)$ is not a slowly varying function 
with $x$, but has a spatial momentum of order 1, hence oscillates wildly. 
The good slowly varying function to move is $C'_{\th_s}(x,y)$ defined by: 
\be
C_{\th_s}(x,y) \; =\; { e^{ir_s(x-y)}\over (2\pi)^2 } \;
\int d^3k\; e^{i(k-r_s)(x-y)}\; C_{\th_s}(k) := e^{ir_s(x-y)} \;
  C'_{\th_s}(x,y) .
\ee
The expression (\ref{local}) can also be obtained defining
\be
\tau_g^*\; C'_{\th_1}(x_1,y_1)\; C'_{\th_2}(x_2,y_2)\; =\; 
  C'_{\th_1}(x_1,y_1)\; C'_{\th_2}(x_1,y_2).
\ee

\paragraph{Choice of the reference vertex.}
The choice of $x_1$ as fixed vertex instead of $x_2$ 
is arbitrary. In this paper we use the rule that most 
simplifies notations and calculations (not exactly the same as in [DR2]).  
For each $g_i\in D({\cal C},P)$ we chose 
as reference vertex the one hooked to the half-line $h^{(1)}_i=h^{root}_i$, 
$v^{(1)}_i$ with position $x^{(1)}_i$. The moved vertex is then 
$x^{(2)}_i$. 
This rule implies that tree lines have never both ends moved, and that
the root vertex $x_1$, which is essential in spatial integration, 
is always fixed.

In the following we will denote by 
$D_t({\cal C},P)$, $D_l({\cal C},P)$, $D_e({\cal C},P)$
the subgraphs in  $D({\cal C},P)$ for which the moved line 
is tree, loop or external respectively. 
 
\subsection{Effective Constants}

At each vertex $v$ we 
can now resum the series of all counterterms obtained applying $\tau_g$ 
to all $g\in D({\cal C},P)$
(for different classes ${\cal C}$, processes $P$ and perturbation orders ${\bar n}$) 
that  have the same set of  external lines as $v$ itself. In this way
we obtain an effective coupling constant which depends on the scale 
$\La(w_{i_v})$  of the highest tree line hooked to the vertex $v$. 
This is automatically true for a two point vertex
(and in fact would also be true as in [DR2] for a four point vertex
because tadpoles are zero by [DR1], Lemma 2).
Each counterterm is now a function 
\bqa
F_{\th_1,\th_2}(y_1,y_2) &=& \int d^3x_1 \;
C_{\th_1}(x_1,y_1)\; C_{\th_2}(x_1,y_2) \;
\left [ \int d^3x_2\; g(x_1,x_2) \; e^{ir_2(x_2-x_1)}
\right ]\no\\
 &=&  \int d^3x_1 \;
C_{\th_1}(x_1,y_1)\; C_{\th_2}(x_1,y_2) \; \hat{g}(-r_2)
\eqa
where we applied the translational invariance of $g$. Now remark that 
$ \hat{g}(k)$ is invariant under rotations of the spatial component $\vec{k}$
of $k$ as the free propagator depends only on the absolute value of $\vec{k}$.
Therefore 
\be
\hat{g}(-r_2)\;=\; \hat{g}(-\pi T,|\vec{r}_2|) \;= \;\hat{g}(-\pi T,1) 
\ee 
is  independent from $\th_1$ and $\th_2$. 

\begin{theor}
If we apply to each two point subgraph $g\in D({\cal C},P)$, 
for any class ${\cal C}$ and process $P$, 
the operator
$(1-\tau_g)+\tau_g= R_g+\tau_g$, (\ref{conv1r}) 
can be written as
\bqa
\lefteqn{\Ga^{\La\La_0}_{2p}(\phi_1^{\La_T},...,\phi_{2p}^{\La_T})=
\sum_{{\bar n}\geq 1} \frac{\la^n}{n!n'!} }\label{conv2r}\\
&&
\sum_{CTS}
\sum_{u-\tree}\sum_{\cal L}\sum_{E\Om}\sum_{{\cal C}}\sum_{J P}
\vep(\tree, \Om)
\int_{w_T\le w_{{\cal A}(i)} \le  w_{i}\le 1}\prod_{q=1}^{{\bar n}-1}  dw_q 
\no\\
&&\prod_{h\in L\cup\tree_L\cup E}
\left\{ [{\scriptstyle {4 \over 3} \La^{-{1\over 2}}(w_{j_{h,n_h}})}]
\int_0^{2\pi}  d\th_{h,n_h}
[ {\scriptstyle{4 \over 3}\La^{-{1\over 2}}(w_{j_{h,n_h-1}})}]
\int_{\Si_{j_{h,n_h}}}
d\th_{h,n_h-1} \right .\no\\
&&...\;\;[{\scriptstyle {4 \over 3}\La^{-{1\over 2}}(w_{j_{h,1}})}]
\left .\int_{\Si_{j_{h,2}}}
d\th_{h,1}
\left [\prod_{r=2}^{n_{h}}
\chi^{\th_{h,r}}_{\al_{j_{h,r}}}(\th_{h,1})\right ]
 \right \}\no\\
&& \prod_{g_i|\ i=r\ {\rm or}\atop |eg_i({\cal C})|\leq 10}
\Upsilon\lp\th_i^{root},\{\th_{h,r(i)}\}_{h\in eg^\ast_i}
 \rp \;
\prod_{v\in V\cup V'}  \Upsilon\lp\th_{h_v^{root}},
\{\th_{h,n_h}\}_{ h\in H^\ast(v)} \rp\no\\
&&\int
d^3x_1...d^3x_{\bar n}\;\;
\phi^{\La_T}_1(x_{i_{1}},\th_{e_1,1})\;...\;
\phi^{\La_T}_{2p}(x_{j_{p}},\th_{e_{2p},1})
\left [\prod_{v\in V'} \de\mu^{\La(w_{i_v})}_\La(\la) \right ]\no\\
&&
\prod_{g_i\in D({\cal C},P)} R_{g_i} \left \{
\left[\prod_{q=1}^{{\bar n}-1}  C^{w_q}
( x_{q}, {\bar x}_{q}, \th_{h,1})\; \right]\right .\no\\
&&\left . 
\prod_{l_{fg}\in P} 
\left [ {\cal M}_{f,g}'({\cal C},E ,\{\th_{a,1}\})\; \right ]
\left [ \prod_{j_q\in J^1} 
\int_0^1 ds_{j_q}\right ] \det {\cal M}'({\cal C},E,\{\th_{a,1}\},\{s_{j_q}\})
 \right \} \ ,
\no\eqa
where $\de\mu^{\La(w)}_\La(\la)$, the effective constant defined by:
\be
\de\mu^{\La(w)}_\La(\la) \;= \;\hat{\Si}^{\La(w)}(-r_2) \;= 
\;\int d^3x_2 \;
\Si^{\La(w)}(0,x_2) \;e^{ir_2x_2} \ ,
\ee
is independent of the choice of the angular component of  $\vec{r}_2$.
This effective constant is the vertex function $\Ga_2^{1\La(w)}$  
for an effective theory with IR parameter $\La(w)$, and 
bare counterterm $\de\mu^1_\La$. 
Furthermore $\de\mu^{\La(w)}_\La(\la)$ is analytic in $\la$ and is 
bounded by
\be
\left |\de\mu^{\La(w)}_{\La}(\la)\right | 
\leq K\; |\la| \;(\La(w)-\La) \label{bound2point}
\ee
for some constant $K$.
The renormalized $\de\mu_{ren}(\La)$ is then the vertex 
function for an effective theory with IR parameter $\La(0)=\La$
\be
\de\mu_{ren}(\La) = \de\mu^{\La}_{\La}(\la) = 0.
\ee  
Finally the first and second derivatives of the self-energy 
$\hat \Sigma (k)$\footnote{Recall that the self-energy 
is the sum of all non-trivial one-particle-irreducible 
two point subgraphs.}
are uniformly bounded:
\[ \left | {\partial \over \partial k_{i} } \hat 
\Sigma _{|{k_{0}={\pi\over\beta},
e(\vec k)=0}} \right | \le K |\lambda |^2  \quad ; \quad  \left |\left |
\ {\partial^{2}\over \partial k_{i}\partial k_{j} }
\hat \Sigma (k) \  \right|  
\right |_{\infty} \le K 
\]
where i and j take values 0,1,2, and $K$ is some constant. These bounds
are proved in Appendix B.

\end{theor}       
\paragraph{Proof} The first part of the 
theorem actually consists in a reshuffling of  perturbation
theory, and can be proved by standard combinatorial arguments as in [R].
The only difficulty that is not in [R] is to prove that the parameter $w$ 
of the effective constant always corresponds to the highest tree line of the 
vertex: as we said above this is obvious for two point vertices.
The second part of the theorem, that is the analyticity of $\de\mu$
and the bound (\ref{bound2point}), corresponds in statistical 
mechanics to the problem of fixing the bare mass in such a way that the 
renormalized mass is zero. This is a standard problem, now well understood.
For instance, a proof in the case of the critical $\phi_4^4$ model,
can be found in [FMRS] and [GK]. For completeness we recall the 
arguments of the proof in Appendix A. Finally the bound of the first
and second derivatives of the self-energy allows a Taylor expansion around
the Fermi surface which proves Fermi liquid behavior [S1]; they would be false
in $d=1$, were Luttinger liquid behavior is known to occur [BGPS]-[BM]. 
\qed

\subsection{Convergence of the Effective Expansion}

\begin{theor}
Let $\vep>0$ and $\La_0=1$ be fixed. 
The series (\ref{conv2r}) is absolutely convergent for 
$|\la|\leq c$ and  
\be
c \leq \frac{1}{K_2|\log (T /K_1)|}
\ee
for some constants $K_1$, $K_2$. This convergence is uniform in $\La$, then 
the IR limits of the vertex functions 
$\Ga^{\La_0}_{2p}= \lim_{\La\rightarrow 0} 
\Ga^{\La\La_0}_{2p}$
exist, they are analytic in $\la$ in a disk of radius $c$, and they
obey the bounds 
\bqa
&&\hspace{-0.8cm}|\Ga^{\La_0}_{2p>4}(\phi_1^{\La_T},..., \phi_{2p}^{\La_T})|
 \leq \\
&& 
 K_0\; {\scriptstyle ||\phi_1||_{_1}
\prod_{i=2}^{2p} ||\hat{\phi}_i||_{\infty,2} }
{T^{{7p\over 2} -{1\over 2}}\over 2p-4} \ [K_1(\vep)]^p\  \lp p!\rp^2 \;
K(c,T)
\; e^{-(1-\vep)\La_T^{1\over s} d^{1\over s}_{\tree}(\Om_1,...,\Om_{2p})}\no
\eqa
\bqa
&& \hspace{-1.8cm}|\Ga_{4}^{\Lazero}(\phi_1^{\La_T},..,\phi_4^{\La_T})|
 \leq  \\
&& K'_0 (\vep)\;
{\scriptstyle ||\phi_1||_{_1}
\prod_{i=2}^{4} ||\hat{\phi}_i||_{\infty,2} }
T^{13\over 2}  \; \
K(c,T)
\; e^{-(1-\vep)\La_T^{1\over s}d^{1\over s}_{\tree}(\Om_1,...,\Om_{4})}\no
\eqa
\be
|\Ga_{2}^{\Lazero}(\phi_1^{\La_T},\phi_2^{\La_T})|\leq
  K''_0 (\vep)\;
{\scriptstyle ||\phi_1||_{_1}
 ||\hat{\phi}_2||_{\infty,2} }
T^2 \ 
K(c,T)
\; e^{-(1-\vep)\La_T^{1\over s}d^{1\over s}_{\tree}(\Om_1,\Om_{2})}
\ee
where $\Om_i$ is the compact support of $\phi_i$, $K_1(\vep)$, $K'_0(\vep)$
and $K''_0(\vep)$ are functions of $\vep$ only, 
$d_{\tree}(\Om_1,...\Om_{2p})$ is defined as in [DR1], Theorem 2, 
$K(c,T)$ is a function which tends to 0 when $c\rightarrow 0$, and 
\be
||\hat{\phi}_i||_{\infty,2} := \lp||\hat{\phi}_i||_\infty +
||\hat{\phi}'_i||_\infty 
+||\hat{\phi}"_i||_\infty \rp.\label{phinorm}
\ee
\end{theor}

This Theorem (that is a generalization of [DR1], Theorem 2) 
means that one can build in a constructive sense the infrared
limit of the Fermi liquid at a 
finite temperature higher than some exponentially
small function of the coupling constant simply by summing up 
perturbation theory. 

The rest of this paper is devoted to the proof of that theorem.

\subsection{Lines interpolation}

\begin{figure}
\centerline{\psfig{figure=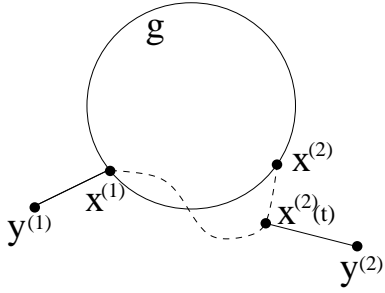,width=5cm}}
\caption{Line interpolation}\label{fermi8}
\end{figure}

Before performing any bound we must study the action of  
$\prod_{g\in D_{\cal C}} R_g^*$. 
For each $g_i\in D_{\cal C}$ the action of   $R_g^*$
on the external lines of $g_i$ is 
\bqa
&&\hspace{-0.5cm}
 R_{g_i}^* \;C_{\th_1}(x^{(1)},y^{(1)})\; C_{\th_2}(x^{(2)},y^{(2)})\; \no\\
&&= C_{\th_1}(x^{(1)},y^{(1)})\;
\left [ C_{\th_2}(x^{(2)},y^{(2)}) - e^{i r_2(x^{(2)}-x^{(1)})}\;
 \; C_{\th_2}(x^{(1)},y^{(2)})\right ]\no\\
&& =  C_{\th_1}(x^{(1)},y^{(1)})\;e^{ir_2x^{(2)}}\;
\left [ C_{\th_2}(x^{(2)},y^{(2)})\; e^{-ir_2x^{(2)}} - 
 \; C_{\th_2}(x^{(1)},y^{(2)})\; e^{-i r_2x^{(1)}}\right ]\no\\
 && = C_{\th_1}(x^{(1)},y^{(1)})\;  e^{i r_2x^{(2)}} \;\int_0^1 dt\; 
\frac{d}{dt} \left [   C_{\th_2}(x^{(2)}(t),y^{(2)})  e^{-i r_2 x^{(2)}(t)}
\right ]
\eqa
where we applied a first order development on
$ C_{\th_2}(x^{(2)},y^{(2)})\; e^{-ir_2x^{(2)}}$ and   
$x^{(2)}(t)$ is any differentiable path with
$x^{(2)}(0)=x^{(1)}$ and $x^{(2)}(1)=x^{(2)}$. 
The external line hooked to $x^{(2)}$ has then been hooked to the point $x(t)$ 
(see Fig.\ref{fermi8}) and has now propagator:
\be
 C^m_{\th_2}(x^{(2)}(t),y^{(2)}) :=   e^{i r_2x^{(2)}}\;
\frac{d}{dt} \left [   C_{\th_2}(x^{(2)}(t),y^{(2)})  
e^{-i r_2 x^{(2)}(t)}\right ].
\ee

The easiest choice for the 
path is a linear interpolation between $x^{(1)}$ and $x^{(2)}$: 
\be
x^{(2)}(t)= x^{(1)} + t(x^{(2)}-x^{(1)})\ . 
\label{loopin1}
\ee
This is actually the kind of path we will take if the moved line is a loop or 
an external one.
The interpolated line can then be written as
\bqa
 &&\hspace{-0.8cm} C^m_{\th_2}(x^{(2)}(t),y^{(2)}) = 
   e^{ir_2(x^{(2)}-x_2(t))} (x^{(2)}-x^{(1)})^\mu   \label{loopin2}\\
&& \left [
-i r_2 +\p{x^{(2)}(t)}\right ]_\mu  \left [  
C_{\th_2}(x^{(2)}(t),y^{(2)})\right ] \no\\
&&\hspace{-0.4cm}  = e^{ir_2(x^{(2)}-x^{(2)}(t))} 
\int d^3k\; e^{ik(x^{(2)}(t)-y^{(2)})} \left [
i (x^{(2)}-x^{(1)})  (k - r_2) \right ]   
   C_{\th_2}(k)\ .\no
\eqa 
When applied to a tree line, this interpolation does not ``follow 
the tree'' as the point $x(t)$ in general no longer hooks to some point 
on a segment  corresponding to a tree line. This leads to some
difficulties when integrating over spatial positions. To avoid this we take 
$x(t)$ as the path in the tree joining $x^{(2)}$ to $x^{(1)}$,
as in [DR2]. This path has 
in general $q$ lines with  vertices $x_0,...x_q$ with the conditions
$x_0=x^{(1)}$ and $x_q=x^{(2)}$. Remark that, with this rule,  
the renormalization 
at higher scales modifies the tree used for renormalization at lower 
scales. We will define below the modified tree by an induction process. 
The interpolated line can then be written as
\bqa
&& e^{i r_2x^{(2)}}
\left [ e^{-i r_2 x^{(2)}}C_{\th_2}(x^{(2)},y^{(2)}) - e^{-i r_2 x^{(1)}}
 C_{\th_2}(x^{(1)},y^{(2)})\right ]\no\\ 
&&= \ \
\sum_{j=1}^q \int_0^1 dt\; 
 C^m_{\th_2}(x_j(t),y^{(2)})\ , 
\label{interp1}\eqa
where we defined 
\bqa
&& C^m_{\th_2}(x_j(t),y^{(2)}) \;=\; 
e^{ir_2x^{(2)}} \; \frac{d}{dt} \left [  C_{\th_2}(x_j(t),y^{(2)}) 
e^{-ir_2x_j(t)}
\right ] \label{interp2}\\
&&\qquad = e^{ir_2(x^{(2)}-x_j(t))} \; (x_j-x_{j-1})^\mu  \left [
-i r_2 +\p{x_j(t)}\right ]_\mu  \left [  
C_{\th_2}(x_j(t),y^{(2)})\right ] \no\\
&&\qquad = {e^{ir_2(x^{(2)}-x_j(t))}\over (2\pi)^2} 
\; \int d^3k\; e^{ik(x_j(t)-y^{(2)})}\; \left [
i (x_j-x_{j-1})  (k - r_2) \right ] \;  
   C_{\th_2}(k)\no 
\eqa 
and 
\be
x_j(t)= x_{j-1} + t(x_j-x_{j-1})\ .\label{interp3}
\ee

\subsubsection{Second order expansion} 

The renormalizing factor is $(k-r_2)(x_j-x_{j-1})$, 
or $(k-r_2)(x^{(2)}-x^{(1)})$. The size of 
$(k-r_2)$ is fixed by the cut-off of the propagator $C_{\th_2}$:
\bqa
(k-r_2)_0 \;  & \simeq & \La(w_{i_2})\quad \;\;
\leq \La(w_{{\cal A}(i)}) \no\\
(k-r_2)_{r(r_2)} &\simeq &  \La(w_{i_2})\quad \;\;
\leq \La(w_{{\cal A}(i)})\no\\
(k-r_2)_{t(r_2)} &\simeq &  \La^{1\over 2}(w_{i_2})\; z_i \leq  
\La^{1\over 2}(w_{{\cal A}(i)}) \; z_i 
\eqa
where $(k-r_2)_{r(r_2)}$ is the spatial component on the direction $\vec{r}_2$ 
and $(k-r_2)_{t(r_2)}$ is the spatial component on the direction  ortogonal 
to $\vec{r}_2$.
Remark that the size of the tangential component $(k-r_2)_{t(r_2)}$ 
is the size of the
finest sector of the propagator $C_{\th_2}$; as we have said in the precedent 
subsection we have cut its finest sector scale 
$ \La^{1\over 2}(w_{i_2}) $  in a smaller sector, to improve the 
renormalizing factor. 

On the other hand $(x_j-x_{j-1})$  is bounded using a fraction of the 
exponential 
decay of tree line propagators and give the scale factors (we will perform the 
detailed calculation in the following):
\bqa
(x_j-x_{j-1})_0 \; &\simeq& \La^{-1}(w_{t(i)})\no\\
(x_j-x_{j-1})_{r(r_j)} &\simeq&  \La^{-1}(w_{t(i)})\no\\
(x_j-x_{j-1})_{t(r_j)} &\simeq&  \La^{-{1\over 2}}(w_{t(i)})\ . 
\eqa
$(x^{(2)}-x^{(1)})$  give the same factors, as it can be written as 
$\sum_j(x_j-x_{j-1})$.   
One sees immediately that the components 
$(k-r_2)_0 (x_j-x_{j-1})_0 $ and $(k-r_2)_{r(r_2)} (x_j-x_{j-1})_{r(r_2)}$ 
give the factor $\La(w_{{\cal A}(i)})\La^{-1}(w_{t(i)})$
that we need to renormalize, 
but $(k-r_2)_{t(r_2)} (x_j-x_{j-1})_{t(r_2)}$  gives only 
$\La^{1\over 2}(w_{{\cal A}(i)})z_i\La^{-1}(w_{t(i)})$ that is not sufficient. 
This is the main difficulty, announced in the Introduction,
that we met in this paper: when trying to renormalize in phase space with 
anisotropic sectors, the internal decay of the tree
does not necessarily match the external sector scales. 
To solve this problem we expand to second order, by $
\int_0^1 dt \; F'(t) = F'(0) + \int_0^1 dt\; (1-t)\; F''(t)$,
and we prove that the first order term which gives the bad power counting 
factor is actually zero. Then we optimize the bound obtained 
with respect to $z_i$.
Indeed this second order Taylor formula gives for loop and external lines
\bqa
&& \hspace{-0.8cm}  \;\int_0^1 dt\; 
\frac{d}{dt} \left [   C_{\th_2}(x^{(2)}(t),y^{(2)})  
e^{-i r_2 x^{(2)}(t)}\right ]=\no\\
&&(x^{(2)}-x^{(1)})^\mu \p{x^{(1)\mu}}  \left [
C_{\th_2}(x^{(1)},y^{(2)})  e^{-i r_2 x^{(1)}}\right ] 
+  (x^{(2)}-x^{(1)})^\mu (x^{(2)}-x^{(1)})^\nu\no\\
&& \int_0^1 dt\; (1-t)
\p{x^{(2)\mu}(t)}  \p{x^{(2)\nu}(t)}  
 \left [
C_{\th_2}(x^{(2)}(t),y^{(2)})  e^{-i r_2 x^{(2)}(t)} \right ] 
\eqa
where we applied $x^{(2)}(0)=x^{(1)}$. 
For tree lines  we have
\bqa
&& \hspace{-0.5cm} \sum_{j=1}^q \int_0^1 dt\; 
\frac{d}{dt} \left [  C_{\th_2}(x_j(t),y^{(2)}) e^{-ir_2x_j(t)}
\right ] =
\sum_{j=1}^q  
\left [  \frac{d}{dt} C_{\th_2}(x_j(t),y^{(2)}) e^{-ir_2x_j(t)}
\right ]_{\big | t=0} \no\\
&& \quad +\; \sum_{j=1}^q \int_0^1 dt\; (1-t)\;
\frac{d^2}{dt^2} \left [  C_{\th_2}(x_j(t),y_2) e^{-ir_2x_j(t)}
\right ].
\eqa
The last sum on the right hand of the equation is a second order term:
\be
\sum_{j=1}^q  (x_j-x_{j-1})^{\mu} (x_j-x_{j-1})^{\nu} \hspace{-0.1cm}
\int_0^1 dt (1-t)
 \p{x^{\mu}_{j}(t)} \p{x^{\nu}_{j}(t)} 
\left [  C_{\th_2}(x_j(t),y^{(2)}) e^{-ir_2x_j(t)}\right ].\label{interp4}
\ee
The first sum on the right hand of the equation contains a first order and a 
second order term:
\bqa
&&  \hspace{-0.5cm} \sum_{j=1}^q  (x_j-x_{j-1})^{\mu}\; 
\p{x^{\mu}_{j-1}} \left [  C_{\th_2}(x_{j-1},y^{(2)}) e^{-ir_2x_{j-1}}\right ]
=\label{interp5} \\
&& \quad (x^{(2)}-x^{(1)})^{\mu}
\p{x^{\mu}_{1}} \left [  C_{\th_2}(x^{(1)},y^{(2)}) e^{-ir_2x^{(1)}}\right ]+ 
 \sum_{j=1}^q  (x_j-x_{j-1})^{\mu} \sum_{k=1}^{j-1}  \no\\
&& \quad \left [
\p{x^{\mu}_{k}} \left [  C_{\th_2}(x_{k},y^{(2)}) e^{-ir_2x_{k}}\right ] -
\p{x^{\mu}_{k-1}} \left [  C_{\th_2}(x_{k-1},y^{(2)}) e^{-ir_2x_{k-1}}\right ] 
\right ]\no
\eqa
where  we applied $x_j(0)=x_{j-1}$. The first term is the same first order 
term we obtain
for loop or external lines,  while 
the second  term can be written as:
\bqa
&&
 \hspace{-0.8cm}\sum_{j=1}^q  (x_j-x_{j-1})^{\mu}\; \sum_{k=1}^{j-1} 
\int_0^1 dt\;  \frac{d}{dt}\; 
\p{x^{\mu}_{k}(t)} 
\left [  C_{\th_2}(x_{k}(t),y^{(2)}) e^{-ir_2x_{k}(t)}\right ]
\\
&& \hspace{-0.8cm} =  \sum_{j=1}^q  \sum_{k=1}^{j-1} (x_j-x_{j-1})^{\mu}
(x_k-x_{k-1})^{\nu} \hspace{-0.1cm}\int_0^1  \hspace{-0.1cm}dt 
\p{x^{\mu}_{k}(t)} \p{x^{\nu}_{k}(t)} 
\left [  C_{\th_2}(x_{k}(t),y^{(2)}) e^{-ir_2x_{k}(t)}\right ]\no
\eqa
and gives a second order term that adds to (\ref{interp4}). 

\begin{lemma}
The contribution coming from the component orthogonal to
$\vec{r}_2$  of the first order term
\bqa
\lefteqn{\int d^3x^{(1)} d^3x^{(2)}\; g(x^{(1)},x^{(2)}) \; 
C_{\th_1}(x^{(1)},y^{(1)})}\\
&& e^{ir_2(x^{(2)}-x^{(1)})} 
(x^{(2)}-x^{(1)})_{t(r_2)}  
\left [ - i r_2 + \p{x^{(1)}} \right ]_{t(r_2)} C_{\th_2}(x^{(1)},y^{(2)}) \no
\eqa
is zero.
\end{lemma}
\paragraph{Proof}
The complete first order term is
\bqa
&&\hspace{-1cm}\int d^3x^{(1)} d^3x^{(2)}\; g(x^{(1)},x^{(2)}) \; 
C_{\th_1}(x^{(1)},y^{(1)})\no\\
&& e^{ir_2(x^{(2)}-x^{(1)})} 
(x^{(2)}-x^{(1)})^{\mu} i \lp -r_2 +{1\over i} \p{x^{(1)}} 
\rp_{\mu} C_{\th_2}(x^{(1)},y^{(2)}) \no\\
&&\hspace{-0.5cm}= \left [ \int d^3x^{(1)}  C_{\th_{1}}(x^{(1)},y_1) 
\lp -r_2 +{1\over i} \p{x^{(1)}}\rp_{\mu} C_{\th_2}(x^{(1)},y^{(2)}) \right ] \no\\
&&\left [ i\int d^3x^{(2)}  g(0,x^{(2)}) e^{ir_2x^{(2)}} x^{(2)\mu} 
\right ]\ ,
\eqa
where we applied the translational invariance of $g(x^{(1)},x^{(2)})$. Now 
\bqa
&& i\int d^3x_2 \; g(0,x^{(2)}) e^{ir_2x^{(2)}} x^{(2)\mu} =
 \p{r_{2\mu}} \int d^3x^{(2)}  g(0,x^{(2)}) e^{ir_2x^{(2)}}   \no\\
&&\qquad  = -\left [\p{k_{\mu}} \hat{g}(k)\right ]_{\big |k=-r_2}.
\eqa
To compute the expression
we take the two spatial axes on the directions parallel and orthogonal to 
$-\vec{r}_2$.
Then $\vec{k}=-\vec{r}_2$ means $k_1=1$, $k_2=0$ or in radial coordinates 
$\rho = 1$ and $\th=0$. As we said before, $\hat{g}(k)$ 
depends only on the zero component $k_0$ and on the module 
of the spatial vector $\rho$:   
\be
\p{\th} \hat{g}(k) = 0  \quad \forall \th \ .
\ee
Now, applying  
\be
{\partial \hat{g}(k)\over \partial k_i}
= {\partial \hat{g}(k)\over  \partial \rho}{\partial \rho\over \partial k_i}  
+
{\partial \hat{g}(k)\over  \partial \th} {\partial \th\over  \partial k_i} 
\ee
for $i=1,2$ and the relations:
\be
 \p{k_1}\rho
= \frac{k_1}{\rho}
\qquad , \qquad 
\p{k_2}\rho= 
\frac{k_2}{\rho}\ ,
\ee
we obtain 
\be
\left [{\partial \hat{g}(k)\over  \partial k_0}\right]_{\big |k=-r_2}  
\neq  0 \quad , \quad 
\left [{\partial \hat{g}(k)\over  \partial k_{r(r_2)}}\right]_{\big |k=-r_2} 
  \neq  0\quad , \quad 
\left [{\partial \hat{g}(k)\over \partial  k_{t(r_2)}}\right]_{\big |k=-r_2}  
 =  0\ .
\ee
This ends the proof.
\qed

\paragraph{Choice of $z_i$}

After putting the dangerous first order term to zero we are left with
the bound
\bqa
&&\La^{1\over 2}(w_{t(i)}) \frac{1}{z_i} 
\left [ { \La(w_{{\cal A}(i)}) \over\La(w_{t(i)})}
+  { \La^2(w_{{\cal A}(i)}) \over\La^2(w_{t(i)})}
+ { \La(w_{{\cal A}(i)})\; z^2_i\over\La^2(w_{t(i)})}
+ {\La^{3\over 2}(w_{{\cal A}(i)})\; z_i\; 
\over \La^2(w_{t(i)})} \right ]\no\\
&& \leq 
\La^{1\over 2}(w_{t(i)}) 
 { \La(w_{{\cal A}(i)}) \over\La(w_{t(i)})}
\left [\frac{2}{z_i} + { z_i\over\La(w_{t(i)})}
+ {\La^{1\over 2}(w_{{\cal A}(i)}) 
\over \La(w_{t(i)})} \right ]
\eqa
where $t(i)$ is the band index of the lowest tree line in the path joining 
the two external vertices of $g_i$,
the global factor $\La^{1\over 2}(w_{t(i)})$ comes from a gain
in sector sum, as explained in Sec. III.4 and in the 
second line we have extracted the renormalization factor and 
bounded $[1+ \La(w_{t(i)})/ \La(w_{{\cal A}(i)})]\leq 2$.
To optimize the bound we study the function
\be
f(z_i) = \left [\frac{2}{z_i} + { z_i\over b}
+ c \right ] 
\ee
This function has a minimum at 
$z_i=\sqrt{2b}= \sqrt{2}\La^{1\over 2}(w_{t(i)})$ whose value is
\be
\left [\frac{2\sqrt{2}}{\La^{1\over 2}(w_{t(i)})} 
+ {\La^{1\over 2}(w_{{\cal A}(i)}) 
\over \La(w_{t(i)})}  \right ]  \leq 
\frac{1}{ \La^{1\over 2}(w_{t(i)}) }  \left [
2\sqrt{2} +  {\La^{1\over 2}(w_{{\cal A}(i)}) 
\over \La^{1\over 2}(w_{t(i)})}  \right ]  \leq 
{ K \over \La^{1\over 2}(w_{t(i)})} 
\ee
This bad factor is compensated by the gain on the sector sum.

\section{Main bound}
\resetequ

Now we have all we need to perform the bound. We introduce absolute 
values inside the sums and integrals. As in [DR1], tree line propagators are
used to perform spatial integrals and the loop propagator is bounded through a
Gram inequality. The difference is that now some propagators 
(tree, loop or external) have been moved, and bear one or two derivatives, 
hence giving  a different scaling factor. Furthermore some loop propagators 
have been taken out of the determinant, 
and there are some additional distance factors to bound, coming from the
renormalization factors. 

\subsection{Loop lines}

For each $g_i\in D_l({\cal C},P)$ 
the interpolation (\ref{loopin1}) applies to the 
determinant, or to a matrix element that has been extracted. 
The distance factors and the integral over $t$ are taken out of the 
determinant by multi-linearity. Then we apply Gram inequality as in 
section IV.1 of [DR1]. Loop lines in $P$ are bounded by a Schwartz inequality
\be
|< F_f(x_f),G_g(x_g) > |\leq ||F_f|| \;||G_g||.
\ee
The interpolated half-line functions  $F_f$ or $G_g$  will have some 
factors $(k-r_f)^\mu$ or $(k-r_g)^\mu$  (actually the two ends of a matrix 
element could be both interpolated), that modify the estimation of
their norms $||F_f||$, $||G_g||$. 
For $f\in L$ being  the interpolated line for the subgraph $g_i$, 
 each $(k-r_f)_0$  and $(k-r_f)_{r(r_f)}$  
adds a factor $(\al^{-{1\over 2}})^2$ in the integral  [DR1](IV.14), 
while each $(k-r_f)_{t(r_f)}$ 
adds  a factor $\left[\La^{1\over 2}(w_{M(f,{\cal C})})\La^{1\over 2}(w_{t(i)})
\right]^2$ as we are integrating  $|F_f|^2$.  Hence, 
for each $g_i\in D_l({\cal C},P)$ the contribution to the bound at the 
first order is
\be
\lp |x_i^{(2)}-x_i^{(1)}|_0 + |x_i^{(2)}-x_i^{(1)}|_{r(r_{a_i})}\rp
\La^{1\over 4}(w_{M(h^{(2)}_i,{\cal C})})
\left [\La^3_{M(h^{(2)}_i,{\cal C})} -\La^3_{{\cal A}(m(h^{(2)}_i,{\cal C}))} 
\right ]^{1\over 2}.
\ee 
At the second order it is given by three terms:
\be
\La^{1\over 4}(w_{M(h^{(2)}_i,{\cal C})})
\left [\La^5_{M(h^{(2)}_i,{\cal C})} -\La^5_{{\cal A}(m(h^{(2)}_i,{\cal C}))} 
\right ]^{1\over 2}
\ee
for the distance factor
\be
\lp |x_i^{(2)}-x_i^{(1)}|_0 + |x_i^{(2)}-x_i^{(1)}|_{r(r_{2})}\rp^2 \ ,
\ee
\be
\La^{1\over 4}(w_{M(h^{(2)}_i,{\cal C})})
\left [\La^3_{M(h^{(2)}_i,{\cal C})} -
\La^3_{{\cal A}(m(h^{(2)}_i,{\cal C}))} \right ]^{1\over 2}
\La^{1\over 2}(w_{M(h^{(2)}_i,{\cal C})})\La^{1\over 2}(w_{t(i)}) 
\ee
for the distance factor
\be
\lp |x_i^{(2)}-x_i^{(1)}|_0 + |x_i^{(2)}-x_i^{(1)}|_{r(r_{2})}\rp \;
|x_i^{(2)}-x_i^{(1)}|_{t(r_{2})}\ , 
\ee
and 
\be
\La^{1\over 4}(w_{M(h^{(2)}_i,{\cal C})})
\left [\La_{M(a,{\cal C})} -\La_{{\cal A}(m(a,{\cal C}))} \right ]^{1\over 2}\;
\La(w_{M(h^{(2)}_i,{\cal C})})\La(w_{t(i)}) 
\ee
for 
\be
|x_i^{(2)}-x_i^{(1)}|^2_{t(r_{2})}\ .
\ee
Then, the loop determinant times the product of extracted loop propagators
is bounded by the usual term
\[
 \prod_{a\in L} \; \La^{3\over 4}_{M(a,{\cal C})} 
\left [1-\frac{\La_{{\cal A}(m(a,{\cal C}))}}{\La_{M(a,{\cal C})}}
 \right ]^{1\over 2}
\]
(where we applied the relations $\sqrt{1-x^3\over 1-x} \leq \sqrt{3}$ and
 $\sqrt{1-x^5\over 1-x} \leq \sqrt{5}$ for $x\leq 1$)
times the terms coming from renormalization: 
\bqa
&&
 \prod_{g_i\in D_l({\cal C},P)} 
\left \{ 
\lp |x_i^{(2)}-x_i^{(1)}|_0 + |x_i^{(2)}-x_i^{(1)}|_{r(r_{2})}\rp
\La(w_{{\cal A}(i)})
\right .
\\
&& 
+\lp |x_i^{(2)}-x_i^{(1)}|_0 
+ |x_i^{(2)}-x_i^{(1)}|_{r(r_{2})}\rp^2
\La^2(w_{{\cal A}(i)}) \no\\
&& +\lp |x_i^{(2)}-x_i^{(1)}|_0 + |x_i^{(2)}-x_i^{(1)}|_{r(r_{2})}\rp \;
|x_i^{(2)}-x_i^{(1)}|_{t(r_{2})} 
\La^{3\over 2}(w_{{\cal A}(i)})\La^{1\over 2}(w_{t(i)}) \no\\
&&\left .
+|x_i^{(2)}-x_i^{(1)}|^2_{t(r_{2})} 
\La(w_{{\cal A}(i)})\La(w_{t(i)}) \right \}\no\\
&&\leq 
 \prod_{g_i\in D_l({\cal C},P)} 
\left \{ 
\lp |x_i^{(2)}-x_i^{(1)}|\rp
\La(w_{{\cal A}(i)}) + |x_i^{(2)}-x_i^{(1)}|^2 
\La(w_{{\cal A}(i)})\La(w_{t(i)}) \right \}.\no
\eqa

\subsection{External lines}
It is easy to see that, when some external test function is moved, 
the bound obtained in [DR1], section IV.2, becomes
\be
 ||\phi_1||_{_{1}}
 \left [\prod_{i=2}^{2p}   ||\hat{\phi}_i||_{\infty,2} \right ]
\lp\La^{5\over 2}_T\rp^{(2p-1)}
\ee
 multiplied by the factor
\bqa
&&
 \prod_{g_i\in D_e({\cal C},P)} 
\left \{ 
\lp |x_i^{(2)}-x_i^{(1)}|_0 + |x_i^{(2)}-x_i^{(1)}|_{r(r_2)}\rp
\La(w_{{\cal A}(i)}) 
\right .
\no\\
&& 
+\lp |x_i^{(2)}-x_i^{(1)}|_0 
+ |x_i^{(2)}-x_i^{(1)}|_{r(r_2)}\rp^2
\La^2(w_{{\cal A}(i)}) \no\\
&& +\lp |x_i^{(2)}-x_i^{(1)}|_0 + |x_i^{(2)}-x_i^{(1)}|_{r(r_2)}\rp \;
|x_i^{(2)}-x_i^{(1)}|_{t(r_2)} 
\La^{3\over 2}(w_{{\cal A}(i)}) \La^{1\over 2}(w_{t(i)}) \no\\
&&\left .
+|x_i^{(2)}-x_i^{(1)}|^2_{t(r_2)} 
\La(w_{{\cal A}(i)})\La(w_{t(i)}) \right \}\no\\
&&\leq \left \{ \lp |x_i^{(2)}-x_i^{(1)}|\rp \La(w_{{\cal A}(i)}) +
|x_i^{(2)}-x_i^{(1)}|^2 \La(w_{{\cal A}(i)})\La(w_{t(i)}) \right \}\ .
\eqa
where $||\hat{\phi}_i||_{\infty,2}$ has been defined in (\ref{phinorm}).

\subsection{Tree lines.}

\begin{figure}
\centerline{\psfig{figure=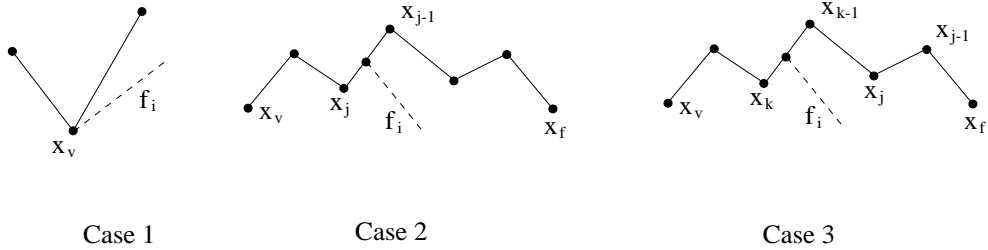,width=13cm}}
\caption{Three possible updatings of the tree}\label{fermi18} 
\end{figure}

As we said,  interpolated tree lines  are moved along the connection 
between the external vertices of any graph provided by the tree. But, as the 
tree itself is modified by renormalization, this process has to be inductive, 
starting from the smallest graph and going down towards the biggest. 
We take for this construction the same rules as in [DR2], with some 
simplifications
as we do not treat four point subgraphs. 
Remark that only the renormalization of subgraphs in $D_t({\cal C},P)$ 
can modify the tree.
Our induction creates  progressively a new tree $\tree({\cal J})$. 
To describe it,
we number the subgraphs  in $D_t({\cal C},P)$
in the order we meet then $g_1,...g_r$. At the stage $1\leq p\leq r$, 
before renormalization of $g_p$, the tree is called  
$\tree({\cal J}_{p-1})$.
Then we interpolate the external line of   $g_p$ following the unique path in
 $\tree({\cal J}_{p-1})$ connecting the two external vertices of 
$g_p$. Then we update ${\cal J}$ and $\tree$. 
We define ${\cal J}_p={\cal J}_{p-1}$ 
for the first order term, as the propagator
hooks to the reference vertex,  
${\cal J}_p={\cal J}_{p-1}\cup \{j\}\cup  \{k\}$ for the
second order term, where $j$ and $k$ are the indices of the lines of 
$\tree({\cal J}_{p-1})$ chosen by renormalization. 
Finally we update the tree according to 
Fig.\ref{fermi18}.

In the following we will call $D^1_t({\cal C},P)$ the set of subgraphs
with the interpolated line fixed by ${\cal J}$  on the reference vertex
(hence giving a first order term), and 
$D^2_t({\cal C},P)$  the set of subgraphs
with the interpolated line fixed by ${\cal J}$ 
on some tree line $x_k-x_{k-1}$,
$k\leq j$
(hence giving a second  order term).

\subsubsection{Spatial decay}
It is easy to see that the interpolated  line 
for the subgraph $g_i$ has the same spatial
decay as the non interpolated one [DR1](IV.24), times a factor 
\be
\lp |x_i^{(2)}-x_i^{(1)}|_0 + |x_i^{(2)}-x_i^{(1)}|_{r(2)}\rp
\La(w_{{\cal A}(i)})
\ee
if $g_i\in D^1_t({\cal C},P)$. If $g_i\in D^2_t({\cal C},P)$ we have 
three multiplying factors depending on the components of the scaling factors
\be
|x_{j}-x_{j-1}|_\mu |x_{k}-x_{k-1}|_\nu.
\ee
If $\mu$ and $\nu\in (0,r(r_2))$ we have the multiplying factor 
$\La^2(w_{{\cal A}(i)})$. If $\mu$, or $\nu$ 
is $t(r_{2})$ and the other belongs to $(0,r(r_{2}))$    
we have the factor 
$\La^{3\over 2}(w_{{\cal A}(i)}) \La^{1\over 2}(w_{t(i)})$. 
Finally,  if 
$\mu$ and $\nu= t(r_{2})$ we have the factor 
$\La(w_{{\cal A}(i)})\La(w_{t(i)})$.

Before going on we take a fraction $(1-\vep)$ of the exponential decay
to ensure the decay between the test function supports of Theorem 3
as in [DR1](IV.25). Of the  remaining 
decay  a fraction ${\vep\over 2}$ will be used   to bound the distance factors
and the other to perform spatial integrals. 

\subsubsection{Bounding distance factors}

For each renormalized subgraph $g_i$ we have to bound one or two
distance factors, depending if it belongs to $D^1({\cal C},P)$ 
or $D^2({\cal C},P)$, which are the subsets of
subgraphs that give a first order or a second order term respectively. 
These sets can be cut in turn into  $D^m_l({\cal C},P)$,  
$D^m_e({\cal C},P)$ and  $D^m_t({\cal C},P)$, $m=1,2$,
for loop, external and tree lines respectively moved. 
Then we have to bound the quantity
\bqa
&& A(x,{\cal J},\tree)=
\prod_{g_i\in D^1({\cal C},P)} |x_i^{(2)}-x_i^{(1)}|
\prod_{g_i\in D^2_t({\cal C},P)} |x_j-x_{j-1}| |x_k-x_{k-1}|\no\\ 
&& 
\prod_{g_i\in D^2_l({\cal C},P)\cup D^2_e({\cal C},P)} 
|x_i^{(2)}-x_i^{(1)}|^2
\prod_{l\in \tree({\cal J})} e^{-a{\vep \over 2} 
(|{\bar x}_l-x_l|\La(w_l))^{1\over s}}\no
\eqa
where we have taken the same spatial decay (actually the worst) for 
all directions.
For each loop or external line the difference $|x_i^{(2)}-x_i^{(1)}|$ 
can be bounded, applying several 
triangular inequalities, by the sum over the tree lines on the 
unique path in $\tree({\cal J})$ connecting  $x_i^{(2)}$ to $x_i^{(1)}$. 

We observe that the same tree line $l_j$ can appear in several 
paths connecting different pairs of points  $x_i^{(2)}$, $x_i^{(1)}$. 
Using the same fraction of its exponential
decay many times might generate some unwanted factorials as
$\sup_x x^n \exp(-x) = (n/e)^n$. To avoid this problem we define 
$D_j$ as the set of subgraphs $g_i\in D({\cal C},P)$ that use the 
tree distance $|{\bar x}_{l_j}-x_{l_j}|$ and we apply the relation
\be
 e^{-a{\vep \over 2} [|{\bar x}_{l_j}-x_{l_j}|\La(w_{l_j})]^{1\over s}}
\leq e^{-a{\vep \over 2} [|{\bar x}_{l_j}-x_{l_j}|]^{1\over s}
\sum_{g_i\in D_j}[\La(w_{t(i)})^{1\over s} -\La(w_{{\cal A}(i)})^{1\over s} ]}
\ee
With this expression a different decay factor is used for each subgraph. 
Now applying this result and the inequality $x e^{- (x)^{1\over s}}\leq s! $ 
we prove the bound:
\bqa
\sup_x |A(x,{\cal J},\tree)| &\leq & K(s)^{\bar n} \prod_{g_i\in D^1({\cal C},P)} 
\La(w_{t(i)})^{-1} 
\left [1 -\lp{\La(w_{{\cal A}(i)})\over  \La(w_{t(i)})}\rp^{1\over s}
 \right ]^{-s}\no\\
&& \prod_{g_i\in D^2({\cal C},P)} \La(w_{t(i)})^{-2}
\left [1 -\lp {\La(w_{{\cal A}(i)})\over  \La(w_{t(i)})}\rp^{1\over s}
 \right  ]^{-2s}
\eqa
where $K(s)$ is some function of $s$. 
The remaining differences are dangerous as they appear with a negative 
exponent. This happens  because  in this continuous
formalism one has to perform renormalization even when the differences 
between internal and external scales of subgraphs are arbitrary small. 
The solution of this problem is given by loop lines factors. 
Indeed any renormalized subgraph has necessarily internal loop lines,
which give small factors when the differences  between internal and external
scales of subgraphs become arbitrarily small.
By Lemma 9 in [DR2] we know that, for each $g_i\in D({\cal C},P)$ 
there are at least two loop lines internal to $g_i$ which satisfy 
$\La(w_{M(a,{\cal C})})\leq\La(w_{t(i)})$ and 
$\La(w_{{\cal A}(m(a,{\cal C}))})\geq \La(w_{{\cal A}(i)})$. 
Then for each
$g_i\in D^1({\cal C},P)$ we have to bound 
\be
f_1(x) = \frac{ [1 -x ]^2}{[1 - (x)^{1\over s}]^{s}}
\ee
and  for each $g_i\in D^2({\cal C},P)$ 
\be
f_2(x)=\frac{ [1 - x]^2 }
{ [1 - (x)^{1\over s} ]^{2s}}\ ,
\ee
where we defined $x= \La(w_{{\cal A}(i)}) /\La(w_{t(i)})$. 
Remark that $f_1(x)\simeq (1-x)^{2-s}$ for $x\rightarrow 1$ 
while  $f_2(x)\simeq (1-x)^{-2(s-1)}$. Therefore choosing 
$1<s<3/2$, $f_1$ is bounded near $x=1$, and $f_2$ is integrable. 
We bound 
\be
f_1(x) \leq \sup_{x\in [0,1]} f_1(x)
\ee
and we keep $f_2$ to be bounded when the integration over the
parameters $w$ will be performed. 
Finally the factors  $\left [1
-{\La({w}_{{\cal A}(m(a,{\cal C}))})\over \La({w}_{M(a,{\cal C})})
}\right ]^{\frac{1}{2}}$ that are not used are bounded by 1. 
\subsubsection{Sum over ${\cal J}$}
We bound the sum over ${\cal J}$ by taking the  $\sup_{\cal J}$ times 
the cardinal of ${\cal J}$. In [DR2], Lemma 7, it is proved
that  $|{\cal J}|\leq K^{\bar n}$ for some constant $K$. 

\subsubsection{Spatial integration}  To perform spatial integration we 
use the remaining tree line decay 
\be
\prod_{l\in \tree({\cal J})} e^{-a{\vep\over 2}\left [
|(\de x_l)_0 \La(w_l)|^{{1\over s}}+|(\de x_l)_r \La(w_l)|^{{1\over s}}+
 |(\de x_l)_t \La^{{1\over 2}}(w_l)|^{{1\over s}}
\right ]}\ .
\ee
These lines depend in general from the interpolation parameters $t$.
In [DR2] it is proved that spatial integration performed with 
interpolated  tree lines does not depend from the interpolating factor
$t$ and give the same result as integration with the starting tree $\tree$.

Summarizing the results, tree lines are used for several purposes:
extracting the exponential decay between the 
test functions supports,   
bounding distance factors and performing spatial integration. 
The resulting bound is:
\bqa
&& e^{-a\;(1-\vep)\;\La_T^{1\over s}\;  d^{1\over s}_T(\Om_1,...,\Om_{2p}) }
 \prod_{q=1}^{\bar n} \frac{1}{\La^3(w_q)}
\prod_{g_i\in D^1({\cal C},P)} 
\La(w_{t(i)})^{-1} \no\\
&& \prod_{g_i\in D^2({\cal C},P)} \La(w_{t(i)})^{-2}
\left [1 -\lp {\La(w_{{\cal A}(i)})\over  \La(w_{t(i)})}\rp^{1\over s}
 \right  ]^{-2s}\ .
\eqa

\subsection{Sector sum}

We still have to perform the sum over sector choices corresponding 
to [DR1], (IV.28). 
We do it in the same way as in section IV.3 of [DR1]. 
The only difference is that, for a two-point subgraph $g_i$, 
by  momentum conservation, there is no sector choice at all:
\be
\left [{\scriptstyle {4\over 3}\La^{-{1\over 2}}\lp 
w_{j_{h^{(2)}_i,r(i)}}\rp}\right ]
\int_{\Si_{j_{h^{(2)}_i,r(i)+1}}}
d\th_{h^{(2)}_i,r(i)}
\Upsilon\lp\th^{root}_i,\th_{h^{(2)}_i,r(i)} \rp \leq K\ ,
\ee
and, for each $g_i\in D({\cal C},P)$ we have to count the number of
choices for the additional 
refinement for the half-line $h_i^{(2)}$  from 
a sector of size $\La^{1\over 2}(w_{i(h_i^{(2)})})$ into  a sector of size
$\La^{1\over 2}(w_{i(h_i^{(2)})}\La^{1\over 2}(w_{t(i)})$. This costs a factor
$\La^{-{1\over 2}}(w_{t(i)})$. 
This term is dangerous as it is on the denominator. To compensate it we 
extract
a factor $\La^{{1\over 2}}(w_{t(i)})$  from the 
subgraphs $g_j$ of $g_i$ defined 
above. This factor is extracted inductively for $j\in C_i^r$.  
For each subgraph $g_j$ we distinguish two situations: 
\begin{itemize}
\item{}  if $|eg_j({\cal C})|>4$ we insert the identity 
$1 = {\La^{1\over 2}(w_{{\cal A}(j)}) \over \La^{1\over 2}(w_{j})}\;\;
{  \La^{1\over 2}(w_{j}) \over \La^{1\over 2}(w_{{\cal A}(j)})}
$,
where the second factor will be compensated by the convergent 
power counting of the subgraph $g_j$;
\item{} if  $|eg_j({\cal C})|=4$ we observe  (see Lemma \ref{anotherlemma}
below) 
that we have counted one unnecessary
sum over sector  choices and we gain again a factor 
${\La^{1\over 2}(w_{{\cal A}(j)}) / \La^{1\over 2}(w_{j})}$.

\end{itemize} 
Putting together all these terms we obtain the factor we want, namely
 $\La^{{1\over 2}}(w_{t(i)})$, times a factor
\be
\prod_{g_j | j\in C^r_i, |eg_j({\cal C})|>4}
{  \La^{1\over 2}(w_{j}) \over \La^{1\over 2}(w_{{\cal A}(j)})}\ .
\ee

\begin{lemma}
Let the two point subgraph  
$g_i\in D({\cal C},P)$, and  the four point subgraph 
$g_j$, $j\in  J_i$,
be fixed.  
Then the number of sector choices predicted by [DR1], Lemma 6, (IV.31)  
must be modified:
\be
\prod_{m=2}^4
\left [{\scriptstyle {4\over 3}\La^{-{1\over 2}}\lp w_{{\cal A}(j)}\rp}\right ]
\int_{\Si_{j_{h^{(m)}_{j},r(j)+1}}}
d\th_{h_j^{(m)}}
\Upsilon\lp\th_j^{(root)},\{\th_{h_j^{(m)},r(j)}\}_{m=2,3,4}\rp \leq K
\ee
for some constant $K$. \label{anotherlemma}
\end{lemma}
 
\paragraph{Proof}
We observe that $\th_{h_j^{(2)}}$ actually is fixed by the 
momentum conservation for the two external lines of $g_i$ 
on an interval of size   
$\La^{1\over 2}(w_{{\cal A }(i)})$. 

In the following we write explicitly the $q$ dependence: 
 $j=j_q$.  We distinguish then three possible situations.

\begin{figure}
\centerline{\psfig{figure=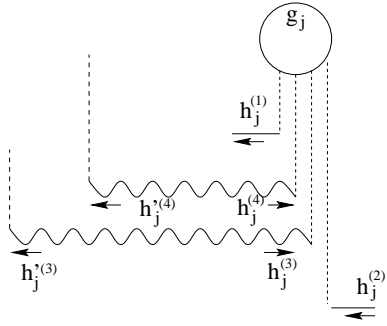,width=5cm}}
\caption{Tree lines are solid, loop lines are wavy; 
the arrows show the  direction of the sector sum}\label{fermi13}
\end{figure}

\paragraph{1.} 
$h_{j_q}^{(3)}$ and $h_{j_q}^{(4)}$  are both loop half-lines 
(see Fig.\ref{fermi13}). 
Then they are contracted to some half-lines  
$h_{j_q}^{'(3)}$ and $h_{j_q}^{'(4)}$, that belong to some sector of size
 $\La^{1\over 2}(w_{M^r\lp h_{j_q}^{'(m)}, {\cal C}\rp }) 
\leq \La^{1\over 2}(w_{{\cal A}(j_q)})$, $m=3,4$. 
Therefore by momentum conservation $\th_{h_{j_q}^{(m)}}$
is restricted  on the sector of $h_{j_q}^{'(m)}$, for  $m=3,4$.

\paragraph{2.} 
$h_{j_q}^{(3)}\in L$ and $h_{j_q}^{(4)}\in t_i$. 
Then we have  two situations.   

\begin{figure}
\centerline{\psfig{figure=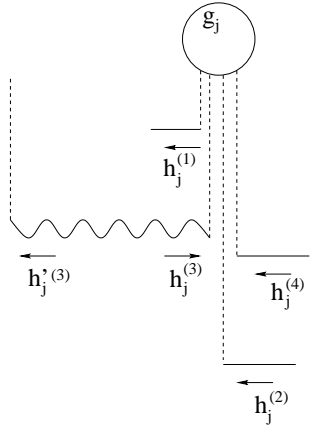,width=4cm}}
\caption{$h^{(3)}_{j}$  
contracts with some element of  $L^R_{j}(i)$ }\label{fermi14}
\end{figure}

When $h^{(3)}_{j_q}$  
contracts with some element of  $L^L_{j_q}(i)$ (see Fig.\ref{fermi14}), 
repeating the argument above, $\th_{h^{(3)}_{j_q}}$ is restricted to 
an interval of width $\La^{1\over 2}(w_{{\cal A}(j_{q})})$, 
and, by momentum conservation ([DR1], Appendix B)
$\th_{h^{(4)}_{j_q}}$ is restricted
to an interval of the same size.

\begin{figure}
\centerline{\psfig{figure=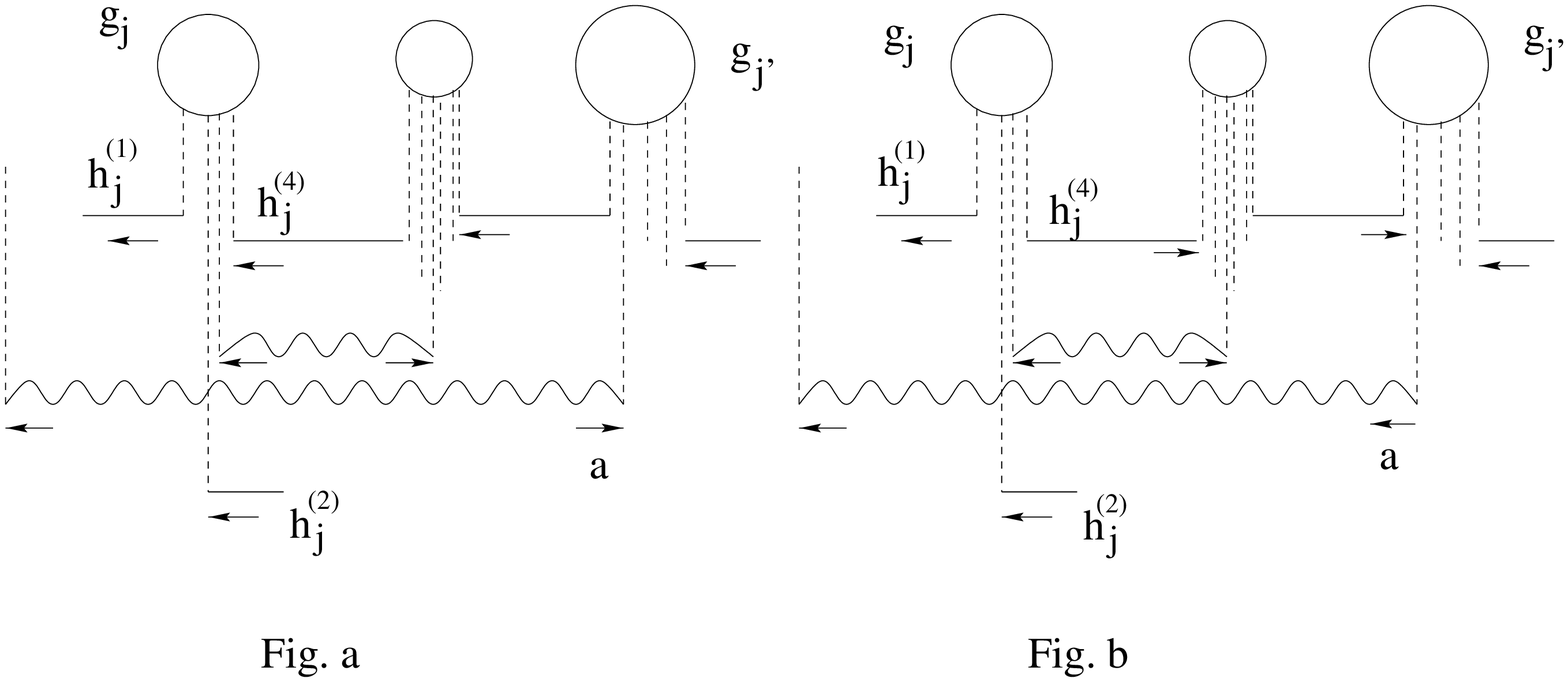,width=14cm}}
\caption{$h^{(3)}_{j_{q}}$  
contracts with some element of  $L^L_{j}(i)$}\label{fermi15}
\end{figure}

When $h^{(3)}_{j_q}$  
contracts with some element of  $L^R_{j_q}(i)$,  there
is a loop line $a$ connecting   $L^L_{j_q}(i)$ with 
$L^R_{j_q}(i)$. This line is external line of some subgraph in 
 $\tree^R_{j_q}(i)$, say $g_{j'}$ (see Fig.\ref{fermi15}a). 
Then we  chose as root half-line for $g_{j'}$ the loop
half-line  $a$ instead of  the tree half-line  $h^{root}_{j'}$ 
and, for all tree lines on the unique path connecting
$v^{root}_{j'}$ to $v^{(4)}_{j_q}$ we can exchange 
$h^L$ and $h^R$ (see Fig.\ref{fermi15}b; the new arrows show the direction
towards this new root). Then $\th_{h^{(4)}_{j_q}}$  is
fixed in an interval of size 
$\La^{1\over 2}(w_{{\cal A}(j_q)})$. 

\begin{figure}
\centerline{\psfig{figure=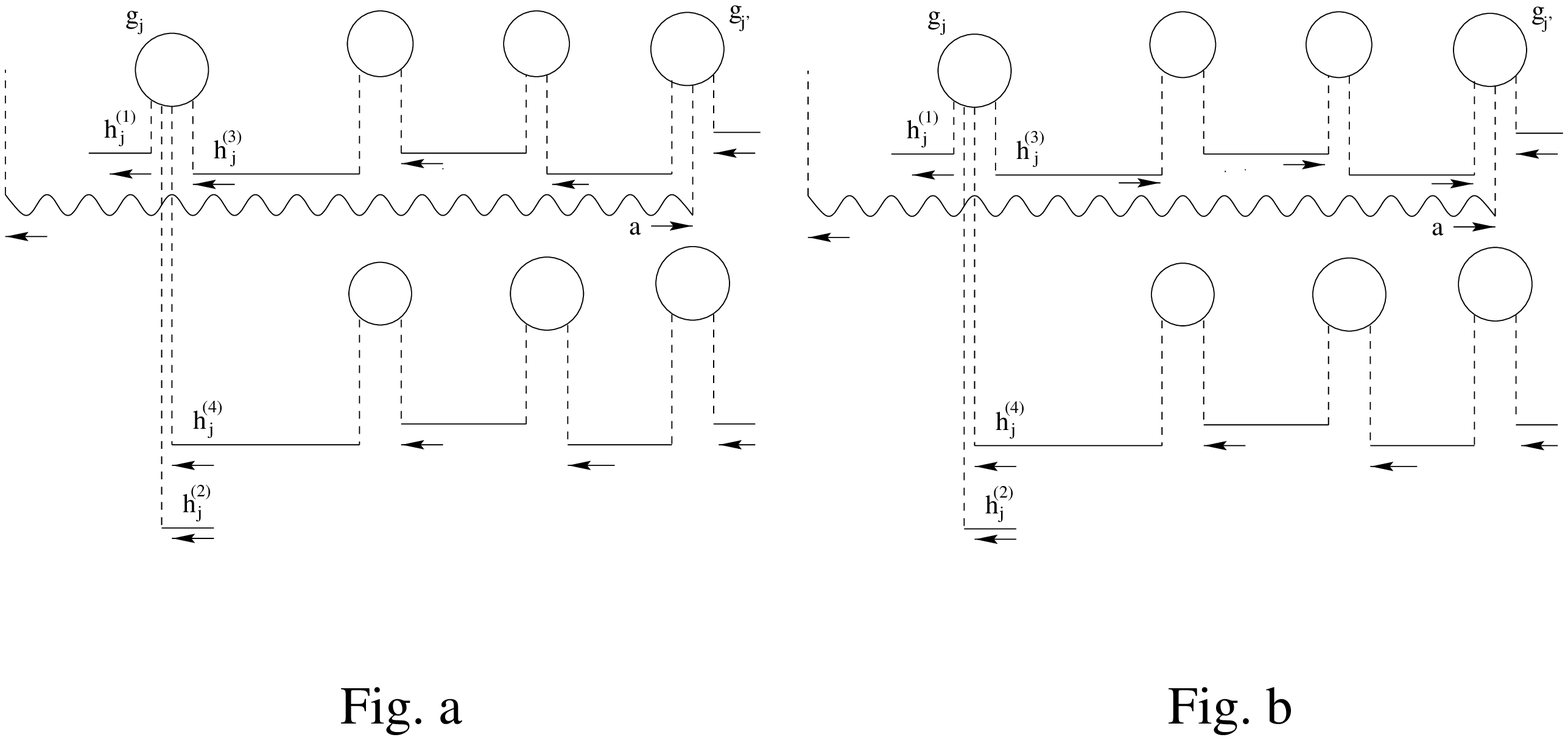,width=14cm}}
\caption{$h^{(3)}_{j_{q}}$ and $h^{(3)}_{j_{q}}$ are tree half-lines,
and there is a loop line  
connecting  $T^{L(3)}_{j_{q}}(i)$ with $T^R_{j_{q}}(i)$ }\label{fermi16}
\end{figure}

\paragraph{3.} 
$h_{j_q}^{(3)}$ and $h_{j_q}^{(4)}\in t_i$.   
Remark that $\tree^R_{j_q}(i)$ is separated into 
two subtrees, $T^{R(3)}_{j_q}(i)$ which  is connected to
$g_{j_q}$ through $l_j^{(3)}$ and $T^{R(4)}_{j_q}(i)$ 
which  is connected to
$g_{j_q}$ through $l_j^{(4)}$. There is a loop half-line $a$ 
hooked to $T^{R(3)}_{j_q}(i)$ or to $T^{R(4)}_{j_q}(i)$, contracting
to some loop half-line in  $L^L_{j_q}(i)$. Let's say  $a$ is
hooked in $T^{R(3)}_{j_q}(i)$ (see Fig.\ref{fermi16}a). 
Then,  repeating the same argument above
(see Fig.\ref{fermi16}b), $\th_{h^{(3)}_{j_q}}$ and $\th_{h^{(4)}_{j_q}}$ are
fixed in an interval of size 
$\La^{1\over 2}(w_{{\cal A}(j_{q})})$. 
This ends the proof.
\qed

\subsection{Integration over the parameters $w_i$}

Putting everything together, we can bound the sum (\ref{conv2r}): 

\bqa
\lefteqn{|\Ga_{2p}^{\La\Lazero}| \leq 
e^{-a\;(1-\vep)\;\La_T^{1\over s}\;  d^{1\over s}_T(\Om_1,...,\Om_{2p}) }
{\scriptstyle ||\phi_1||_{_1}
 \prod_{i=2}^{2p} ||\hat{\phi}_i||_{\infty,2} }}\\
&& K_0\;
\sum_{{\bar n}\geq 1} \frac{c^{\bar n}}{n!n'!} K^{\bar n}
\sum_{CTS}\sum_{u-\tree}\sum_{\cal L}\sum_{\Om\; E}
\sum_{{\cal C}}\sum_{J P}
\int_{w_T\le w_{{\cal A}(i)}
\le  w_{i}\le 1} \prod_{i=1}^{{\bar n}-1} dw_i \no\\
&& \left [ \La^{5\over 2}(w_T)\right ]^{(2p-1)}
\prod_{i=1}^{{\bar n}-1} \frac{1}{\La^3(w_i)}
 \prod_{a\in L}
\;\La^{3\over 4}\lp w_{M(a,{\cal C})}\rp   \prod_{v\in V'}
\La(w_{i_v})\no\\
&&
\prod_{g_i |\  i\neq r\ {\rm or}  \atop |eg_i({\cal C})|\geq 11}\left [
\prod_{h\in eg^\ast_i \atop j_{h,1}=i(h)={\cal A}(i)}
\frac{ \La^{1\over 2}\lp w_{j_{h,2}}\rp}{\La^{1\over 2}\lp w_{j_{h,1}}\rp }
\right ]
\prod_{g_i|\ i=r\ {\rm or} \atop 2\leq |eg_i({\cal C})|\leq 10 }
 \left [
\prod_{h\in I(i)} \frac{\La^{1\over 2}\lp w_{j_{h,r(i)+1}}\rp}
{\La^{1\over 2}\lp w_{j_{h,r(i)}}\rp}
\right ] \no\\
&&
 \La_T^{-{1\over 2}}\;
\prod_{v\in V}\La^{-{1\over 2}}\lp w_{i_v}\rp
\prod_{g_i\in D({\cal C},P)} 
{\La(w_{{\cal A}(i)})\over\La(w_{i}) }
 \no\\
&& \prod_{g_i\in D({\cal C},P)} \left [
\prod_{j\in C_i^r\backslash J_i}
 \lp {\La(w_{i})\over \La(w_{{\cal A}(i)})}\rp^{1\over 2} \right ]  
   \prod_{g_i\in D^2({\cal C},P)} \frac{
\left [1 - {\La(w_{{\cal A}(i)})\over  \La(w_{t(i)})}
 \right  ]^{2}}{
\left [1 -\lp {\La(w_{{\cal A}(i)})\over  \La(w_{t(i)})}\rp^{1\over s}
 \right  ]^{2s}}\ , \no
\eqa
where  we  bounded 
$\de\mu^{\La(w_{i_v})}_\La(\la)\leq K |\la|\La(w_{i_v}) $ and
$|\la|\leq c$. Remark that sector counting for vertex
gives a factor depending from $V$ only, as for two point vertex 
no sum has to be paid.  
Now we can send $\La$ to zero, hence $\La(w)=\sqrt{w}$ as $\La_0=1$.
The bound becomes
\bqa
&&\hspace{-0.8cm} |\Ga_{2p}^{\Lazero}|\leq
K_0\;
e^{-a\;(1-\vep)\;\La_T^{1\over s}\;  d^{1\over s}_T(\Om_1,...,\Om_{2p}) }
{\scriptstyle ||\phi_1||_{_1}
 \prod_{i=2}^{2p} ||\hat{\phi}_i||_{\infty 2}}
\sum_{{\bar n}\geq 1} \frac{c^{\bar n}}{n!n'!} K^{\bar n} \\
&&\hspace{-0.5cm}\sum_{CTS}\sum_{u-\tree}\sum_{\cal L}\sum_{\Om\; E}
\sum_{{\cal C}}\sum_{J P}
\int_{w_T\le w_{{\cal A}(i)}
\le  w_{i}\le 1} \prod_{i=1}^{{\bar n}-1} dw_i 
\prod_{i=1}^{{\bar n}-1} w_i^{-{3\over 2}} \
 \prod_{a\in L}
\; w_{M(a,{\cal C})}^{3\over 8}\;
 \prod_{v\in V'} w_{i_v}^{1\over 2}
\no\\
&&\hspace{-0.5cm}\prod_{v\in V }w_{i_v}^{-{1\over 4}}
\prod_{g_i|\ i\neq r \ {\rm or} \atop  |eg_i({\cal C})|\geq 11}\left [
\prod_{h\in et_i^\ast\cup el_i\atop j_{h,1}=i(h)={\cal A}(i)}
\frac{w_{j_{h,2}}^{1\over 4}}{w_{j_{h,1}}^{1\over 4}}\right ]
\prod_{g_i|\ i=r \ {\rm or} \atop 2\leq  |eg_i({\cal C})|\leq 10 }
 \left [
\prod_{h\in I(i)} \frac{w_{j_{h,r(i)+1}}^{1\over 4}}
{w_{j_{h,r(i)}}^{1\over 4}}
\right ] \no\\
&&\hspace{-0.5cm} w_T^{{5p\over 2}-{3\over 2}}  
 \prod_{g_i \in D({\cal C},P)} 
{w^{1\over 2}_{{\cal A}(i)}\over w^{1\over 2}_{i} }
 \prod_{g_i\in D({\cal C},P)} \left [
\prod_{j\in C_i^r\backslash J_i}
 {w^{1\over 4}_{i}\over w^{1\over 4}_{{\cal A}(i)}} \right ]
  \prod_{g_i\in D^2({\cal C},P)} \frac{
\left [1 - {w^{1\over 2}_{{\cal A}(i)}\over  w^{1\over 2}_{t(i)}}
 \right  ]^{2}}{
\left [1 -\lp {w^{1\over 2}_{{\cal A}(i)}\over  
w^{1\over 2}_{t(i)}}\rp^{1\over s}
 \right  ]^{2s}}\ . \no\eqa
Now we can introduce the variables $\bt_i$ exactly as in section IV.4
of [DR1] (see (IV.38-40)) and obtain in these new coordinates:

\bqa
&&\hspace{-0.8cm} |\Ga_{2p}^{\Lazero}|\leq K_0\;
{\scriptstyle ||\phi_1||_{_1}
 \prod_{i=2}^{2p} ||\hat{\phi}_i||_{\infty,2} }
e^{-a\;(1-\vep)\;\La_T^{1\over s}\;  d^{1\over s}_T(\Om_1,...,\Om_{2p}) }
\sum_{{\bar n}\geq 1} \frac{c^{\bar n}}{n!n'!} K^{\bar n}\label{gammar}\\
&&\hspace{-0.5cm}\sum_{CTS}\sum_{u-\tree}\sum_{\cal L}\sum_{\Om\; E}
\sum_{{\cal C}}\sum_{J P}
\int_{w_T}^1 \prod_{i=1}^{{\bar n}-1} d\bt_i
 \; w_T^{{\bar n}-1}\; w_T^{{5p\over 2}-{3\over 2}}\; 
\prod_{i=1}^{{\bar  n}-1} \bt_i^{-1+(1-{\bar n}_i)}
\no\\
&&\hspace{-0.5cm} 
\prod_{i=1}^{{\bar n}-1}\left [
\lp\prod_{j\in C_i}\bt_i^{3\over 2}\rp w_T^{-{3\over 2}}\right ]
\; \prod_{a\in L}\left [
\;\lp\prod_{j\in C_{M(a,{\cal C})}}\bt_{j}^{-{3\over 8}}\rp w_T^{3\over 8}
\right ] 
\prod_{v\in V'} \left [\lp \prod_{j\in C_{i_v}} 
\frac{1}{\bt^{1\over 2}_j}\rp w_T^{1\over 2}\right ]\no\\  
&&\hspace{-0.5cm} \prod_{g_i|\ i\neq r \ {\rm or}\atop |eg_i({\cal C})|\geq 11}\Bigg[
\prod_{h\in eg_i^\ast\atop j_{h,1}=i(h)={\cal A}(i)}
\lp \prod_{j\in C_{r(i)+1}\backslash C_{r(i)} }
\frac{1}{\bt_{j}^{1\over 4}}\rp
\Bigg] \no\\
&&\hspace{-0.5cm}\prod_{g_i|\ i=r \ {\rm or}\atop 2\leq |eg_i({\cal C})| \leq 10 }
 \Bigg[
\prod_{h\in I(i)}\lp\prod_{j\in C_{r(i)+1}\backslash C_{r(i)} }
\frac{1}{\bt_{j}}\rp^{1\over 4}
\Bigg] \;
\prod_{v \in V}\left [ \lp\prod_{j\in C_{i_v}}\bt_{j}\rp^{1\over 4}
w_T^{-{1\over 4}}\right ]
\no\\
&&\hspace{-0.5cm} \prod_{
 g_i \in D({\cal C},P)} \bt^{1\over 2}_i 
\prod_{g_i\in D({\cal C},P)} \left [
\prod_{j\in C_i^r\backslash J_i}
 {1\over \bt_j^{1\over 4}} \right ]
  \prod_{g_i\in D^2({\cal C},P)} \frac{
\left [1 - \prod_{j\in C_{t(i)}\backslash C_{{\cal A}(i)}} \bt_j 
 \right  ]^{2}}{
\left [1 -\lp\prod_{j\in C_{t(i)}\backslash 
C_{{\cal A}(i)}} \bt_j \rp^{1\over s}
 \right  ]^{2s}}\ . \no
\eqa
where ${\bar n}_i= n_i+n'_i$, and $n_i$, $n'_i$ are respectively
the number of four points and two points vertex in $g_i$. 
Now we compute power counting as in [DR1], section IV.4, and we obtain the 
same expressions, substituting $n$ by ${\bar n}$. 
The only different expressions are
\bqa
\prod_{v\in V} \lp \prod_{j\in C_{i_v}} \bt_j \rp^{1\over 4} 
&=& \prod_{i=1}^{{\bar n}-1} \bt_i^{n_i\over 4}\no\\
\prod_{v\in V'} \lp \prod_{j\in C_{i_v}} \bt_j \rp^{-{1\over 2}}
&=&  \prod_{i=1}^{{\bar n}-1} \bt_i^{-{n'_i\over 2}}
\eqa 
Then we obtain
\bqa
&&\hspace{-0.8cm} |\Ga_{2p}^{\Lazero}|\leq K_0\;
{\scriptstyle ||\phi_1||_{_1}
 \prod_{i=2}^{2p} ||\hat{\phi}_i||_{\infty,2} }
e^{-a\;(1-\vep)\;\La_T^{1\over s}\;  d^{1\over s}_T(\Om_1,...,\Om_{2p}) }
 w_T^{{7p\over 4}-{1\over 4}} \sum_{{\bar n}\geq 1}
\frac{c^{\bar n}}{n!n'!} K^{\bar n}\no\\
&&\hspace{-0.8cm}\sum_{CTS}\sum_{u-\tree}\sum_{\cal L}\sum_{\Om\; E}
\sum_{{\cal C}}\sum_{J P} 
\int_{w_T}^1 \prod_{i=1}^{{\bar n}-1} d\bt_i\; \bt_i^{-1+x_i}
  \prod_{g_i\in D^2({\cal C},P)} \frac{
\left [1 - \prod_{j\in C_{t(i)}\backslash C_{{\cal A}(i)}} \bt_j 
 \right  ]^{2}}{
\left [1 -\lp\prod_{j\in C_{t(i)}\backslash 
C_{{\cal A}(i)}} \bt_j \rp^{1\over s}
 \right  ]^{2s}}\; . \no\\
\eqa
To compute $x_i$ we apply the following relation 
\be
|il_i({\cal C})| = 2n_i +2 -|eg_i({\cal C})|.
\ee
For all $g_i$ with $|eg_i({\cal C})|>4$ such that there exists some 
$g_{i'}\in D({\cal C},P)$ with 
$i\in C_{i'}^r\backslash J_{i'}$, the factor   $x_i$ is given by:
\bqa
x_i &=& {1\over 2}({\bar n}_i-1) -{3\over 8}|il_i({\cal C})| +{1\over 4}n_i
-{1\over 2}n'_i -{1\over 4}(|eg_i({\cal C})|-3) -{1\over 4}\no\\
&=&
{1\over 8} (|eg_i({\cal C})|-6) \quad {\rm if}\ 
4 < |eg_i({\cal C})|\leq 10
 \no\\
x_i &\ge& {1\over 2}({\bar n}_i-1) -{3\over 8}|il_i({\cal C})| +{1\over 4}n_i
-{1\over 2}n'_i -{1\over 4}(|eg_i({\cal C})|-1) -{1\over 4}\no\\
&=& {1\over 8} (|eg_i({\cal C})|-10) \quad {\rm if}\  
|eg_i({\cal C})|> 10\ ,
 \eqa
where the last term $-1/4$ corresponds to the factor extracted to 
perform sector sum in section III.4. 
For the remaining $g_i$ with 
$|eg_i({\cal C})|\geq 4$ we have the usual power counting
\bqa
x_i &=& {1\over 2}({\bar n}_i-1) -{3\over 8}|il_i({\cal C})| +{1\over 4}n_i
-{1\over 2}n'_i -{1\over 4}(|eg_i({\cal C})|-3)\no\\
&=&
{1\over 8} (|eg_i({\cal C})|-4) \quad {\rm if}\ 
4\leq |eg_i({\cal C})|\leq 10
 \no\\
x_i &\ge& {1\over 2}({\bar n}_i-1) -{3\over 8}|il_i({\cal C})| +{1\over 4}n_i
-{1\over 2}n'_i -{1\over 4}(|eg_i({\cal C})|-1)\no\\
&=& {1\over 8} (|eg_i({\cal C})|-8) \quad  {\rm if}\ 
|eg_i({\cal C})|> 10.
\eqa
Remark that in the first situation six-points subgraphs  become
logarithmic divergent, while the other ones still have $x_i>0$; this is a price
to pay for our anisotropic analysis. However by Lemma \ref{exterpow}
$x_i$ is still proportional to the number of 
tree external lines $|et_i|$, which is crucial to perform the sum over 
partial orders. This is the reason why, when introducing classes in [DR1],  
we have selected up to 11 external lines per subgraph.

Finally we consider two-point subgraphs.
For all $g_i\in D({\cal C},P)$ we have
\be
x_i = {1\over 2}({\bar n}_i-1) -{3\over 8}|il_i({\cal C})| +{1\over 4}n_i
-{1\over 2}n'_i +{1\over 2} = {1\over 2}-{1\over 2}=0
\ee
where the term $1/2$ comes from renormalization. 
The corresponding 
power counting is logarithmic in $T$. We have still to consider 
the 1PR two-point subgraphs
$g_i\in  ND({\cal C},P)$, that have $x_i=-1/2$.
Their external momentum at scale ${\cal A}(i)$ is equal to that of 
some internal line $l_j$. Since our Gevrey cutoffs have compact support,
this forces a relation between 
external and internal scales, namely
$\La(w_i)\leq \sqrt{2}\La(w_{{\cal A}(i)})$. This means 
$w_i\leq 2 w_{{\cal A}(i)}$, or equivalently
$\bt_i\geq 1/2$. The corresponding integral is then bounded 
by a constant:
$\int_{1\over 2}^1 d\bt_i \; \bt_i^{-1-{1\over 2}} = 2(\sqrt{2}-1) $.

Now Lemma 8 in [DR1] can be generalized:
\begin{lemma}
For any subgraph $g_i$ ($i\neq r$) with $|eg_i|>6$ we have
\be
x_i\geq {|et_i|\over 88}\    .
\ee
\label{exterpow}
\end{lemma}
\paragraph{Proof}
For $g_i$ with $|eg_i|>4$ and such that there is no
$g_{i'}\in D({\cal C},P)$ with 
$i\in C_{i'}^r\backslash J_{i'}$, Lemma 8 of [DR1] applies directly.
 For $g_i$ with $|eg_i|>6$ and such that there is some
$g_{i'}\in D({\cal C},P)$ with $|eg_i|>6$,
$i\in C_{i'}^r\backslash J_{i'}$,
we have to bound 
$ {1\over 8} (|eg_i({\cal C})|-10)$ for $|eg_i|>10$ or
$ {1\over 8} (|eg_i({\cal C})|-6)$ for  $|eg_i|\leq 10$.
Then we apply the same reasonings as in Lemma 8 of [DR1].
\qed

To complete the bound we must factorize the 
integrals over the $\bt$ parameters as in [DR1]. Some sets of
$\bt_j$   are not independent yet:
\be
 \prod_{g_i\in D^2({\cal C},P)}\left [
 \prod_{j\in C_{t(i)}\backslash C_{{\cal A}(i)}}\int_{w_T}^1 d\bt_j\;
 \bt_j^{-1+x_j} 
 \frac{
\left [1 - \prod_{j\in C_{t(i)}\backslash C_{{\cal A}(i)}} \bt_j 
 \right  ]^{2}}{
\left [1 -\lp\prod_{j\in C_{t(i)}\backslash 
C_{{\cal A}(i)}} \bt_j \rp^{1\over s}
 \right  ]^{2s}}\right ]\ .
\ee
The mixed term has an integrable singularity at the point $\bt_j=1$, 
$\forall j\in C_{t(i)}\backslash C_{{\cal A}(i)}$. 
We decompose the integration domain of each $\bt_j$ into two subsets
$[w_T,1]=I^1\cup I^2$ where $I^1= [w_T,1/2]$ and 
$I^2=[1/2,1]$. The integral above, for a fixed 
$g_i\in D^2({\cal C},P)$  is  written as
\be
 \prod_{j\in C_{t(i)}\backslash C_{{\cal A}(i)}}
 \sum_{m_j=1,2 }
\int_{I^{m_j}} d\bt_j\;
 \bt_j^{-1+x_j} 
 \frac{
\left [1 - \prod_{j\in C_{t(i)}\backslash C_{{\cal A}(i)}} \bt_j 
 \right  ]^{2}}{
\left [1 -\lp\prod_{j\in C_{t(i)}\backslash 
C_{{\cal A}(i)}} \bt_j \rp^{1\over s}
 \right  ]^{2s}}
\label{mixed0}\ee
We distinguish two situations.

\paragraph{1.} If $ m_j=1$ for some $j$, then some $\bt_j\leq 1/2$, and 
the mixed term can be bounded by
$1/[1-(1/2)^{1\over s}]^{2s}$ and taken out of the integral. The 
integrals in (\ref{mixed0}) are then  factorized.  
\paragraph{2.}  If $ m_j=2$ $\forall j$, we have to compute
\be
 \prod_{j\in C_{t(i)}\backslash C_{i}}
\int_{1\over 2}^1 d\bt_j\; 
 \bt_j^{-1+x_j} \int_{1\over 2}^1 d\bt_i\;\bt_i^{-1}\; 
 \frac{
\left [1 - \prod_{j\in C_{t(i)}\backslash C_{{\cal A}(i)}} \bt_j 
 \right  ]^{2}}{
\left [1 -\lp\prod_{j\in C_{t(i)}\backslash 
C_{{\cal A}(i)}} \bt_j \rp^{1\over s}
 \right  ]^{2s}}\ ,
\ee
where $\bt_i$ appears with exponent $-1$ because $g_i$ is a two
point renormalized subgraph. Then $x_i=0$. We perform the change of variable
on $\bt_i$: 
\be
z\;:=\;\bt_i\;c_i\ ,\quad c_i:=\lp 
\prod_{j\in C_{t(i)}\backslash C_{i}}\bt_j\rp\ ,
\ee
and the integral becomes:
\be
\prod_{j\in C_{t(i)}\backslash C_{i}}
\int_{1\over 2}^1 d\bt_j\; 
 \bt_j^{-1+x_j} 
\int_{{c_i\over 2}}^{c_i} 
dz\;z^{-1}\; 
 \frac{
\left [1 - z \right  ]^{2}}{
\left [1 - z^{1\over s} \right  ]^{2s}}\ .
\label{mixed}\ee
We observe that $c_i$ varies on the interval $[2^{-|c_i|},1]$, where we 
defined $|c_i|$ as the number of $\bt_j$ in 
$C_{t(i)}\backslash C_{i}$.  To bound the integral over $z$  
and verify this bound does not depend from $c_i$ we 
distinguish two cases.
\paragraph{a: $c_i\geq {1\over 2}$} then  
\be
\int_{c_i\over 2}^{c_i}  
dz\;z^{-1}\; 
 \frac{
\left [1 - z \right  ]^{2}}{
\left [1 - z^{1\over s} \right  ]^{2s}} \leq 
\int_{c_i\over 2}^{1\over 2}   
dz\;z^{-1}
+ \int_{ 1\over 2}^{c_i}     dz\;z^{-1}\; 
 \frac{
\left [1 - z \right  ]^{2}}{
\left [1 - z^{1\over s} \right  ]^{2s}} \leq K;
\ee
\paragraph{b: $c_i < {1\over 2}$} then  
\be
\int_{c_i\over 2}^{c_i}  
dz\;z^{-1}\; 
 \frac{
\left [1 - z \right  ]^{2}}{
\left [1 - z^{1\over s} \right  ]^{2s}} \leq 
K \int_{c_i\over 2}^{c_i}   
dz\;z^{-1} = K \log 2 \ ,
\ee
where $K$ is a constant.
In both cases the bound does not depend on $c_i$ and
the integrals in (\ref{mixed}) are factorized.

Finally we can bound the integrals over the parameters $\bt_i$:
\bqa
&&\prod_{g_i\in\!\!\!/ ND({\cal C},P)} 
\int_{w_T}^1 d\bt_i\; \bt_i^{-1+x_i}
\leq \prod_{g_i\in D({\cal C}, P)} |\log w_T| 
\prod_{g_i\;|\; |eg_i|=4} |\log w_T|\no\\
&&\qquad \prod_{g_i\in D({\cal C},P)} \left [
\prod_{j\in C_i^r\backslash J_i|\; |eg_j|=6} |\log w_T|  \right ]
  \prod_{g_i\;|\; |eg_i|>6} {1\over x_i}\ .
\eqa
Now, like in [DR1], Lemma 8, we can bound the vertex functions by
\bqa
&&\hspace{-0.8cm} |\Ga_{2p>4}^{\Lazero}|\leq K_0\;
{\scriptstyle ||\phi_1||_{_1}
 \prod_{i=2}^{2p} ||\hat{\phi}_i||_{\infty,2} }
e^{-a\;(1-\vep)\;\La_T^{1\over s}\;  d^{1\over s}_T(\Om_1,...,\Om_{2p}) }
 {w_T^{{7p\over 4}-{1\over 4}}\over 2p-4} \sum_{{\bar n}\geq 1}
\frac{c^{\bar n}}{n!n'!} K^{\bar n}\no\\
&&\hspace{-0.4cm}\sum_{CTS}\sum_{u-\tree}\sum_{\cal L}\sum_{\Om\; E}
\sum_{{\cal C}}\sum_{J P}\sum_{I} 
\prod_{i\neq r}{1\over |et_i|}
\prod_{g_i\in D({\cal C},P)} |\log w_T|
\no\\
&&\hspace{-0.4cm} \prod_{g_i |\; |eg_i|=4}  |\log w_T|
\prod_{g_i\in D({\cal C},P)} \left [
\prod_{j\in C_i^r\backslash J_i|\; |eg_j|=6} |\log w_T| \right ] \ ,
\label{lastbound}
\eqa
where the sum over $I$ gives the choices of the integration domain 
of $\bt_i$ between $I^1$ and $I^2$. The set $I$ has then cardinal 
proportional to $2^{\bar n}$. 
If $2p=4$ or $2p=2$ we substitute an additional factor  $|\log w_T|$
to the global factor 
$1/(2p-4)$ in front of (\ref{lastbound}). 
Now we bound all the sums exactly as in [DR1] (the only difference being
that we are working with ${\bar n}$ instead of $n$). 
Finally we obtain
\bqa
&&\hspace{-3cm} |\Ga_{2p>4}^{\Lazero}|\leq K_0\;
{\scriptstyle ||\phi_1||_{_1}
 \prod_{i=2}^{2p} ||\hat{\phi}_i||_{\infty,2} }
e^{-a\;(1-\vep)\;\La_T^{1\over s}\;  d^{1\over s}_T(\Om_1,...,\Om_{2p}) }\\
&&  {w_T^{{7p\over 4}-{1\over 4}}\over 2p-4} K_1^p\; (p!)^2\;
\sum_{{\bar n}\geq 1}
\frac{1}{n!n'!}\lp c K_2 |\log w_T|\rp^{\bar n}\ , \no
\eqa
\bqa
&&\hspace{-3cm} |\Ga_{2p\leq 4}^{\Lazero}|\leq K_0\;
{\scriptstyle ||\phi_1||_{_1}
 \prod_{i=2}^{2p} ||\hat{\phi}_i||_{\infty,2} }
e^{-a\;(1-\vep)\;\La_T^{1\over s}\;  d^{1\over s}_T(\Om_1,...,\Om_{2p}) }\\
&& w_T^{{7p\over 4}-{1\over 4}} 
\sum_{{\bar n}\geq 1} 
\frac{1}{n!n'!}\lp c K_2 |\log w_T|\rp^{\bar n}\ . \no
\eqa
These sums are convergent for $c  K_2 |\log w_T| < 1$ which achieves the 
proof of Theorem 3.
\vskip 0.5cm

\resetsect 

\renewcommand{\thesection}{\Alph{section}}

\noindent{\Large {\bf Appendix A: Flow of $\de\mu$}}
\medskip

\resetequ

To study the flow of the chemical potential counterterm we 
introduce some definitions. 
We define $\Si[\de\mu,C]$ as the two point vertex function
$\Ga_2(\phi^0_1,\phi^0_2)$ 1PI  and
with at least one internal line, for a theory 
with bare chemical potential counterterm $\de\mu$ and propagator $C$. 
The test functions are  $\phi_i^0(x)=\de(x)$ and $\phi_2^0(x)=e^{ik_Fx}$, 
hence 
the external impulsion is fixed to $k_F$, as near a possible to the 
Fermi surface. Remark that, as the external impulsion is fixed near the 
Fermi surface,  we do not introduce any 
Gevrey cut-off on the test functions, and, when performing 
sector sum, the factor for the root sector $\La_T^{-{1\over 2}}$ does 
not appear.
The two fundamental equations are then
\bqa
\de\mu^1_\La(\la) &=& \Si[\de\mu^1_\La(\la)  ,C^1_\La] \no\\
\de\mu^{\La'}_\La(\la) &=& \de\mu^1_\La(\la)-
\Si[\de\mu^1_\La(\la)  ,C^1_{\La'}]\quad  \La\leq \La'\leq 1
\no\\
 &=& \Si[\de\mu^1_\La(\la)  ,C^1_\La]-
\Si[\de\mu^1_\La(\la)  ,C^1_{\La'}]
 \eqa
These equations are consistent with the BPHZ condition $\de\mu^\La_\La(\la)=0$.
To study the flow we write  $\Si[\de\mu^1_\La(\la)  ,C^1_{\La'}]$
as an expression where the dependence from $\La$ and $\La'$ is explicit:
\bqa
&&\Si[\de\mu^1_\La(\la)  ,C^1_{\La'}]=
\sum_{n+n'\geq 2 \atop n\geq 1} \frac{\la^n}{n!} 
\frac{\lp\de\mu^1_\La\rp^{n'}}{n'!}\label{flow1}\\
&& \sum_{CTS}
\sum_{u-\tree}\sum_{\cal L}\sum_{E\Om}\sum_{{\cal C}}\sum_{J,P}
\vep(\tree, \Om) \lp 2\rp^{{\bar n}-1}
\int_{\La'_T\le \La_{{\cal A}(i)} \le  \La_{i}\le 1}
\prod_{q=1}^{{\bar n}-1} \La_q d\La_q\no\\
&&\prod_{h\in L\cup\tree_L\cup E}
\left\{ [{\scriptstyle {4 \over 3} \La^{-{1\over 2}}(w_{j_{h,n_h}})}]
\int_0^{2\pi}  d\th_{h,n_h}
[ {\scriptstyle{4 \over 3}\La^{-{1\over 2}}(w_{j_{h,n_h-1}})}]
\int_{\Si_{j_{h,n_h}}}
d\th_{h,n_h-1} \right .\no\\
&&...\;\;[{\scriptstyle {4 \over 3}\La^{-{1\over 2}}(w_{j_{h,1}})}]
\left .\int_{\Si_{j_{h,2}}}
d\th_{h,1}
\left [\prod_{r=2}^{n_{h}}
\chi^{\th_{h,r}}_{\al_{j_{h,r}}}(\th_{h,1})\right ]
 \right \}\no\\
&& \prod_{g_i|\ i=r\ {\rm or}\atop |eg_i({\cal C}|)\leq 8}
\Upsilon\lp\th_i^{root},\{\th_{h,r(i)}\}_{h\in eg^\ast_i}
 \rp \;
\prod_{v\in V\cup V'}  \Upsilon\lp\th_{h_v^{root}},
\{\th_{h,n_h}\}_{ h\in H^\ast(v)} \rp\no\\
&&\int
d^3x_1...d^3x_{\bar n}\;\;
\phi_1(x_{i_{1}},\th_{e_1,1})\;
\phi_{2}(x_{j_{1}},\th_{e_{2},1})
\left [\prod_{q=1}^{{\bar n}-1} \frac{1}{\La^4_q} C_{\al=\La^{-2}_q}
( x_{q}, {\bar x}_{q}, \th_{h,1}) \right ]\no\\
&& \left [ \prod_{l_{fg}\in P}
{\cal M}_{f,g}'({\cal C},E ,\{\th_{a,1}\})\right ] 
\;\left [  \prod_{ j_q\in J^1} 
\int_0^1 ds_{j_q} \right ] 
\det {\cal M}'({\cal C},E,\{\th_{a,1}\},\{s_{j_q}\}) \ ,
\no\eqa
where we performed the change of variable $w_i = (\La_i^2-\La^{'2})/(1-\La^{'2})$,
and $\La'_T=\La'$ for $\La'\geq \sqrt{2}\pi T$.
Remark that the $\La'$ dependent factor $1/(1-\La')^{{\bar n}-1}$, coming from
the Jacobian is cancelled by
the corresponding factor $(1-\La')^{{\bar n}-1}$ coming from tree line
propagators. Then $\La'$  appears only in the  $\La_r$ integration and 
in some loop line propagators (those of the loop fields with 
$m(a,{\cal C})=\La'$).  The power counting is performed as usual, passing to
the variables $\bt_i$ defined by $\La^2_i=\La^2_{{\cal A}(i)}/\bt_i $. 
\medskip

We want to prove, by induction, that the property $H(\La)$, defined by
\be
\left | \de\mu_\La^{\La'}(\la)\right | \leq K_1 |\la| (\La'-\La) \qquad 
\forall \La'\geq \La
\ee 
is true $\forall\  0\leq \La\leq 1$. 
We suppose $H(\La)$ is true for a certain $\La$, then we  prove 
Lemma 6 and 7, about the existence and the bound satisfied by the derivative.
 These lemmas ensure that $H(\La)$ is true for all $\La$.
Indeed otherwise  there exists $\La_m>0$ defined as
\be
\La_m = \inf_{\La\in [0,1]} \{\La| H(\La)\ {\rm is\ true}\}
\ee
Then by lemma 7 we can write a Taylor expansion at  first order
\bqa
 |\de\mu_{\La-\vep}^{\La'}(\la) | &\leq  &
|\de\mu_\La^{\La'}(\la) |+
\left |\frac{d}{d\La} \de\mu_\La^{\La'}(\la)\right | \vep + o(\vep)\\
&\leq  & K_1 |\la| (\La'-\La)+
K_3 |\la| \vep + o(\vep)\leq K_1 |\la| (\La'- (\La-\vep))\no
\eqa
for all $\La'\geq \La$. The same bound in the case  
$\La-\vep\leq\La'\leq \La$ is proven in Lemma 8. This result 
contradicts with the definition of $\La_m$, therefore $\La_m=0$.

\begin{lemma}
If $H(\La)$ is true then the derivative   
$\frac{d}{d\La} \de\mu_\La^{1}(\la)$, $\La\leq \La'\leq 1$ 
exists and satisfies the bounds:
\be
\left |\frac{d}{d\La} \de\mu_\La^{1}(\la)\right | \leq K_2 
|\la| \, .\label{flow4}
\ee
\end{lemma} 
\paragraph{Proof} 
If the derivative exists, it satisfies the formal equation:
\be
\frac{d}{d\La} \de\mu_\La^{1} = A + \frac{d}{d\La} \de\mu_\La^{1} \; 
B\label{flow3}
\ee
where $A$ is the expression for $\de\mu_\La^{1}$ with the derivative
$d/d\La$ applied to one propagator, and $B$
is the same expression for $\de\mu_\La^{1}$, but  with one 
special two point vertex with value 1 instead of  $\de\mu_\La^{1}$
(as the derivative of the corresponding factor
has been taken out of the sum in (\ref{flow3}).  
Then, formally, the solution for (\ref{flow3}) is
\be
\frac{d}{d\La} \de\mu_\La^{1} = \frac{A}{1-B} 
\label{RGce}\ee
Now, if we can prove that $A\leq K' \la$ and $B\leq 1/2$, 
we obtain (\ref{flow4}) with $K_2=2K'$. 

\begin{figure}
\centerline{\psfig{figure=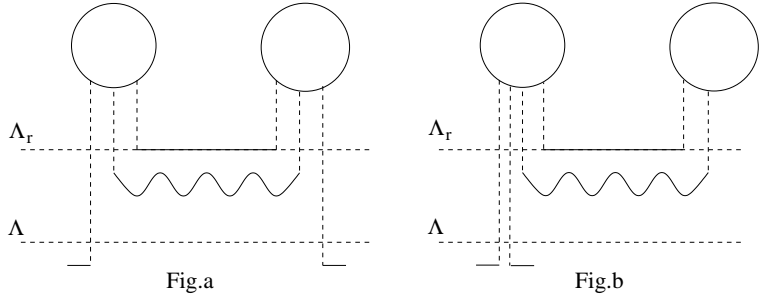,width=10cm}}
\caption{Two possible schema for the lowest band in a 
two point 1PI graph}\label{fermi21}
\end{figure}

\paragraph{Bound on $A$.}
Remark that, as shown in Fig.\ref{fermi21} a,b,   there is
at least one loop line in the first band,  obtained 
in the first case  by 1PI, in the second case by parity of the 
number of external lines for any subgraph.

Therefore  the derivative $d/d\La$ may apply only to a loop line propagator. 
Indeed, if it applies to the $\La_r$ integral, 
the first band width  and the corresponding loop line amplitude are reduced to 
zero. 

The action of the derivative on the loop propagator is given by
\be
\frac{d}{d\La}C^{\La_M}_\La = -\frac{2}{\La^3}\; C_{\al=\La^{-2}}
\ee
Then, when performing the estimations we add the  factor
$\frac{1}{\La_M}\leq\frac{1}{\La_r}=\frac{\bt^{1\over 2}_r}{\La} $. 
\bqa
\lefteqn{|\hat{\Si}| \leq  K_0\;
\sum_{{\bar n}\geq 2} \frac{c^{\bar n}}{n!n'!} K^{\bar n}
\sum_{CTS}\sum_{u-\tree}\sum_{\cal L}\sum_{\Om\; E}
\sum_{{\cal C}}\sum_{J P}\lp 2\rp^{{\bar n}-1} } \label{selfenergy}\\
&& \int_{\La_T\le \La_{{\cal A}(i)} \le  \La_{i}\le 1}
\prod_{q=1}^{{\bar n}-1} \La_q d\La_q 
\prod_{i=1}^{{\bar n}-1} \frac{1}{\La^3(w_i)}
 \prod_{a\in L}
\;\La^{3\over 4}\lp w_{M(a,{\cal C})}\rp   \prod_{v\in V'}
\La(w_{i_v})\; \La_r^{-1}\no\\
&&
\prod_{g_i |\  i\neq r\ {\rm or}  \atop |eg_i({\cal C})|\geq 11}\left [
\prod_{h\in eg^\ast_i \atop j_{h,1}=i(h)={\cal A}(i)}
\frac{ \La^{1\over 2}\lp w_{j_{h,2}}\rp}{\La^{1\over 2}\lp w_{j_{h,1}}\rp }
\right ]
\prod_{g_i|\ i=r\ {\rm or} \atop 2\leq |eg_i({\cal C})|\leq 10 }
 \left [
\prod_{h\in I(i)} \frac{\La^{1\over 2}\lp w_{j_{h,r(i)+1}}\rp}
{\La^{1\over 2}\lp w_{j_{h,r(i)}}\rp}
\right ] \no\\
&&
\prod_{v\in V}\La^{-{1\over 2}}\lp w_{i_v}\rp
\prod_{g_i\in D({\cal C},P)} 
{\La(w_{{\cal A}(i)})\over\La(w_{i}) }
 \no\\
&& \prod_{g_i\in D({\cal C},P)} \left [
\prod_{j\in C_i^r\backslash J_i}
 \lp {\La(w_{i})\over \La(w_{{\cal A}(i)})}\rp^{1\over 2} \right ]  
   \prod_{g_i\in D^2({\cal C},P)} \frac{
\left [1 - {\La(w_{{\cal A}(i)})\over  \La(w_{t(i)})}
 \right  ]^{2}}{
\left [1 -\lp {\La(w_{{\cal A}(i)})\over  \La(w_{t(i)})}\rp^{1\over s}
 \right  ]^{2s}}\ , \no
\eqa
where  we  the factor $\La_T^{{5\over 2}(2p-1)} \La_T^{-{1\over 2}}$ has 
desappeared as the external impulsion (hence the sector too) is fixed.
Remark that the set of renormalized subgraphs $D({\cal C},P)$ does
not contain the global graph $g_r$.  Passing to the variables
$\bt_i={\La^2_{{\cal A}(i)}\over \La_i^2 }$, we obtain 
\be
 |\hat{\Si}|\leq K_0\; \sum_{{\bar n}\geq 1}
\frac{c^{\bar n}}{n!n'!} K^{\bar n}
\sum_{CTS}\sum_{u-\tree}\sum_{\cal L}\sum_{\Om\; E}
\sum_{{\cal C}}\sum_{J P} \;\La_T\;
\prod_{i=1}^{{\bar n}-1}\int_{\La^2_{{\cal A}(i)}}^1  d\bt_i\; \bt_i^{-1+x_i}
 \bt_r \frac{1}{\La_T}
\ee
where the integral limit $\La^2_{{\cal A}(i)}\geq \La_T$. We have 
not written the non factorized terms, that appear in eq. III.28
and come from renormalized
two point subgraphs, as their power counting is not modified at all.
The factor $1/\La_T$ cancels with
the global factor $\La_T$, giving a constant independent from 
$\La_T$. The power counting of $\bt_r$
becomes logarithmic, instead of linearly divergent; 
this is the reason for which we can extract only one coupling constant 
$\la$. Indeed:
\be
A \leq |\la| \sum_{{\bar n}\geq 2} \lp |\la\ln \La|\rp^{{\bar n}-1}
\leq |\la| {|\la\ln \La| \over  1- |\la\ln \La|}\leq K_2 |\la|.
\ee 

\paragraph{Bound on $B$.}
The estimation for $B$ is performed as that for $\de\mu^1_\La$. The 
only difference is that, when a two point subgraph contains the special
insertion, it is not renormalized, as the power counting is logarithmic
instead of linearly divergent. This happens because there is one two
point insertion (the special one) that is not compensated by the
corresponding $\de\mu$ scaling factor, then we have
\be
\int_{\La^2}^1 d\bt_i\; \bt_i^{-1-{1\over 2}+{1\over 2}} \leq |\log\La|
\ee
Of course, the $\bt_r$ power counting becomes logarithmic too, as 
$g_r$ always contains the special insertion. The global 
factor $\La$ is then cancelled by the global factor coming from the
special insertion. Then
\be
B\leq |\la| \sum_{{\bar n}\geq 2} \lp |\la\ln \La|\rp^{{\bar n}-1}
\leq |\la| {|\la\ln \La| \over  1- |\la\ln \La|}\leq K_2 |\la|\leq {1\over 2}
\ee
for $\la$ small enough. 

\paragraph{Existence of the derivative}
We still have to prove that the derivative exists. For that we apply the 
definition
\be
\frac{d}{d\La} \de\mu_\La^{1} = \lim_{\vep\to 0} \frac{1}{\vep}\lp
\Si[\de\mu_{\La-\vep}^{1},C^1_{\La-\vep}] - \Si[\de\mu_{\La}^{1},
C^1_\La]\rp = \lim_{\vep\to 0} \frac{\De_{\La,\vep}^1}{\vep}
\ee
The difference $\De_{\La,\vep}^1$ can be written as
\bqa
&&\hspace{-0.5cm} \De_{\La,\vep}^1 = 
\lp\Si[\de\mu_{\La-\vep}^{1},C^1_{\La-\vep}] - 
\Si[\de\mu_{\La}^{1},C^1_{\La-\vep}]\rp\; +\;
\lp\Si[\de\mu_{\La}^{1},C^1_{\La-\vep}] - 
\Si[\de\mu_{\La}^{1},C^1_\La]\rp\no\\
&&= \sum_{p=1}^\infty  \lp\De_{\La,\vep}^1\rp^p F_p + A_1 + A_2
\label{RGde}\eqa
where $A_1$ is the expression for 
$\Si[\de\mu_{\La}^{1},C^1_{\La-\vep}]$ with 
one loop propagator $C^{\La}_{\La-\vep}$, and $A_2$ is the same expression, 
but this time  with the first band
$\La_r\leq\La$. Finally  $F_p$ is the expression for  
$\Si[\de\mu_{\La}^{1},C^1_{\La-\vep}]$ with $p$ insertions of 
special two point vertex, obtained by substituting the coefficient by 1.
With the same kind of argument as before we can prove that 
$|F_p| \leq |\la|/\La^{p-1}$, $|A_1|\leq \vep |\la| $ and 
$|A_2|\leq \vep^2 |\la|$.
Then we can prove that $\De_{\La,\vep}^1$ exists 
and the derivative takes the form (\ref{flow3}). 
\qed

\begin{lemma}
If $H(\La)$ is true then the derivative   
$\frac{d}{d\La} \de\mu_\La^{\La'}(\la)$
exists and satisfies the bound:
\be
\left |\frac{d}{d\La} \de\mu_\La^{\La'}(\la)\right | \leq K_3 |\la| \, .
\label{flow5}
\ee
\end{lemma} 
\paragraph{Proof}
The proof is a direct consequence of Lemma 6. By the definition of 
$ \de\mu_\La^{\La'}(\la)$ the derivative is given by
\be
\frac{d}{d\La} \de\mu_\La^{\La'} = \frac{d}{d\La} \de\mu_\La^{1}(1 -F)
\ee
where $F$ is the expression for $\Si[\de\mu^1_\La,C^1_{\La'}]$ 
with one special insertion, that means one two point vertex factor
substituted by $1$. As for $B$ in Lemma 6, we can prove that 
\be
|F|\leq  |\la| 
\ee
then 
\be
\left |\frac{d}{d\La} \de\mu_\La^{\La'}\right | 
\leq K_2 |\la| (1+|\la|)) \leq K_3 |\la|. 
\ee
The existence of this derivative is a consequence of Lemma 6,
as 
\be
\De_{\La,\vep}^{\La'} = \De_{\La,\vep}^1 - 
\sum_{p=1}^\infty \lp \De_{\La,\vep}^1\rp^p F'_p
\ee 
where we can prove that $|F'_p|\leq |\la|/\La^{p-1}$.
\qed

\begin{lemma}
If the bound
\be
\left |\de\mu^{\La'}_\La\right |\leq K_1 |\la| (\La'-\La)
\ee
is true for all $\La'\geq \La'_0=\La+\vep$ then it is true for
$\La'_0-\vep'$ for $\vep'<\vep$ small enough.  
\end{lemma}
\paragraph{Proof}
For all $\La'\geq\La'_0$
we can prove (by the same arguments as before) that
\be
\left|\frac{d}{d\La'} \de\mu_\La^{\La'}\right| \leq K_2 |\la|
\ee
Then we can perform a first order Taylor expansion
\bqa
&&\left |\de\mu^{\La'_0-\vep'}_\La\right | 
\leq \left |\de\mu^{\La'_0}_\La\right | +  
\left |\frac{d}{d\La'} \de\mu_\La^{\La'}\right | \vep'+ o(\vep')\no\\
&&\leq K_2 |\la| (\vep+\vep') + o(\vep) + o(\vep')\leq K_1 |\la|(\vep-\vep')
\eqa
for $\vep'$ small enough. Remark that we used the inequality
\bqa
&& \left |\de\mu^{\La'_0}_\La\right | =  \left |\de\mu^{\La+\vep}_\La\right |
\leq \left |\de\mu^{\La+\vep}_{\La+\vep}\right |\no\\
&& + \left |\frac{d}{d\La''} 
\de\mu_{\La''}^{\La'}\right |_{\La'=\La''=\La+\vep} \vep
+ o(\vep)\no\\
&&\leq K_2 |\la| \vep + o(\vep)
\eqa
\qed

Remark that the differential RG equations (\ref{RGce}) is 
simpler than its discretized counterpart (\ref{RGde}). This is an advantage of
the differential version of the RG.
\vskip 1cm

\setcounter{section}{2}

\noindent{\Large {\bf Appendix B: Study of the selfenergy. }}
\vskip 1cm

\resetequ

\centerline{\large {\bf B1: A loop expansion for extracting the 
self energy $\Sigma$}}

\centerline{\\ 
(in collaboration with D. Iagolnitzer and J. Magnen)}
\vskip 1cm

Our tree formula selects connected graphs. But the self energy $\Sigma$ is
the sum over all non trivial two point connected subgraphs which are
furthermore 1PI-irreducible (with respect to the single channel between the
two external points). In this appendix we apply an (unpublished) formula,
due to D. Iagolnitzer and J. Magnen, which proves that this additional
information of 1PI in a single channel can be extracted by expanding
some loops out of the determinant without generating any factorials in the 
bounds.

\paragraph{The arch expansion}

\begin{figure}
\centerline{\psfig{figure=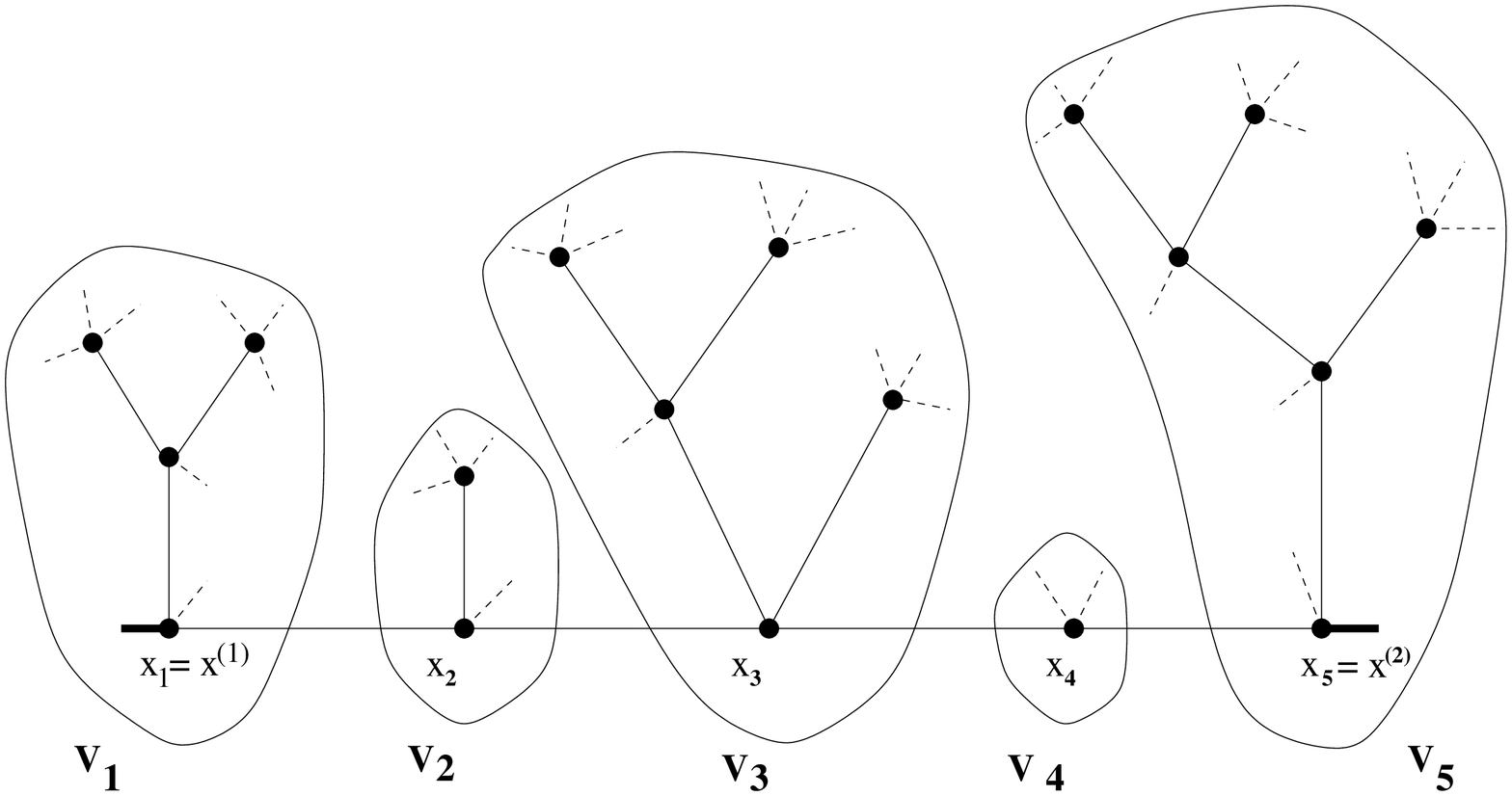,width=10cm}}
\caption{Example of tree, with $p=5$ vertices on
${\cal P}_{1,2}$. Loop fields are dashed, tree lines are solid, 
and the amputated external lines are darker }\label{arch1}
\end{figure}

We consider a two point connected graph with $n$ vertices, equipped with its
spanning tree $\tree$. If the two (amputated) external lines
are hooked to the same vertex, we have a ``generalized tadpole'' which 
is automatically 1PI, hence belongs to the self energy.
In that case no additional expansion is performed.
Otherwise there is a unique non-empty linear path $\cal{P}_{1,2}$ made of 
$p-1\le n-1$ lines in the tree $\tree$
joining the two external vertices $x^{(1)}=x_{1}$ and $x^{(2)}=x_{p}$ through the 
intermediate vertices $x_{2},...,x_{p-1}$. The set $V$ of vertices
of $G$ is then the disjoint union of the sets $V_{j}$, $j=1,...,p$,
where a vertex belongs to $V_{j}$ if and only if the unique path in $\tree$
joining it to the root $x_{1}$ passes through $x_{j}$ but not through 
$x_{j+1}$ (Fig.\ref{arch1}).

We call $F(\{C_{i,j}\})$ the loop determinant of the remaining fields
(it depends on the weakening $w$ parameters, but this is completely irrelevant
in what follows). Expanding completely $F$ would cost $n!$.
But we just want to know if the graph is 1PI, which means 
1PI with respect to the $p-1$ lines of ${\cal{P}}_{1,2}$, since by parity there
cannot be 1 particle reducibility in a 0-2 channel. 
We perform an auxiliary expansion {\`a} la Brydges-Battle-Federbush,
and we call it the ``arch expansion''. This means
that we first test if some vertex of $V_{1}$ is
linked to some vertex of  $V_{k_{1}}$, with $k_{1}>1$.
This is done by introducing the interpolation parameter 
$0\leq s_1\leq 1$ and defining
\bqa
C_{ij}(s_1) & := & s_1\; C_{ij}\quad {\rm if}\quad 
i\in V_{1},j \not\in V_{1}\no\\
            &  := &   C_{ij}\quad {\rm otherwise} 
\eqa

Then we can write
\be
F(\{C_{i,j}\}) = F(\{C_{i,j}(s_1)\})_{\big |_{s_1=1}} = 
 F(\{C_{i,j}(s_1)\})_{\big |_{s_1=0}} + \int_0^1 \; ds_1\; 
\frac{d}{ds_1}F(s_1)
\ee 
\begin{figure}
\centerline{\psfig{figure=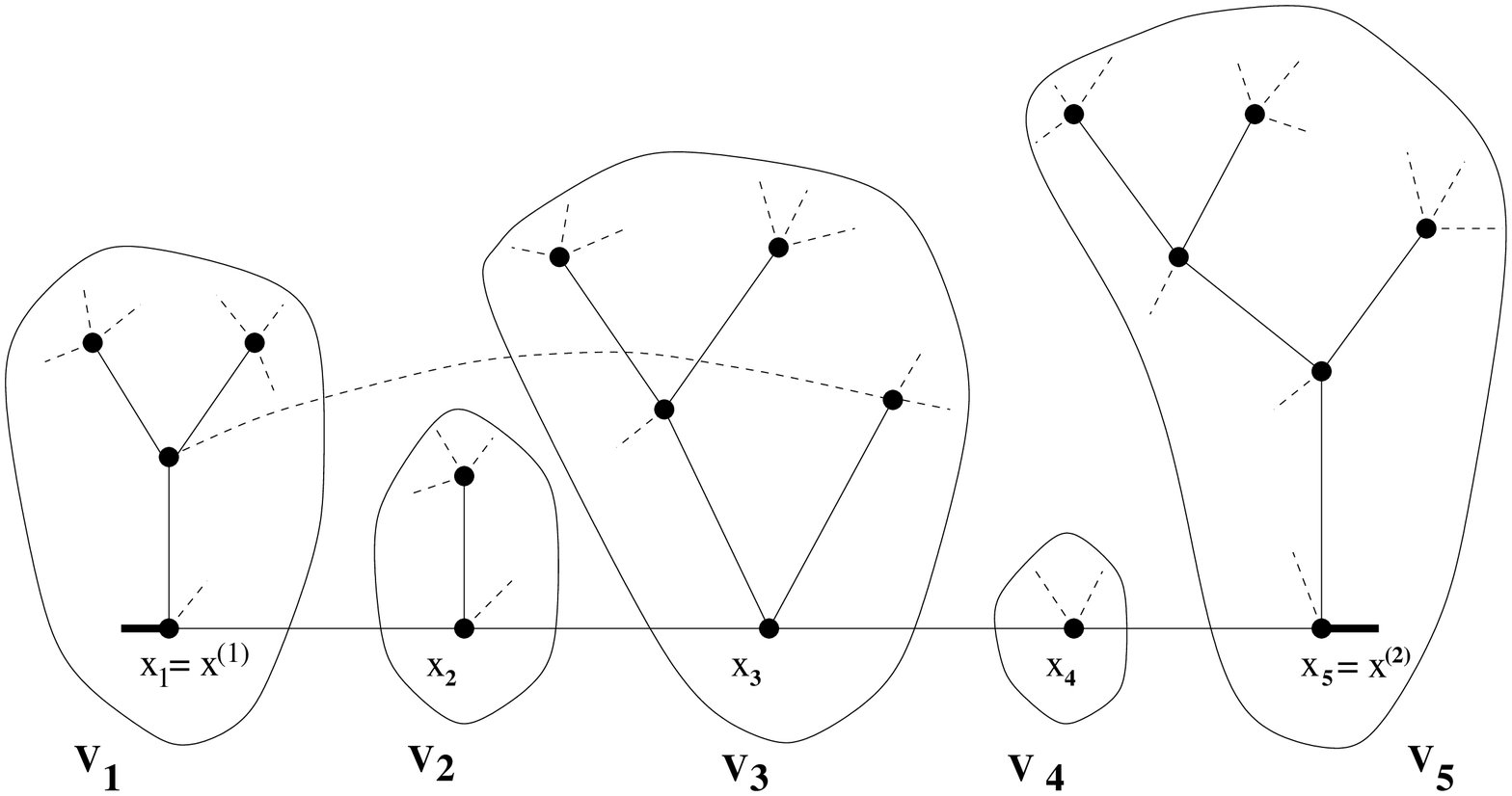,width=10cm}}
\caption{Extraction of one loop line from $V_1$. The loop line is dashed.}
\label{arch2}
\end{figure}

\begin{figure}
\centerline{\psfig{figure=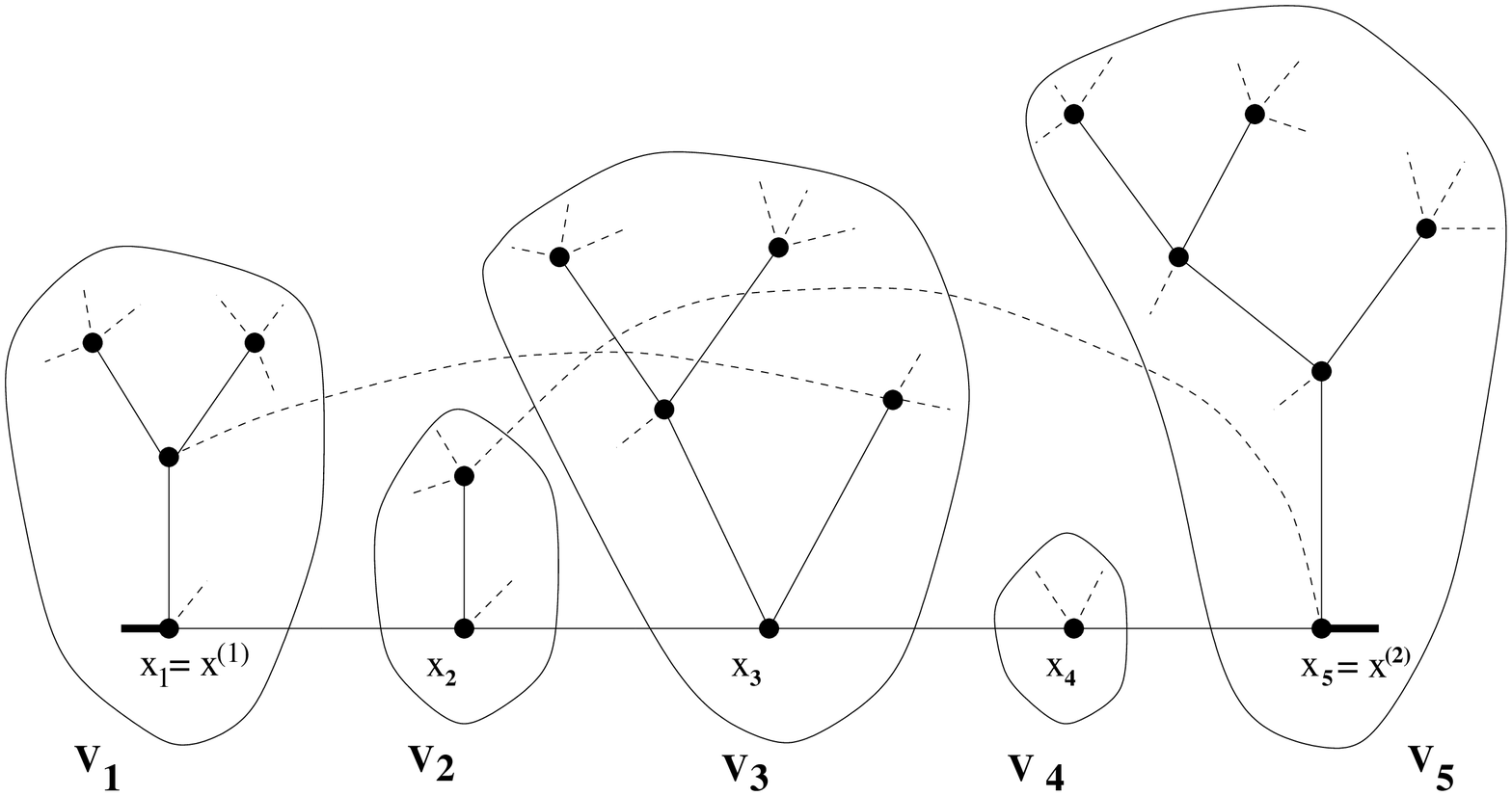,width=10cm}}
\caption{Extraction of two loop lines.}\label{arch3}
\end{figure}

The first term $s_1=0$ means that the graph is 1PR (by cutting the first
line of ${\cal{P}}_{1,2}$ as no loop line connects $V_1$ to its complement). 
Otherwise we derive an explicit loop line out of the determinant,
which connects a vertex of $V_{1}$ to a vertex of $V_{k_{1}}$, for some 
$k_{1}>1$ (see Fig.\ref{arch2}, where we have $k_{1}=3$).
If $k_{1}=p$ we are done since the full graph is 1PI. Otherwise
we repeat the same procedure, but bewteen $\cup_{l=1}^{k_{1}} V_{l}$
and its non empty complement, introducing a second
interpolation parameter $0\leq s_2\leq 1$:

\bqa
C_{ij}(s_1,s_2) & := & s_2\; C_{ij}(s_1)\quad 
        {\rm if}\quad i \in \cup_{l=1}^{k_{1}} V_{l} ,j
\not\in\cup_{l=1}^{k_{1}} V_{l}\no\\
            &  := &   C_{ij}(s_1)\quad {\rm otherwise} 
\eqa
Then we can write
\bqa
F_1(\{C_{ij}(s_1)\})& =& F_1(\{C_{i,j}(s_1,s_2)\})_{\big |_{s_2=1}} = \no\\
&=& F(\{C_{i,j}(s_1,s_2)\})_{\big |_{s_2=0}} + \int_0^1 \; ds_2\; 
\frac{d }{ds_2}F_1(s_1,s_2) 
\eqa 
Once again the first term at $s_{2}=0$ means that the block 
$\cup_{l=1}^{k_{1}} V_{l}$  is 
not linked to its complement by any loop line, and the graph is 1PR
across the line number $k_{1}$ of ${\cal{P}}_{1,2}$.
The second term corresponds to extract a new loop line
(see Fig.\ref{arch3}) and   can be written again as

\be
\int_0^1 ds_2 \; \frac{d }{ds_2}F_1(s_1,s_2)= 
\sum_{i_2 \in \cup_{l=1}^{k_{1}} V_{l} \atop k_{2}>k_{1}\ ; \
j_2\in V_{k_{2}}} 
\int_0^1 ds_2 \; \frac{\partial  }{\partial s_2}C_{i_2j_2}(s_1,s_2) \; 
\frac{\partial }
{\partial C_{i_2j_2}}F_1(s_1,s_2) 
\ee
Remark that 
\bqa
\frac{\partial  }{\partial s_2}C_{i_2j_2}(s_1,s_2) &=&  C_{i_2j_2}
\quad {\rm if} \quad i_2 \in \cup_{l=2}^{k_{1}} V_{l}\no\\
 &=&  s_1 C_{i_2j_2}
\quad {\rm if} \quad i_2 \in V_{1}
\eqa

We repeat this procedure until we reach $V_{p}$ with a loop line. 
Then, if the process stops at the q-th step we have the expression
\bqa
A(q,n)&=&\sum_{1<k_{1}<...<k_{q}=p} \;
\sum_{i_1 \in V_{1},j_1 \in V_{k_{1}}}\;
\sum_{i_2\in \cup_{l=1}^{k_{1}}V_{l},j_2 \in V_{k_{2}} }
...\sum_{i_q\in \cup_{l=1}^{k_{q-1}}V_{l},j_q\in V_{k_{q}}=V_{p}}
\no\\ 
&&\int_0^1 ds_1...\int_0^1 ds_q\; 
 \frac{\partial  }{\partial s_1} C_{i_1j_1}(s_1)\; \frac{\partial
    }{\partial s_2}C_{i_2j_2}(s_1,s_2) \;...\;  \no\\ 
&&\quad \frac{\partial  }{\partial s_q} C_{i_qj_q}(s_1,s_2,...,s_q)
\;\; \frac{\partial^q F_1(s_1,s_2,...s_q) }
{\partial C_{i_1j_1}\partial C_{i_2j_2}...\partial C_{i_qj_q} } 
\eqa

To extract the exact expression for the derived propagators
we introduce some notations. We call $l(V)$ the number of loop fields
hooked to the vertices in $V$, $W_i$ the set of vertices
from where the line $l_i$ may start from,  and $m_i$ the number of
loop fields where $l_i$ may contract without crossing more than
one arch:
\bqa
W_i &=& V_1\cup V_2\cup...\cup V_{k_{i-1}}\no\\
m_i &=& l(W_i\backslash W_{i-1})
\eqa
Remark thar $m_1=l(V_1)$ and 
$m_2=l(V_1\cup V_2\cup...\cup V_{k_1}\backslash V_1)$. Now we observe that
the interpolated propagator is given by
\be
C_{i_rj_r}(s_1,...s_{r}) = s_{r'+1} s_{r'+2}... s_{r} C_{i_rj_r}
\ee 
if  $i_r\in m_{r'}$. Then the derivative just takes away the
factor $s_{r}$. Remark that the remaining determinant 
satisfies all properties of the initial one, in particular,
as the interpolation respects positivity, 
a Gram inequality  can be applied.

Hence the functional $F$ has been developed as follows
\be
F(\{C_{ij}\}) = \sum_{\cal L F} \prod_{I\in{\cal L F} } A_I
\ee
where ${\cal L F}$ 
is a set of subsets $I$ of the path $\{x_1,..,x_n\}$ which form 1PI clusters and
\be
A_I = \sum_{q=1}^{n_i/2} A(q,n_i)  
\ee
where $n_i\leq n$ is the number of vertices belonging to $I$
and $q$ is the number of loop lines ensuring 1PI. 
Now we want to prove that $A_I\leq K$. As the functional and 
the propagators can always be bounded by a constant, the problem 
is to prove the following lemma:

\begin{lemma}
The sum over all possible arch systems that connect $p$ points in such a way
to obtain a 1PI block does not develop a factorial, in other words:
\be
\sum_{q=1}^{p}\; \sum_{1<k_{1}<...<k_{q}=p}\;
\sum_{j_r \in V_{r}\atop r=1,...,q} 
\int_0^1 ds_1...\int_0^1 ds_q\;  
\sum_{i_r \in W_{r}\atop r=1,...,q}  a(s_1,...,s_q,i_1,...,i_q)\leq K^n
\ee
where $a$  is the function we  obtain after bounding the determinant and
the propagators by a constant.
\end{lemma}
\paragraph{Proof.} 
We start observing that
\be
\sum_{i_r \in W_{r}\atop r=1,...,q}  a(s_1,...,s_q,i_1,...,i_q)\leq 
 \prod_{r=1}^{q} a_{r}(s_{1},...,s_{r-1})  
\ee
where $a_r$ is defined inductively by
$a_{1}= m_{1} $ and
$a_{r}(s_{1},...,s_{r-1}) = m_{r}+s_{r-1} a_{r-1}(s_{1},...,s_{r-2})$.
To see this we remark that  we have $m_1$ choices
to choose $i_1$.  In the same way, we have $m_2$ choices to choose
$i_2$  if it does not hook to $V_1$.  
If it does hook to $V_1$, we have $m_1=a_1$  choices, but we also have a 
factor $s_1$.  Remark that this is an overestimate, as, once fixed $i_1$ 
we have only $m_1-1$ choices for $i_2$.
Hence we have 
\be
\int_{0}^{1} \prod_{1}^{q} ds_{r} \prod_{r=1}^{q} 
a_{r}(s_{1},...,s_{r-1})  \le e^{\sum_{r=1}^{q} m_{r}}.
\ee
This is indeed obvious if we use inductively the fact
that for $a>0, b>0$, $\int_{0}^{1} (as +b ) ds \le (1/a)e^{a +b}$.
Now, as  $m_{r}= l(W_r\backslash W_{r-1})$, we have
\be
\sum_{r=1}^{q} m_{r}\leq  \sum_{i=1}^{p} l(V_i)  < 4n
\ee

Finally we prove that
\be
\sum_{q=1}^{p/2} \;\sum_{1<k_{1}<...<k_{q}=p}\;
\sum_{j_r \in V_{r}\atop r=1,...,q} 1\leq K^n 
\ee
Actually 
\be
\sum_{j_r \in V_{r}\atop r=1,...,q} 1 = \sum_{r=1}^q l(V_{k_r}) < 4n
\ee
and 
$\sum_{1<k_{1}<...<k_{q}=p} 1 $
corresponds to the number of partitions of $\{1,...,p\}$
into $q$ intervals, hence is bounded by $ 2^p \leq 2^n$.
This ends the proof. \qed
\vskip 0.5cm

\paragraph{Selfenergy} Now, we can apply the arch formula to
the two point vertex function and extract the following expression for
the selfenergy.
\bqa
\lefteqn{\Si^{\La}(\phi_1,\phi_{2})=
\sum_{{\bar n}\geq 1} \frac{\la^n}{n!} 
\frac{\lp\de\mu^1_\La\rp^{n'}}{n'!}
 \sum_{CTS}
\sum_{u-\tree}\sum_{\cal L}\sum_{E\Om}\sum_{{\cal C}}\sum_{J,P}
\sum_{L_e} }
\\
&&
\vep(\tree, \Om)\int_{w_T\le w_{{\cal A}(i)} \le  w_{i}\le 1}
\prod_{q=1}^{{\bar n}-1} dw_q
\int_{0}^1
\prod_{q=1}^{|L_e|} ds_q 
\no\\
&&\prod_{h\in L\cup\tree_L\cup E}
\left\{ [{\scriptstyle {4 \over 3} \La^{-{1\over 2}}(w_{j_{h,n_h}})}]
\int_0^{2\pi}  d\th_{h,n_h}
[ {\scriptstyle{4 \over 3}\La^{-{1\over 2}}(w_{j_{h,n_h-1}})}]
\int_{\Si_{j_{h,n_h}}}
d\th_{h,n_h-1} \right .\no\\
&&...\;\;[{\scriptstyle {4 \over 3}\La^{-{1\over 2}}(w_{j_{h,1}})}]
\left .\int_{\Si_{j_{h,2}}}
d\th_{h,1}
\left [\prod_{r=2}^{n_{h}}
\chi^{\th_{h,r}}_{\al_{j_{h,r}}}(\th_{h,1})\right ]
 \right \}\no\\
&& \prod_{g_i|\ i=r\ {\rm or}\atop |eg_i({\cal C}|)\leq 8}
\Upsilon\lp\th_i^{root},\{\th_{h,r(i)}\}_{h\in eg^\ast_i}
 \rp \;
\prod_{v\in V\cup V'}  \Upsilon\lp\th_{h_v^{root}},
\{\th_{h,n_h}\}_{ h\in H^\ast(v)} \rp\no\\
&&\int
d^3x_1...d^3x_{\bar n}\;\;
\phi^{\La_T}_1(x_{i_{1}},\th_{e_1,1})\;...\;
\phi^{\La_T}_{2p}(x_{j_{p}},\th_{e_{2p},1})\no\\
&&\left[\prod_{q=1}^{{\bar n}-1}  C^{w_q}
( x_{q}, {\bar x}_{q}, \th_{h,1}) \right]
 \left [ \prod_{l_{fg}\in P\cup L_e}
{\cal M}_{f,g}'({\cal C},E ,\{\th_{a,1}\},\{s_q\} )\right ] \no\\
&&
\left [  \prod_{ j_q\in J^1} 
\int_0^1 ds_{j_q} \right ] 
\det {\cal M}'({\cal C},E,\{\th_{a,1}\},\{s_{j_q}\}, \{z_q\} ) \ ,
\no\eqa
where we took $\phi_1(x)= \de(x)$ and 
   $\phi_2(x)= e^{-ixk}$, to obtain $\hat{\Si}(k)$. 
$L_e$ is the set of loop lines
extracted to ensure 1PI,  $s_q$ is the set of 
interpolation parameters used to extract them while
$P$ is the set of loop lines extracted in Sec.II.1.  With this expression
we can perform the same bound as for the vertex function $\Ga$, as
the additional sums do not generate any factorial.
The only difference is that, when performing sector counting,
the real external impulsion is not always near the Fermi surface.
This does not change the counting lemmas, as this impulsion
is fixed with a precision $T$.

\vskip 1cm

\centerline{\large {\bf B2:  First Derivative of the Selfenergy at the Fermi
Surface}}
\vskip 0.5cm

The bound on the first derivative below already proves
that our system is not a Luttinger liquid [S1][BGPS][BM].

We want to prove that the first order derivative of the selfenergy
computed at the impulsion $k_F$ is bounded by
\be
\left | \partial_{k_\al} \hat{\Si}(k)_{\big | k_F} \right | \leq |\la|^2 \ M_1
\ee
for $\al=0,1,2$ and for all $\la$ and $T$ satisfying
$|\la\ln T|\leq M_0$, where $M_0$ and $M_1$ are some constants.   
The derivative actually corresponds to the 
multiplication by a factor $x-y$ in position space
\be
\partial_{k_\al} \hat{\Si}(k)_{\big | k_F} = \int d^3x \; e^{ik(x-y)} \de(x) 
(x-y)_\al \Si(x,y)
\ee
Then we can perform power counting as usual, the only difference being 
an additional factor $1/\La(w_{t(r)})$, where
$t(r)$ is the band index of the lowest tree line in the path joining
the two external points $x$ and $y$. Nevertheless, as the 
two point function $\Si$ itself is not renormalized, the factor
$\La^{1\over 2}(w_{t(r)})$ coming from loop contractions is
not consumed. Then we are left with the factor 
\be
{1\over \La^{1\over 2}(w_{t(r)})} = 
\lp\prod_{j\in C_{t(i)}} \bt_j^{1\over 4}\rp  
{1\over \La^{1\over 2}_T} 
\ee
We remark that, by 1PI, all $j\in C_{t(r)}$, except for the last one
$j=r$, correspond to a subgraph with at least four external legs. 
Then a factor $\bt_j^{1\over 4}$ just makes their power counting even more 
convergent. The last subgraph gives
\be
\La_T \int_{\La_T^2}^1 d\bt_r \; 
\bt_r^{-1-{1\over 2}} \; \bt_r^{1\over 4} {1\over \La^{1\over 2}_T} 
\leq K 
\ee
Hence the derivative is bounded by 
\be
\left | \partial_{k_\al} 
\hat{\Si}(k) \right |_{k_F} \leq 
\sum_{{\bar n}=2}^{\infty} \lp K|\la|^{\bar n}|\ln T|\rp^{{\bar n}-2}
\leq |\la|^2 M_1 \label{fderiv}
\ee
for $|\la| |\ln T|\leq 1/K = M_0$. The extraction of two coupling 
constants from the sum does not affect the convergence as 
there are at most ${\bar n}-2$ subgraphs logarithmic divergent.
Actually, there are  ${\bar n}-1$ subgraphs, and one of them, 
$g_r$ does not give a logarithm, as shown in the equation above.
This ends the proof. \qed

\vskip 1cm

\noindent{\large {\bf B3: Second Derivative of the Selfenergy}}
\vskip 0.5cm

The bound on the second derivative is the one which proves really 
``Fermi liquid behavior'' [S1].

We want to prove that the second order derivative of the selfenergy
computed at any impulsion $k$ is bounded by
\be
\left |\partial_{k_\al}  \partial_{k_\bt} 
\hat{\Si}(k) \right | \leq  \ M_3
\ee
for $\al,\bt=0,1,2$ and for all $\la$ and $T$ satisfying
$|\la\ln T|\leq M_0$, where $M_0$ and $M_3$ are some constants. 
Applying a double  derivative in impulsion space corresponds to  
multiply by a factor $|x-y|^2$ in position space
\be
\partial_{k_\al}\partial_{k_\bt} \hat{\Si}(k) = 
\int d^3x \; e^{ik(x-y)} \de(x) 
(x-y)_\al (x-y)_\bt \Si(x,y)
\ee
This time we have the bad factor $\La^{-2}(w_{t(r)})$, then the 
factor $\La^{1\over 2}(w_{t(r)})$ extracted from sector sum is not enough 
to assure the bound. Actually we need to extract a second 
factor  $\La^{1\over 2}(w_{t(r)})$. 
It turns out that this is almost possible but not quite.
One can  only extract  
$\La^{1\over 2}(w_{t(r)}) \left |\ln\La(w_{t(r)})\right |$, using
the so-called ``volume effect''. This explains the absence of 
any $\la$ in the final bound (\ref{fbound}). 
Indeed the second $\la$ is also consumend 
since $g_r$ becomes logarithmic.

\paragraph{Extracting a second factor $\La^{1\over 2}(w_{t(r)})$.}

\begin{figure}
\centerline{\psfig{figure=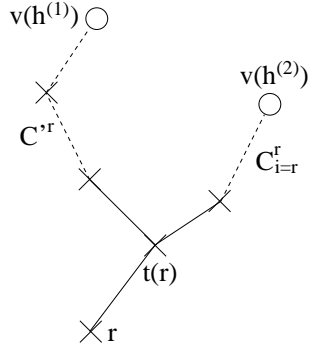,width=4cm}}
\caption{example of ${C'}^{r}$.}\label{selfen1}
\end{figure}

When extracting loop lines in Sec.II.1, we introduced the chain $C_i^r$
($i=r$ in this case),
joining the dot vertex  $v_{h^{(2)}}$ to
the cross vertex just above $t(r)$ (see Fig.\ref{fermi9}),
where $h^{(1)}$ is the root external half-line of
the self-energy and $h^{(2)}$ is the other one. 
Now we introduce the equivalent 
chain ${C'}^{r}$ for  $h^{(1)}$,  joining 
the dot vertex  $v_{h^{(1)}}$ to
the cross vertex just above $t(r)$ (see Fig.\ref{selfen1}),

Remark that, for all $j\in {C'}^{r}$ we must have 
$|eg_j({\cal C})|\geq 4$, by 1PI. As for $C^r$ we call $J'_r$
the set of  $g_j$ in the chain  ${C'}^{r}$ with $|eg_j({\cal C})|=4$,
ordered from the lowest and going up. For each $g_j$, with   $j\in J'_r$,
we call $h_j^{(1)}$ the real external half-line $h^{(1)}$, 
$h_j^{(2)}$ the tree half-line going towards the external
vertex  $v_{h^{(2)}}$, $h_j^{(3)}$ and $h_j^{(4)}$ the two remaining
external half-lines. The tree line $l^{(2)}_j$ (corresponding to 
$h_j^{(2)}$) cuts the tree in two connected components.
By analogy with the definitions in Sec.II.1, we call    
${\tree'}^L_{j}$ the component containing the vertex
 $v_{h^{(1)}}$, and  ${\tree'}^R_{j}$ the component containing the vertex
 $v_{h^{(2)}}$. Remark that $g_j$ belongs to 
  ${\tree'}^L_{j}$. Finally we call ${L'}^L_{j_q}$ the set of internal 
loop half-lines of $g_{j_{q-1}}$, hooked to   ${\tree'}^L_{j_q}$,
that may connect somewhere in  ${\tree'}^R_{j_q}$, and have not
been already contracted. In the same way
we introduce  ${L'}^R_{j_q}$.  

Now we extract the factor we need inductively for $j\in {C'}^{r}$. 
For each subgraph $g_j$ we distinguish two situations:

\begin{itemize}
\item{} if $|eg_j({\cal C})|>4$ we extract the factor 
${ \La^{1\over 2}(w_{{\cal A}(j)})\over\La^{1\over 2}(w_{j})}$
from the convergent power counting of $g_j$, and we pass to 
the cross above in the chain;
\item{}  if $|eg_j({\cal C})|=4$ we test the number of loop lines 
connecting ${\tree'}^L_{j_q}$ to ${\tree'}^R_{j_q}$.
Remark that this number must be always even, and cannot be zero
by 1PI. If there are two loop lines, we know, by Lemma 9, that
a finer estimation of the sector volumes gives an additional
factor  
$\La^{1\over 2}(w_{{\cal A}(j)})\left |\ln \La(w_{{\cal A}(j)})\right |$.
Then we stop the induction. If there are four or more, we 
observe (Lemma 10) that  we have counted one unnecessary sum on sector
choices and we gain the factor
${ \La^{1\over 2}(w_{{\cal A}(j)})\over\La^{1\over 2}(w_{j})}$. Then we
pass to the following cross in the chain.
\end{itemize}
Putting together all these terms we obtain the factor we wanted times
a logarithm.
\be
\La^{1\over 2}(w_{t(r)})\left | \ln \La(w_{t(r)})\right |
\label{sectorc}
\ee

\paragraph{Extracting loop lines.}
We consider the four point subgraph $g_{j}$  on the chain.
We distinguish three situations.

\paragraph{ 1.} If $h^{(3)}_{j}$ and $h^{(4)}_{j}$ are both loop
half-lines we contract them developing the determinant
(see eq.(\ref{loop1})). As in Sec.II.1,
the number of choices is bounded by $|{L'}^{R}_{j}|^2$.

\paragraph{ 2.} If $h^{(3)}_{j}$ is a tree half-line and  
$h^{(4)}_{j}$ is a loop one, we contract $h^{(4)}_{j}$ by 
developing the determinant. If it contracts to 
 ${\tree'}^R_{j}$, then we have to extract, applying
several times  the formula eq.(\ref{test1}-\ref{test2}), one or three
loop lines
joining  ${\tree'}^L_{j}$ with  ${\tree'}^R_{j}$
 (depending if there are two or more loop lines joining 
 ${\tree'}^L_{j}$ with  ${\tree'}^R_{j}$).
If  $h^{(4)}_{j}$  contracts to 
 ${\tree'}^L_{j}$, then we have to extract  two or four additional
loop lines (depending if there are two or more loop lines
joining  ${\tree'}^L_{j}$ with  ${\tree'}^R_{j}$).
In any case the number of choices is bounded by 
$|{L'}^R_{j}|^4|{L'}^L_{j}|^5$.

\begin{figure}
\centerline{\psfig{figure=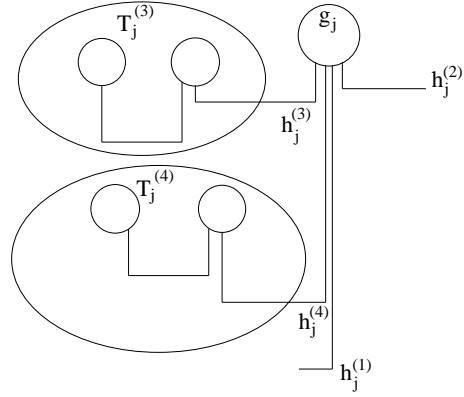,width=6cm}}
\caption{example of  ${T'}^{(3)}_{j}$ and ${T'}^{(4)}_{j}$}\label{selfen3}
\end{figure}

\paragraph{ 3.} If $h^{(3)}_{j}$ and $h^{(4)}_{j}$ are both tree
half-lines, then we  call ${T'}^{(3)}_{j}$ the 
subtree connected to $g_{j}$ through  $h^{(3)}_{j}$,
and   ${T'}^{(4)}_{j}$ the 
one connected to $g_{j}$ through  $h^{(4)}_{j}$ (see Fig.\ref{selfen3}).
In the same way we define   ${L'}^{(3)}_{j}$ and  ${L'}^{(4)}_{j}$
(${L'}^L_{j}= {L'}^{(3)}_{j}\cup {L'}^{(4)}_{j}$). 
Then we apply eq.(\ref{test1}-\ref{test2}) several times, until we extract
two or four loop lines joining  ${\tree'}^L_{j}$ with  ${\tree'}^R_{j}$.
Finally, if there are four loop lines we perform an additional analysis.
If  four or two loop lines extracted  are hooked to   ${T'}^{(3)}_{j}$,
then we apply eq.(\ref{test1}) once more to extract a loop line
joining ${T'}^{(4)}_{j}$ to ${T'}^{(3)}_{j}$ or to ${\tree'}^R_{j}$
(there must be one by 1PI, and by the parity of the number of
external lines).  In any case the number of choices is bounded
by  $|{L'}^{(3)}_{j}|^5 |{L'}^{(4)}_{j}|^5|{L'}^L_{j}|^5$. 

\paragraph{Number of choices.} Applying Lemma 1 and 2 
we see that the remaining 
determinant still satifies a Gram inequality, and the number of choices
to extract the loop lines is bounded by $K^{\bar n}$.

Now we prove the following lemmas.

\begin{lemma}
Let $g_j$ be a four point subgraph on the chain ${C'}^{r}$. If
there are  only two loop lines $l_{lj}^{(1)}$ and $l_{lj}^{(2)}$,  
connecting  ${\tree'}^R_{j}$ to  ${\tree'}^L_{j}$, 
then the power counting has an additional
 volume factor
$\La^{1\over 2}(w_{{\cal A}(j)}) \left |\ln \La(w_{{\cal A}(j)})\right |$.
\end{lemma}

\begin{figure}
\centerline{\psfig{figure=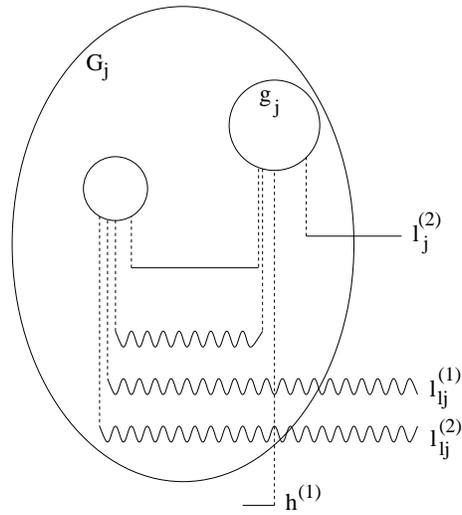,width=6cm}}
\caption{Example of a non quasi-local four point subgraph $G_j$.}
\label{selfen2}
\end{figure}

\paragraph{Proof}
As  there are only two loop lines , then  ${\tree'}^L_{j}$
actually is a four point subgraph $G_j$ (but not necessarily quasi-local)
with external lines $l_1(G_j)=h_j^{(1)}$,  $l_2(G_j)=l_j^{(2)}$,  
$l_3(G_j)=l_{lj}^{(1)}$ and $l_4(G_j)=l_{lj}^{(2)}$. 
For an example see Fig.\ref{selfen2}.

Now we refine the sectors of  $l_i(G_j)$, $i=1,...,4$
 (that may be of different sizes
$\La^{1\over 2}(l(G_j))\leq \La^{1\over 2}(w_{{\cal A}(j)})$)
in smaller sectors multiplying their size by
$\La^{1\over 2}(w_{{\cal A}(j)})$. Remark that, when 
$\La^{1\over 2}(l_i(G_j))= \La^{1\over 2}(w_{{\cal A}(j)})$ $\forall i$,
this means we are passing to isotropic sectors.
By Lemma 2, Sec.II, in [FMRT]  the new sector sum costs 
\be
 \frac{1}{\La^{1\over 2}(w_{{\cal A}(j)})} 
\left |\ln \La(w_{{\cal A}(j)})\right |
\ee 
Remark that the bad  factor  
$\La^{-{1\over 2}}(w_{{\cal A}(j)})$ coming from the 
worse spatial decay of the tree line is compensated by the good factor
$\La^{1\over 2}(w_{{\cal A}(j)})$ from the smaller 
volume in impulsion space.
Loop lines are not used for spatial integration, 
therefore  their smaller volume factor in impulsion space
is not consumed. Ech loop line 
gives therefore a net bounus 
$\La^{1\over 2}(w_{{\cal A}(j_q)})$. Finally we are left with the
factor
\be
  \La(w_{{\cal A}(j_q)}) \La^{-{1\over 2}}(w_{{\cal A}(j_q)}) 
\; \left | \ln \La(w_{{\cal A}(j_q)}) \right |= 
\La^{1\over 2}(w_{{\cal A}(j_q)})\; \left |\ln \La(w_{{\cal A}(j_q)})
\right |
\ee
\qed

When there are at least 
four loop lines joining ${\tree'}^L_{j_q}$ with  ${\tree'}^R_{j_q}$,
 the following lemma proves that
 we have paid one unnecessary sector refinement. 

\begin{lemma}
Let the four point subgraph $g_{j}$ on the chain ${C'}^{r}$ and 
the four loop lines
$l_{lj}^{1}$, ...$l_{lj}^{4}$, joining  ${\tree'}^L_{j}$ with
 ${\tree'}^R_{j}$, be fixed. 
Then the number of sector choices predicted by [DR1], Lemma 6, (IV.31)  
must be modified:
\be
\prod_{m=2}^4
\left [{\scriptstyle {4\over 3}\La^{-{1\over 2}}\lp w_{{\cal A}(j)}\rp}\right ]
\int_{\Si_{j_{h^{(m)}_{j},r(j)+1}}}
d\th_{h_j^{(m)}}
\Upsilon\lp\th_j^{(root)},\{\th_{h_j^{(m)},r(j)}\}_{m=2,3,4}\rp \leq K
\ee
for some constant $K$. 
\end{lemma}
\paragraph{Proof}
The proof is quite similar to that of Lemma 4, but slightly more complicated,
as this time the sector of the external tree line on the path joining 
$x^{(1)}$ with $x^{(2)}$ must be summed.
We distinguish two cases.

\begin{figure}
\centerline{\psfig{figure=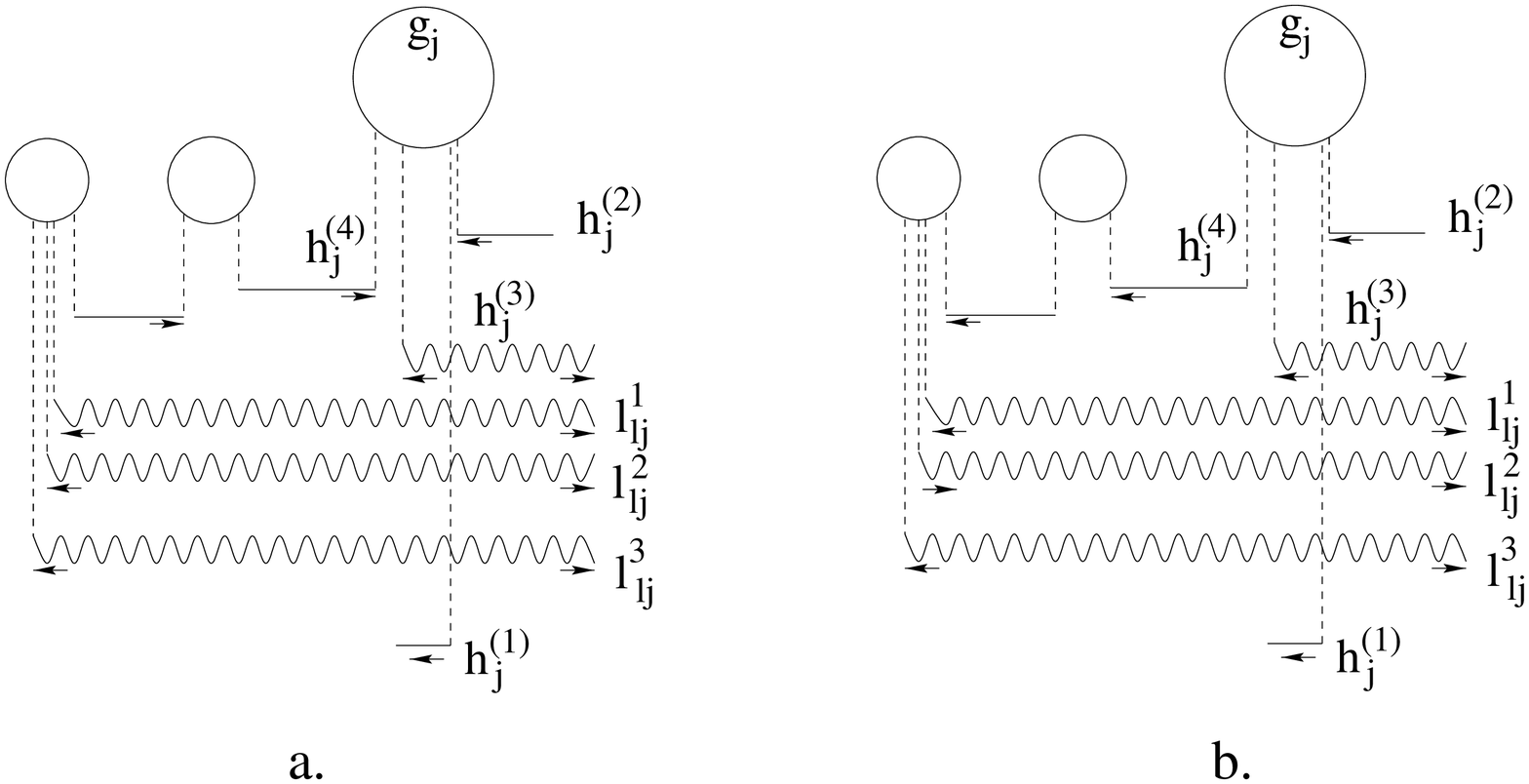,width=10cm}}
\caption{a. shows the usual sector counting and b. shows the new sector
  counting taking $l_{lj}^{2}$ as root.}\label{selfen4}
\end{figure}

\paragraph{1.} 
If  $h_{j}^{(3)}$ is a loop half-line
and  $h_{j}^{(4)}$  is a tree one,
we know there are at least three loop lines (different
from  $h_{j}^{(3)}$), called $l_{lj}^{1}$, $l_{lj}^{2}$, $l_{lj}^{3}$,   
joining 
${\tree'}^R_{j}$ to ${\tree'}^L_{j}$ 
(see Fig.\ref{selfen4}a). One of these lines, say $l_{lj}^{1}$,  may have been
used to gain a sector sum on the other chain $C^r$ 
(in Lemma \ref{anotherlemma}).
At least one of the two remaining loop lines, say $l_{lj}^{2}$,  
 has  been paid in refining sectors. This was not necessary, as
its  sector is fixed  by impulsion 
conservation along the loop line.
Then this line  can be chosen as a new root to perform sector counting 
(see Fig. \ref{selfen4}b).
This permits to fix the sector of $h_{j}^{(4)}$.
 The sector of $h_{j}^{(3)}$ is fixed by impulsion 
conservation along the loop line.

\begin{figure}
\centerline{\psfig{figure=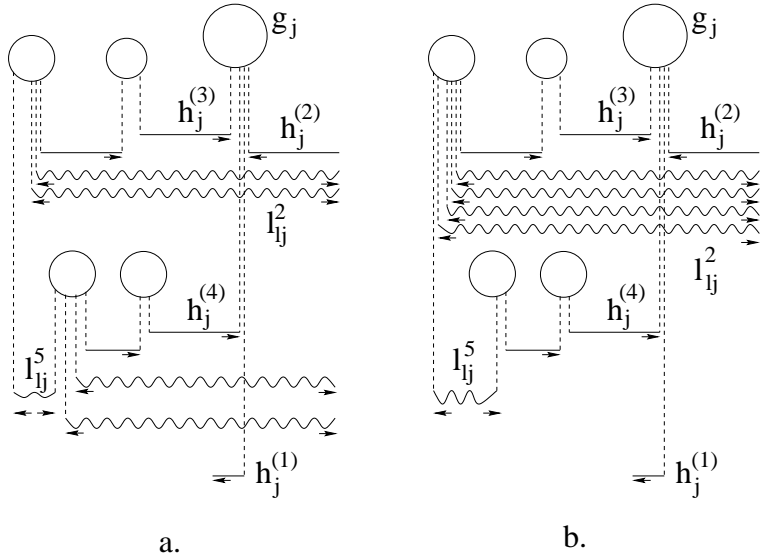,width=10cm}}
\caption{a.:  two loop lines hook to  ${T'}^{(3)}_j$
and two to  ${T'}^{(4)}_j$. b.: four loop lines hook to  ${T'}^{(3)}_j$
and none to  ${T'}^{(4)}_j$. }\label{selfen5}
\end{figure}

\paragraph{2.}    
If  $h_{j}^{(3)}$ and  $h_{j}^{(4)}$  are both 
tree half-lines, then we know there are   
four loop lines $l_{lj}^{i}$, $i=1,..,4$, 
 connecting  ${\tree'}^L_{j}$ with ${\tree'}^R_{j}$. Now we have three 
situations, shown on Fig.\ref{selfen5}.

\begin{figure}
\centerline{\psfig{figure=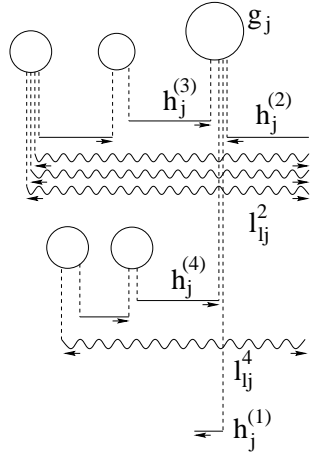,width=4cm}}
\caption{Three loop lines hook to  ${T'}^{(3)}_j$
and only one to  ${T'}^{(4)}_j$ }\label{selfen6}
\end{figure}

\begin{itemize}
\item{} $l_{lj}^{1}$,  $l_{lj}^{2}$ and $l_{lj}^{3}$
are  hooked to ${T'}^{(3)}_j$ and only $l_{lj}^{4}$ is hooked 
to ${T'}^{(4)}_j$ (see Fig.\ref{selfen6}). One of the first three lines,
 say $l_{lj}^{1}$,  may have been
used to gain a sector sum on the other chain $C^r$, then,
among  $l_{lj}^{2}$ and $l_{lj}^{3}$) we choose
as new root for sector counting in ${T'}^{(3)}_j$ the one that has been 
summed in usual sector refinement. 
In ${T'}^{(4)}_j$ we choose as new root the unique loop
line $l_{lj}^{4}$. 

\item{} $l_{lj}^{1}$ and  $l_{lj}^{2}$ are  hooked to ${T'}^{(3)}_j$ 
while   $l_{lj}^{3}$ and $l_{lj}^{4}$ are hooked 
to ${T'}^{(4)}_j$ (see Fig.\ref{selfen5}a).  Then there is a fifth loop line 
$l_{lj}^{5}$ hooked to ${T'}^{(4)}_j$, and we repeat the same argument
above, exchanging ${T'}^{(4)}_j$ with ${T'}^{(3)}_j$. 

\item{} All the four loop lines are hooked to ${T'}^{(3)}_j$ 
(see Fig.\ref{selfen5}b).
 Then there is a fifth loop line 
$l_{lj}^{5}$ hooked to ${T'}^{(4)}_j$, and we repeat the  same argument
above.
\end{itemize}

In any case, the two sectors, for $h_{j}^{(3)}$ and $h_{j}^{(4)}$, are 
fixed. Then,  as the sector of $h^{(1)}$ is always fixed, three sectors
are known, hence the fourth one too and there is no sector refinement to pay. 
Then we gain the factor
${ \La^{1\over 2}(w_{{\cal A}(j)})\over\La^{1\over 2}(w_{j})}$ and
iterate the process.
\qed
\medskip

\paragraph{Final bound.} Inserting all these results, and
performing power counting we find the bound
 \be
\left |\partial_{k_\al}  \partial_{k_\bt} 
\hat{\Si}(k) \right | \leq  \sum_{{\bar n}=2}^\infty 
\lp |\la\ln T|\rp^{\bar n} \leq M_3
\label{fbound}\ee
for $|\la\ln T|\le 1/K=M_3$. Remark that there is no factor $\la^2$
as in eq.(\ref{fderiv}), as there are two additional logarithms,
 coming  one from the power counting of the subgraph $g_r$, and the other
from the bound eq.(\ref{sectorc}) on sector counting. 
\vskip 1cm

\medskip
\noindent{\bf Acknowledgements}
\medskip

We thank V. Mastropietro for a careful reading of this paper
which has lead us to add the appendix on the flow of $\de\mu$.
We are also extremely grateful to M. Salmhofer who explained to us
the importance of proving the bounds on the derivatives of the self energy
to distinguish between Fermi and Luttinger liquids. Finally we thank
J. Magnen for useful discussions.

\medskip
\noindent{\large{\bf References}}
\medskip

\vskip.1cm

\noindent [BGPS] G.Benfatto,  G.Gallavotti, A.Procacci, B.Scoppola:
 {\em Commun. Math. Phys.} {\bf 160}, 93 (1994).

\noindent [BM] F.Bonetto, V.Mastropietro:
{\em Commun. Math. Phys.} {\bf 172}, 57 (1995).

\noindent [DR1] M. Disertori and V. Rivasseau, 
Interacting Fermi liquid in two dimensions
at finite temperature, Part I: Convergent Attributions,
to appear.
\vskip.1cm

\noindent [DR2] M. Disertori and V. Rivasseau,
Continuous Constructive Fermionic Renormalization, preprint (1998),
to appear in Annales Henri Poincar{\'e}.
\vskip.1cm

\noindent[FMRS] J. Feldman, J. Magnen, V. Rivasseau and R. S{\'e}n{\'e}or,
Construction of infrared $\phi_4^4$ by a phase space expansion, 
Comm. Math. Phys. {\bf 109}, 437 (1987).
\vskip.1cm

\noindent [FMRT] J. Feldman, J. Magnen, V. Rivasseau and E. Trubowitz,
An infinite Volume Expansion for Many Fermion Green's Functions,
Helv. Phys. Acta {\bf 65}.
(1992) 679
\vskip.1cm

\noindent [FT1]  J. Feldman and E. Trubowitz, 
Perturbation theory for Many Fermion
Systems, Helv. Phys. Acta {\bf 63} (1991) 156.
\vskip.1cm

\noindent [FT2]  J. Feldman and E. Trubowitz, The flow of an Electron-Phonon
System to the Superconducting State, Helv. Phys. Acta {\bf 64}
(1991) 213.
\vskip.1cm

\noindent [GK] K. Gawedzki, A. Kupiainen, Massless $\phi_4^4$ theory:
rigorous control of a renormalizable asymptotically free model, 
Comm. Math. Phys. {\bf 99}, 197 (1985). 

\noindent [R] V. Rivasseau, From perturbative to constructive renormalization,
Princeton University Press (1991).
\vskip.1cm

\noindent [S1] M. Salmhofer,
Continuous renormalization for Fermions and Fermi liquid theory,
Commun. Math. Phys.{\bf 194}, 249 (1998).


\end{document}